\documentclass[12pt]{elsarticle}

\makeatletter
\def\ps@pprintTitle{%
	\let\@oddhead\@empty
	\let\@evenhead\@empty
	\let\@oddfoot\@empty
	\let\@evenfoot\@oddfoot
}
\makeatother

\usepackage[english]{babel}
\usepackage[margin=1in]{geometry}
\usepackage{graphicx}
\usepackage{tabto}
\usepackage{subcaption}
\usepackage{amsmath}
\usepackage{hyperref}
\usepackage{url}
\usepackage{xcolor}

\journal{Information Sciences}

\begin{document}

	\begin{frontmatter}
		
		
		
		\title{Scaling Laws And Statistical Properties\\of The Transaction Flows And Holding Times of Bitcoin}
		
		
		\author[1]{Didier Sornette \corref{cor1} }
		\author[2,1]{Yu Zhang \corref{cor1} }
		\cortext[cor1]{Co-first and corresponding authors. \\Email address: dsornette@ethz.ch (Didier Sornette), zhangyu@ifi.uzh.ch (Yu Zhang).}
		
		\affiliation[1]{organization={Institute of Risk Analysis, Prediction and Management (Risks-X), Southern University of Science and Technology (SUSTech)},
			city={Shenzhen},
			postcode={518055}, 
			country={China}}
		
		\affiliation[2]{organization={Blockchain and Distributed Ledger Technologies Group, Department of Informatics, University of Zurich},
			city={Zurich},
			postcode={8051}, 
			country={Switzerland}}
		
		\begin{abstract}
			We study the temporal evolution of the holding-time distribution of bitcoins and find that the average distribution of holding-time is a heavy-tailed power law extending from one day to over at least $200$ weeks with an exponent approximately equal to $0.9$, indicating very long memory effects. We also report significant sample-to-sample variations of the distribution of holding times, which can be best characterized as multiscaling, with power-law exponents varying between $0.3$ and $2.5$ depending on bitcoin price regimes. We document significant differences between the distributions of book-to-market and of realized returns, showing that traders obtain far from optimal performance. We also report strong direct qualitative and quantitative evidence of the disposition effect in the Bitcoin Blockchain data. Defining age-dependent transaction flows as the fraction of bitcoins that are traded at a given time and that were born (last traded) at some specific earlier time, we document that the time-averaged transaction flow fraction has a power law dependence as a function of age, with an exponent close to $-1.5$, a value compatible with priority queuing theory. We document the existence of multifractality on the measure defined as the normalized number of bitcoins exchanged at a given time.
		\end{abstract}

		
	\end{frontmatter}
	
	
	
	\vskip 1cm
	
	\section{Introduction}
	
	Financial economics is often considered to be blessed with an abundance of data, which include
	market data (prices, volumes, and the historical performance of stocks, bonds, and other securities at many time scales including tick level),
	macroeconomic data (broad economic indicators like GDP, inflation rates, employment figures, and interest rates),
	fundamental data (company-specific information such as earnings, balance sheets, and profit margins),
	alternative data (social media sentiment, weather patterns, or satellite imagery),
	qualitative data (management quality, brand strength, or market trends), and so on.
	However, these data are grossly incomplete. Consider transaction data. In stock markets, the full history of 
	all transactions is not publicly available. While aggregate data like trading volumes, closing prices, and market capitalization 
	are accessible, the detailed history of who bought or sold, how much, and when, is typically not.
	This has limited the possibility for researchers to understand the decision-making processes and
	the underlying mechanisms at the origin of price formation and financial investment properties.
	
	The introduction of Blockchain technology represents a pivotal transformation, since the 
	complete transaction history is recorded in the Blockchain.
	Indeed, Blockchain technology, as exemplified by Bitcoin, provides a publicly accessible, 
	complete record of all transactions ever made. This level of transparency allows for a comprehensive 
	analysis of transaction flows and investor behaviors over time.
	In blockchain systems, each unit of the asset (a bitcoin) can be traced through its entire transaction history, 
	providing insights into how it has been used, moved, and held over time.

	Building on the fact that all atomic transaction data is recorded on the blockchain, we use the on-chain data 
	to analyze the behavior of investors in the Bitcoin blockchain by tracing their transaction data, 
	such as transacting time, transacting amount, holding times, and so on.
	Such information is not accessible in any other financial market. Combining with bitcoin's price data allows us 
	to explore the behavior patterns of investors. 
	
	Bitcoin and other so-called cryptocurrencies have 
	evolved into a distinct asset class for financial investors. Investors now see cryptocurrencies as a means to diversify portfolios, 
	hedge against market volatility, and gain exposure to new technological developments like blockchain. 
	This shift is driven by the unique characteristics of cryptocurrencies: their decentralized nature (at least in theory), the potential for high returns, 
	and their increasing integration into the broader financial landscape. However, this comes with high volatility and regulatory uncertainties, 
	making them a speculative and high-risk investment option. Their progressive adoption by institutional investors and inclusion 
	in investment funds further legitimize their status as an emerging asset class. Therefore, the insights that we obtain here 
	are likely to also apply to other financial markets. 
	
	The paper is organized as follows. Section 2 presents a simple mathematical formulation in terms of the number of bitcoins last exchanged
	at a specific previous time that are still held at a later given time. This quantity is the building block from which everything derives.
	Section 3 describes our methodology based on how Bitcoin data is organized on its blockchain, how Bitcoin transactions run, and how these transaction data 
	can be used to extract variables useful to explore the behavior of bitcoin traders.
	Section 4 analyzes the age (or holding time) distribution for six distinct time points, corresponding to three distinct bubble peaks and three subsequent crash troughs.
	Section 5 compares the distributions of Book-to-Market returns and of realized returns.
	Section 6 examines the dependence of age-dependent transaction flows on holding times. 
	Section 7 introduces a measure, the normalized number of bitcoins exchanged at a given time, and shows that it has multifractal properties.
	Section 8 concludes by summarising our main findings.

	\section{Mathematical Formulation}\label{sec:model}
	
	At time $t$ (present time), we consider all bitcoins (BTC) and give each one a timestamp defined as the time $\tau \leq t$ of the last transaction. Then, we can interpret $\tau$ as the ``birth date'' of the corresponding bitcoins (or fraction of BTC)
	from the point of view of the new owner and $t-\tau$ as its age or holding time. At the next timestep $t+\Delta t$, some BTC (or fraction of BTC) will ``die'', i.e. they will change hands. In our representation, this transaction corresponds to the ``death'' of the corresponding exchanged amount of BTC and its immediate rebirth, amounting to putting back the age counter to $0$.
	
	We perform our analysis in discrete times and  define the following quantities and expressions:
	\begin{enumerate}
		\item $\Delta t$: Discrete time interval, where we take $\Delta t=1$ for simplicity. It can be 10 minutes, one day, or several days and will be specified below.
		\vspace{1.5mm}
		
		\item $n_{\tau}(t)$: The amount of BTC last exchanged at the specific time $ \tau<t $ that are still held at $ t $ (i.e not  yet sold again).
		\vspace{1.5mm}
		
		\item $N(t)$: The total number of BTC bought at any time up to $ t $ that are still held at $ t $ (i.e. not yet sold again). It is also the total number of bitcoins mined up to time $ t $. \vspace{1.5mm}
		
		\item $\pi_{\tau\rightarrow t}(t) $:  The probability that a bitcoin, which is born at $ \tau<t $ and held until $t-\Delta t$, is exchanged at time $ t $ (i.e., in discrete time, this is the probability for a bitcoin to be exchanged during the interval $[t-\Delta t, t]$). By the frequentist interpretation of probability, it is the fraction of bitcoins that are born at $ \tau $, are held until $t-\Delta t$, and then exchanged at time $ t $.  We can refer to  $ \pi_{\tau\rightarrow t}(t) $ as the death rate per (discrete) time interval $ \Delta t $ at time $ t $ of the cohort of bitcoins born at time $ \tau $ and still held at $t-\Delta t$. \vspace{1.5mm}
		
		\item $V(t)$:  The total number of bitcoins that have been transferred at time $ t $, in other words, it is the total transaction volume at time $t$.
		
		\item $S(t)$:  The number of newly mined BTC at time $t$.
	\end{enumerate}
	
	By definition, we have:
	\begin{equation}\label{eq:n-tau-t}
		n_{\tau}(t)=\big(1-\pi_{\tau\rightarrow t}(t)\big)n_{\tau}(t-\Delta t) \quad\quad \forall \tau<t
	\end{equation}
	This equation is just an expression of the number of bitcoins born at time $ \tau $ that ``survive'' from $ t-\Delta t$ to $ t $ with probability $ 1-\pi_{\tau\rightarrow t}(t) $. Observing the quantity $ n_{\tau}(t) $ over all past transaction times $ \tau<t $ yields the age (or holding time) distribution of existing bitcoins at time $t$.
	
	For $\tau = t$, we define $ n_{t}(t) $ as representing all BTC born earlier than $ t $ that die at time $ t $ and are immediately 
	reborn at this time:
	\begin{equation}\label{eq:n-t-t}
		n_t(t)=V(t) + S(t)
	\end{equation}
	where
	\begin{equation}\label{eq:v-t}
		V(t)=\sum_{\tau<t}\pi_{\tau\rightarrow t}(t)n_{\tau}(t-\Delta t)
	\end{equation}
	is the total transaction volume in units of BTC at time $ t $. 
	The total volume $V(t)$ includes all those coins that have been bought and resold once (or more), including those born before $t-\Delta t$ and those born within the most recent interval $[t-\Delta t,t]$.  
	Thus, $ n_t (t) $ is equal to the total number $ V(t) $ of bitcoins that have been transferred at time $ t $ (transaction volume at time $ t $) plus the source term $ S(t) $ of newly mined BTC that are released at time $ t $. 
	
	By definition, the total number of mined coins is given by
	\begin{equation}\label{eq:N-t}
		N(t)=\sum_{t'\leq t}S(t')
	\end{equation}
	The times $ t' $ are the mining times up to $ t $, at which new blocks have been mined (i.e. approximately every 10 minutes by construction). Equation \eqref{eq:N-t} can be recursively rewritten as: 
	\begin{equation}\label{eq:N-t-rec}
		N(t)=N(t-\Delta t)+S(t)
	\end{equation}
	Furthermore, $ N(t) $ can be expressed as
	\begin{equation}\label{eq:N-t-sum}
		N(t)=\sum_{\tau\leq t}n_{\tau}(t)~,
	\end{equation}
	which expresses that the total number of mined bitcoins in the system up to and including time t is the sum over the current age distribution
	of all bitcoins still being held at time $t$.
	
	Equations (\ref{eq:n-tau-t}-\ref{eq:N-t-sum}) express the conservation of existing bitcoins under exchanges and the growth of their number via mining. This excludes taking into account losses of tradeable bitcoins, attributable to dead wallets, i.e. bitcoin ``accounts'' for which the secret access private key was lost and the corresponding ``trapped'' BTC are not tradeable anymore (given the assumption that there is no way to recover the private key accessing these particular bitcoins). 
	
	From Equation \eqref{eq:n-tau-t}, we have:
	\begin{equation}\label{eq:pi-tau-t}
		\pi_{\tau\rightarrow t}(t) = 1 - \dfrac{n_\tau(t)}{n_\tau(t-\Delta t)}
	\end{equation}
	The death rate (or rate of transactions per time interval) $ \pi_{\tau\rightarrow t}(t) $ at time $ t $ of the cohort of bitcoins born at time $ \tau $ is the fundamental quantity of interest, which we want to investigate from the data. Observing the quantity $ \pi_{\tau\rightarrow t}(t) $ over all past buying times $\tau$'s
	for various times $t$ yields the distribution of transition probabilities from time $ \tau$ to these $t$'s for all $\tau$'s.
	
	This mathematical formulation underlines the existence of exact analogies with the epoch-dependent death rate of the human population. We expect $ \pi_{\tau\rightarrow t}(t) $ to depend both on age $ t-\tau $ and on birth time $ \tau $. The dependence on age could be non-monotonous as for humans, where the death rate is relatively high for newborn infants, then decreases until their teenage, remains low, and then increases in an accelerated way into old age \footnote{https://ourworldindata.org/grapher/annual-death-rate-by-age-group}. For bitcoins, newly exchanged bitcoins could be part of a speculative trading process where the owner is attentively following the market dynamics and is eager to cash in. When a bitcoin has been held for quite some time, the owner may divert her attention to other interests and the bitcoin may sit idle for a long while, making its death rate decrease with longevity. The propensity to sell one's bitcoin (and buy) is also
	likely to depend on the instantaneous as well as recent price dynamics, volume and reports on standard and social media. 
	More specifically, one can expect that a bitcoin owner forms a desire to cash in gains that depend on the difference between
	the present price and the price at the time when the bitcoin was bought.
	One can thus expect an influence of the price dynamical structure over the distribution of ages $t-\tau$ onto the propensity to sell. Vice-versa, 
	implemented sell orders will impact the price. 
	These are some of the hypotheses that one could formulate and that can be tested by analyzing the data set.
	
	The dependence on birth time $ \tau $ is likely to reflect an anchoring of the price at which a given amount of bitcoin was exchanged. This behaviour is associated with the so-called disposition effect \cite{shefrin1985disposition}. Thus, $ \pi_{\tau\rightarrow t}(t)$ is likely to reflect the memory of the bitcoin price at time $ \tau $ in comparison with the price at time $ t $. Thus, it should be dependent on the buying and selling 
	prices, $ p(\tau) $ and  $ p(t) $ respectively.  This assumption can be formulated as
	\begin{equation}\label{eq:pi-f-ret}
		\pi_{\tau\rightarrow t}(t)=f(t-\tau, r_{\tau\rightarrow t})
	\end{equation}
	where $ f(\cdot) $ is some empirically determined function and 
	\begin{equation}\label{eq:log-ret}
		r_{\tau\rightarrow t}=\ln\dfrac{p(t)}{p(\tau)}
	\end{equation}
	is the log-return of the price over the period $ [\tau,t] $. Expression (\ref{eq:pi-f-ret}) assumes that buy and sell orders 
	are only dictated by the return $ r_{\tau\rightarrow t}$, in addition to the age $t-\tau$. The intuition is that the transaction decision is based on the potential gain (or loss) of that bitcoin, where the anchoring on the buying price impacts the attractiveness (or repulsion) to realize a profit (or loss). The disposition effect is based on the observation that investors tend to realize profits too early and at small gains while, on the contrary, they realize losses too late and thus incur large losses. On average, this negatively biases the long-term profit of their trades \cite{BarberisWei09}. Origins of this phenomenon are sought in the domain of behavioral economics (prospect theory) \cite{kai1979prospect}, which explains the disposition effect as a psychological trait of investors, who have a propensity to be impatient and risk-averse when holding unrealized profitable positions. In contrast, investors are more risk-takers when facing losses, willing ``to gamble their way out of misery" when holding unrealized losses.
	
	More sophisticated formulations for the transition probability can be considered, for instance
	\begin{equation}\label{eq:pi-f-ret-v}
		\pi_{\tau\rightarrow t}(t)=f(t-\tau, r_{\tau\rightarrow t},V(t))
	\end{equation}
	which is a function of both the return and the volume (Eq.\eqref{eq:v-t}) at time $ t $.
	Generalizing, $ \pi_{\tau\rightarrow t}(t) $ may be a function of not only the prices, time, and volume, but additionally of other state variables at $ \tau $ and $ t $.
	This has to be determined empirically.

	\section{The Bitcoin Blockchain}\label{sec:method}
	
	\subsection{A brief introduction to Bitcoin}
	
	A blockchain is a transparent and decentralized public ledger or database that allows anyone to validate and verify transactions. It is distributed across and maintained by a large number of different nodes (computers) in contrast to it being held by a single authority or party. The goal of the technology behind cryptocurrencies such as Bitcoin is to make peer-to-peer payment (transaction) available without centralized authority. Blockchain miners verify these transactions and bundle them into a block, then broadcast this block to other nodes in the node network. If other nodes confirm the block, then this block will be appended to the end of the blockchain as the latest block. The transaction data on the blockchain is public, which means that everyone can get access to these transaction data without getting permission from any central authority.
	
	In practice, Bitcoin network is much more concentrated, in fact extremely hierarchical, than in theory \cite{nakamoto2008peer} or believed by non-experts.
	Indeed, the network is held by a strong hierarchy, composed of a few leading maintainers, a few dominating mining pools leaders, 
	`whales', a few news outlets and a few financial institutions. This is quite similar to the classic structure:
	maintainers $\to$ lawmakers and government, mining pool leaders $\to$ large companies CEOs/board members, 
	whales $\to$ rich and powerful people, new outlets, and financial institutions.  The maintainers are like ``religious zealots"
	motivated by a solid belief (quasi-religious) system, understood not at 
	an equalitarian system, but rather as a new utopian movement aiming a creating a new class \cite{Mannheim29},
	similarly to the epoch when the industry-owner class took over the land-owner class during the two first industrial revolutions.
	
	There have been many different blockchain platforms so far, like Bitcoin, Ethereum, Monero, Dogecoin, and so on, but Bitcoin is one of the earliest and most successful blockchain platform. On 2018-10-31, the author or group with the pseudonym Satoshi Nakamoto published the white paper entitled `Bitcoin: A Peer-to-Peer Electronic Cash System' \cite{nakamoto2008peer}. Then they built the Bitcoin network in January 2009.  Bitcoin, i.e. the entire Bitcoin network, includes the underlying blockchain technology, the protocol, the decentralized ledger system, and the overall infrastructure that enables the functioning of the digital currency bitcoin. Bitcoin is based on blockchain, a distributed and decentralized ledger technology employing cryptographic hashing, sequentially adding blocks to form immutable chains of bitcoin transactions. At the core of its operation is a consensus mechanism called proof-of-work, whereby individuals with specialized hardware --referred to as `miners'--compete to receive bitcoins from solving complex mathematical puzzles \cite{nakamoto2008peer,Bohmeetal15,Ornes1z9}.
	
	The period extending from the beginning of 2009 to 2010 can be taken as the period of `proof of concept' of Bitcoin. During this period, individuals, especially computer scientists or developers who initially encountered this technology, were only curious about this technology and some might make transactions and mine blocks just for fun, being unaware of the remarkable popularity and resounding success it would ultimately achieve. In May 2010, programming developer Laszlo Hanyecz paid 10,000 bitcoins (BTC) to buy two pizzas, which is the first event in the Bitcoin world built as a relationship with the real world and made people realize that bitcoin may be valuable. For example, Ref.~\cite{di2023rise} found that bitcoin users become suddenly rational in transactions after this event by analyzing the transaction data (the number of transactions in each block, the number of inputs and outputs in each transaction).
	
	In Bitcoin blockchain, anyone can participate in mining activities and transactions by running a node or a light node. The only necessity is to generate the private-public key pairs for these activities by their wallet, which is very easy. People do not go to any central authority, like a bank or exchange, to apply to open an account with their personal documents. Furthermore, anyone can generate as many pairs of private-public keys as desired. The private-public key pair is generated based on the arithmetic operations on an elliptic curve. The private key is randomly generated by the wallet which corresponds to the password of the account. The public key, which is like an account number that can be seen by the public, is generated from the elliptic curve multiplication of private keys. Anyone can generate Bitcoin account without being censored by a third centralized party.
	
	Bitcoin began trading in July 2010, after Laszlo Hanyecz's first order was paid using bitcoins, at the price \$0.0008 which is another historic event because its trading provided burgeoning investing opportunities to make a profit for investors.  Though Bitcoin is called cryptocurrency, it is different from fiat currencies in failing to function well as a medium of exchange, a store of value, and a unit of account \cite{YERMACK201531} because its price fluctuates too much. Bitcoin is different from a commodity currency because it has no intrinsic value. Bitcoin is also different from a firm's stocks because there are no firms' businesses that back up its price. It has been taken as an investment vehicle \cite{hong2017bitcoin} since its appearance in centralized exchanges. For example, by analyzing Bitcoin's on-chain transaction data, Ref.\cite{baur2014bitcoin} finds that bitcoins are used mainly as speculative investments due to the high volatility of its price but not as purchasing goods and services. 
	
	From the beginning of its first trade to the present, bitcoin has experienced several price peaks and troughs. In fact, its price dynamics have been
	characterized by a succession of massive bubbles and crashes (see e.g. \cite{gerlach2019dissection,wheatley2019bitcoin}), 
	which is not surprising given the absence of a firm fundamental value anchor and its resulting highly speculative value.
	For example, at the time of writing, its last peak price was over \$65,000  reached in November 2021 with a corresponding peak market cap of \$1.27 trillion. 
	This peak has been followed by a long ``crypto-winter'' regime, perhaps ending with the appearance of a new bullish regime since the beginning of 2023. Because of its potential profit opportunities and 
	supposed platform inspiring innovation in the domain of payments and other financial products, Bitcoin attracts numerous investors.

	\subsection{Methodology to Analyse Bitcoin Transactions}
	
	On the Bitcoin Blockchain \cite{nakamoto2008peer}, transactions are stored with a recursive dependence. Every single transaction is recorded since the creation of the genesis block by Satoshi Nakamoto in 2009. By August 2022, the total number of transactions already approached 755 million. Transactions are packed and released in so-called blocks one by one according to chronological order, and these sequential blocks form the blockchain that contains the entire ledger history. In this section, we outline the steps required to unravel this encrypted data set and extract the relevant quantities of interest that are described in Section \ref{sec:model}. 
	
	The Bitcoin client ``Bitcoin Core'' summarizes packs of blocks as numbered ``blk0000X.dat''-files\footnote{The blocks stored inside the files are not necessarily ordered. Each block is assigned a hash string and it contains the information about its predecessor's hash string. In this way, the chain of blocks is assembled by matching and ordering blocks according to their hash and predecessor-hash value.}. Each file has a size of 128MiB (``X'' in the name stands for the file number). These encrypted files can be accessed openly by anyone. We downloaded all these ``blk" files that correspond to the history from January 2009 to August 2022 and formed the data set that is analyzed in this paper. These data are already parsed using the library Blocksci (Version 0.7.0) \cite{kalodner2020blocksci}.
	After storing it on our server, we then apply this library to get our desired Bitcoin data. 
	
	
	The individual transactions are recorded in the block body, They are structured and identified according to certain schemes (code scripts) that correspond to different transaction types. The various types of transactions fulfill different functionalities. Currently, the most common transaction type on the Bitcoin blockchain is ``Pay to PubKey Hash'' (P2PKH). In essence, the vast majority of transactions\footnote{There can be special cases of Bitcoin transactions, as the potential design rules are quite unconstrained. Thus, there is a small fraction of exceptions.}
	carry (amongst others) several inputs and outputs, as well as the hash of the transaction, which is referred to as the transaction ID (txid)
	\footnote{A transaction ID (txid), also known as a transaction hash, serves as an identifying number for each transaction within a blockchain. It is a distinct sequence of characters that can be authenticated and included in the blockchain. Once a transaction is appended to the blockchain, it is given its own exclusive transaction ID.}. For each transaction, a set of attributes \{time, inputs, outputs, txid\} can be collected and parsed to human-readable output\footnote{We use the Blocksci to parse the Bitcoin transaction data.}. Appending the list of transactions of each block in the right block order (i.e sorting in time) results in a ``block-free'' ledger. 
	
	Bitcoin transactions carry one or more inputs and outputs. Usually, the outputs are Bitcoin addresses to send coins to. Each output also carries the amount of BTC to send, expressed in Satoshis ($ 1 $ BTC $ = 10^{8}$ Satoshi), as well as its number of appearance in the list of outputs (the vector entry parameter $v_{out}$). An output can thus be characterized by the attributes \{output address, value, $ v_{out} $\}. 
	
	As transaction inputs, i.e. as the source of the coins to be spent, the sender of a transaction references one or several so-called \textit{unspent transaction outputs} (UTXO). A UTXO is nothing but the output of another transaction that was previously sent to some Bitcoin user, in this case, the sender of the current transaction, and not yet spent by that same user. As it is unspent, it can be used as input in a new transaction. In this way, the sender of a transaction ``forwards" his bitcoins received from old transactions to others. By the usage of the correct private key, which is typically only known to the sender, the UTXO is digitally unlocked. In this process, the sender publicly demonstrates the right to further spend the amount of coins linked to the UTXO. The main task of miners is to verify these inputs, inscribe verified transactions to the blockchain, and mark the involved UTXOs that were used by the sender as spent such that they cannot be double-spent. Miners compete for the task of creating blocks in a computational race for the next block, to collect the associated block reward that is given to the miner with each block as compensation.
	
	When a number of coins $ Y $ shall be transferred between two individuals, the spender creates a transaction that specifies the address(es) of the receiver(s) as output(s). As input(s), the sender references the txid(s) of valid UTXO(s) to spend. Each UTXO is uniquely identified and referenced through its txid, which is given by the cryptographic hash of the corresponding transaction script. It is not possible to spend a UTXO only partly. Instead, a UTXO always needs to be forwarded as a whole. Nevertheless, the sum of the values of all UTXO used in a transaction $ X = \sum_i X_i $ (where $ X_i $ is the number of coins linked to the $ i $'th UTXO)) can be arbitrarily split into multiple output amounts $ Y_i $, again. This ensures that the creator of a transaction can send arbitrary amounts of money to many addresses. 
	
	More generally, a transaction with $n_{in}$ inputs ($ X=\sum_1^{n_{in}} X_i $) and $ n_{out} $ outputs ($ Y = \sum_0^{n_{out}-1}Y_{v_{out}} $) will consume $n_{in}$ UTXO(s) and create $ n_{out} $ new UTXO(s), directed to the output addresses. In future transactions, these UTXOs will be referenced by the txid of the new transaction and its output vector number $v_{out}$. Furthermore, $ F = Y - X $ are the implied fees paid (voluntarily) by the creator of a transaction to the miner of the block that 
	will contain the transaction. Fees are directed to the miner together with the block reward in the so-called coinbase transaction which is the first transaction of each block and there is no input(s), but only output(s).
	
	If the sum of the UTXOs utilized for a transaction is larger than what the sender intends to transfer ($ X > Y $), the excess amount $ X-Y $ must be redirected back to the sender. The sender solves this by specifying his address as an additional output to send coins to, i.e. the sender transfers part of the coins back to himself. Such a transaction will then have two outputs $ Y_1 = Y $ (sender to receiver), $ Y_2 = X-Y $ (sender to sender) which is called a change transaction. As the sum of UTXO's $ X $ typically does not exactly match the total amount of $ Y $ that the sender intends to transfer to others, the majority of transactions involve a change output. 
	
	The primary quantity of interest that we extract from the parsed transaction data is the age distribution $ n_\tau(t)\ \forall \tau$ (see Eq.\eqref{eq:n-tau-t})
	for various times $t$. Knowledge of this distribution allows us to derive all other quantities, the transition probability distribution $ \pi_{\tau\rightarrow t}(t) $ (see Eq.\eqref{eq:pi-tau-t}), the total number of mined coins $ N(t) $ (see Eq.\eqref{eq:N-t}) as well as the transaction volume $V(t) $ (see Eq.\eqref{eq:v-t}). Other distributions and statistics will be derived below based on these variables. 
	
	To obtain the age distribution $ n_\tau(t)\ \forall\tau\leq t$ at a given time $t$, the number of coins that were last exchanged at the corresponding time must be computed. In other words, every bitcoin must be assigned an age (i.e. the present time minus the timestamp of its last transaction time) and the corresponding ID and time of its last transaction. This task amounts exactly to monitoring the set of outstanding UTXOs at each point in time, as a UTXO is essentially an unspent amount of coins, last exchanged at time $\tau$, i.e. the time of the creation of the UTXO. Therefore, at time $t$, the sum of all UTXOs created at $\tau$ (bitcoins born at $\tau$) is exactly $ n_{\tau}(t) $. Iterating over all $ \tau\leq t $ yields the full age distribution at time $t$. 
	
	On the blockchain, transactions do not carry a timestamp. Instead, a transaction time, the mining time of the block that the transaction is inscribed in (which is encoded in the block header), is the closest approximation of the transaction time. Thus, all transactions within a single block will be assigned the same time\footnote{The smallest resolution of the transaction time will therefore be ``block time'', i.e. approximately 10 minutes. As the time between two block minings is stochastic, sampling the transaction times between subsequent blocks will result in irregular time steps (of approx. 10 min duration). On the small scale of minutes to hours, this may cause inaccuracies. By upsampling the involved quantities, i.e. aggregating them over a fixed discrete time interval $ \Delta t \gg 10 $min, these inaccuracies will not affect our analysis.}. For this study, transaction times are further up-sampled to a larger time interval
	\begin{equation}\label{eq:delta-t}
		\Delta t = 1 \text{ day}
	\end{equation}
	All transaction times $ \tau $ are rounded up (to avoid forward-looking) to the next sampling time: 
	\begin{equation}\label{eq:sampling}
		\tau \leftarrow t_{s,i} \quad\quad\text{ if } \tau\in(t_{s,i}-\Delta t,t_{s,i}] \quad\quad\text{ where } t_{s,i}=t_{s,0}+i\Delta t \quad\quad\text{ for } i=1,2,3,...
	\end{equation}
	where $ t_{s,0} = $ $ 1970 $-$ 01 $-$ 01 $ is a standard timestamp reference date. The choice of a larger interval decreases the amount of computation. 

	\section{Age Distribution \label{trhj3yrbgq}}
	
	\subsection{Definition of Six Remarkable Times}
	
	Based on the methodology outlined in Section \ref{sec:method}, the age distribution $n_{\tau}(t) $ is computed in time steps of size $ \Delta t = 1$ day. 
	The analysis presented below uses data until 2020 in some cases and until August 2022 in other cases. 
	Table \ref{t:times} gives the set of times $t$'s at which the age distributions are constructed.
	
	\begin{table}[ht]
		\centering
		\begin{tabular}{|c|c|l|l|}
			\hline
			\textbf{Nr.}& \textbf{Date}	& \textbf{Event}	& \textbf{Price (\$)}			\\\hline\hline
			1 			& $t_1=$ 2013-04-04 	& Bubble peak		& 135.00 					\\\hline
			2 			& $t_2=$ 2013-07-04 	& Crash trough		& 79.40 					\\\hline
			3 			& $t_3=$ 2013-11-28 	& Bubble peak		& 1012.98 					\\\hline
			4 			& $t_4=$ 2015-01-15 	& Drawdown trough	& 210.47 					\\\hline
			5 			& $t_5=$ 2017-12-21 	& Bubble Peak		& 15600.01 					\\\hline
			6 			& $t_6=$ 2019-01-31 	& Drawdown trough		& 3413.11 					\\\hline
		\end{tabular}
		\caption{\label{t:times} The six times at which the distributions $n_{\tau}(t)$ are calculated and analysed in the following. The column ``Event" 
			refers to a major peak or trough along the bitcoin price time series.
			The last column is the close price in USD/BTC on the event day.}
	\end{table}
	
	Over the period from 2009 to 2020, let us mention particularly the following regimes that Bitcoin developed \cite{gerlach2019dissection,wheatley2019bitcoin}: (i) the early 2013 bubble and crash, (ii) the late 2013 bubble (iii) a long drawdown phase until 2015, (iv) the strong bubble behavior
	exhibited by the major cryptocurrencies from mid 2017 until December 2017 (and the following crash) and (v) the large following drawdown
	bottoming on 31st January 2019.
	The analysis times in Table \ref{t:times} are set as closely as possible to the corresponding peak and trough events.

	\subsection{Age Distributions at Six Remarkable Times}
	
\begin{figure}[!ht]
		\centering
		\subfloat[Subfigure 1 list of figures text][]{
			\includegraphics[width=0.48\textwidth]{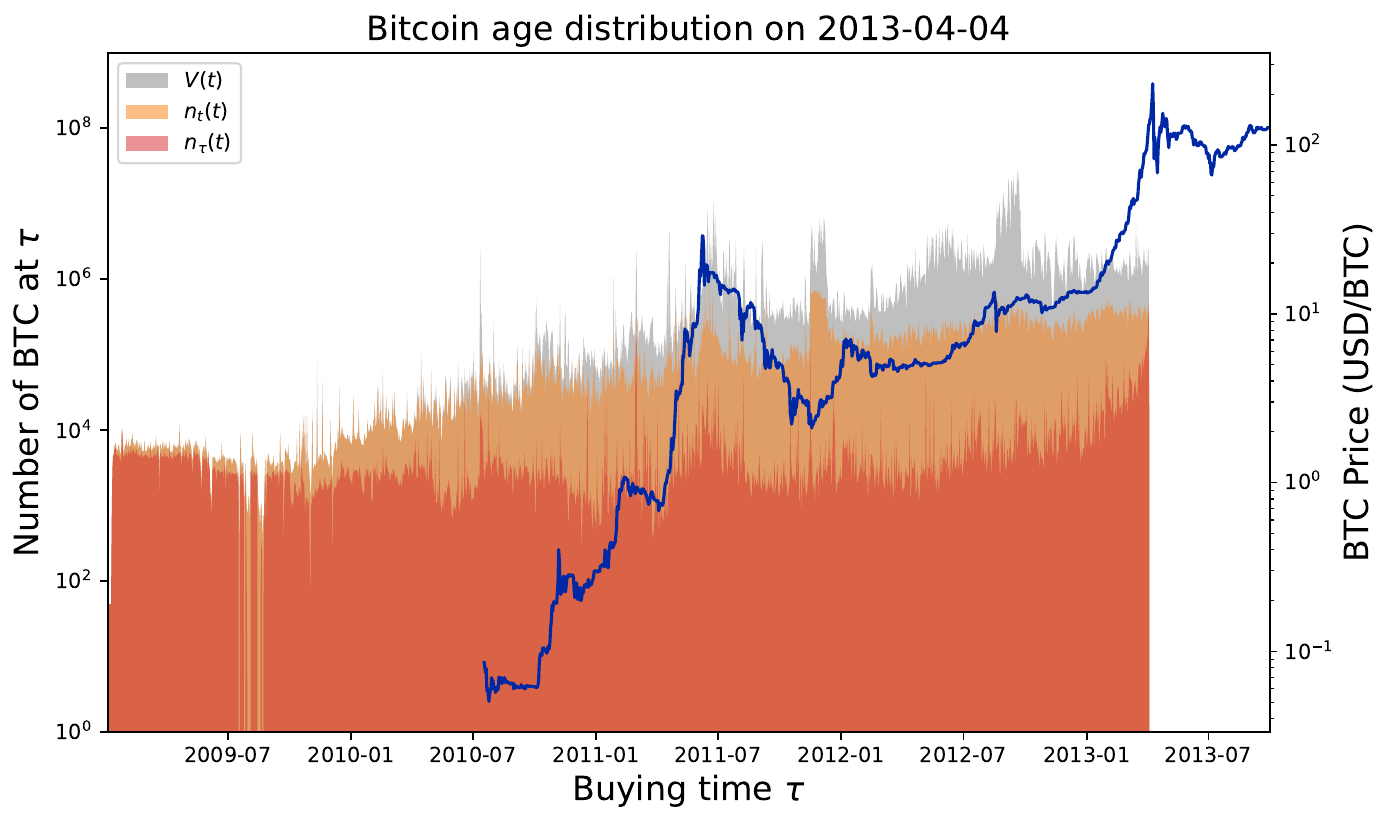}
			\label{fig:btc-age-subfig1}}
		\subfloat[Subfigure 2 list of figures text][]{
			\includegraphics[width=0.48\textwidth]{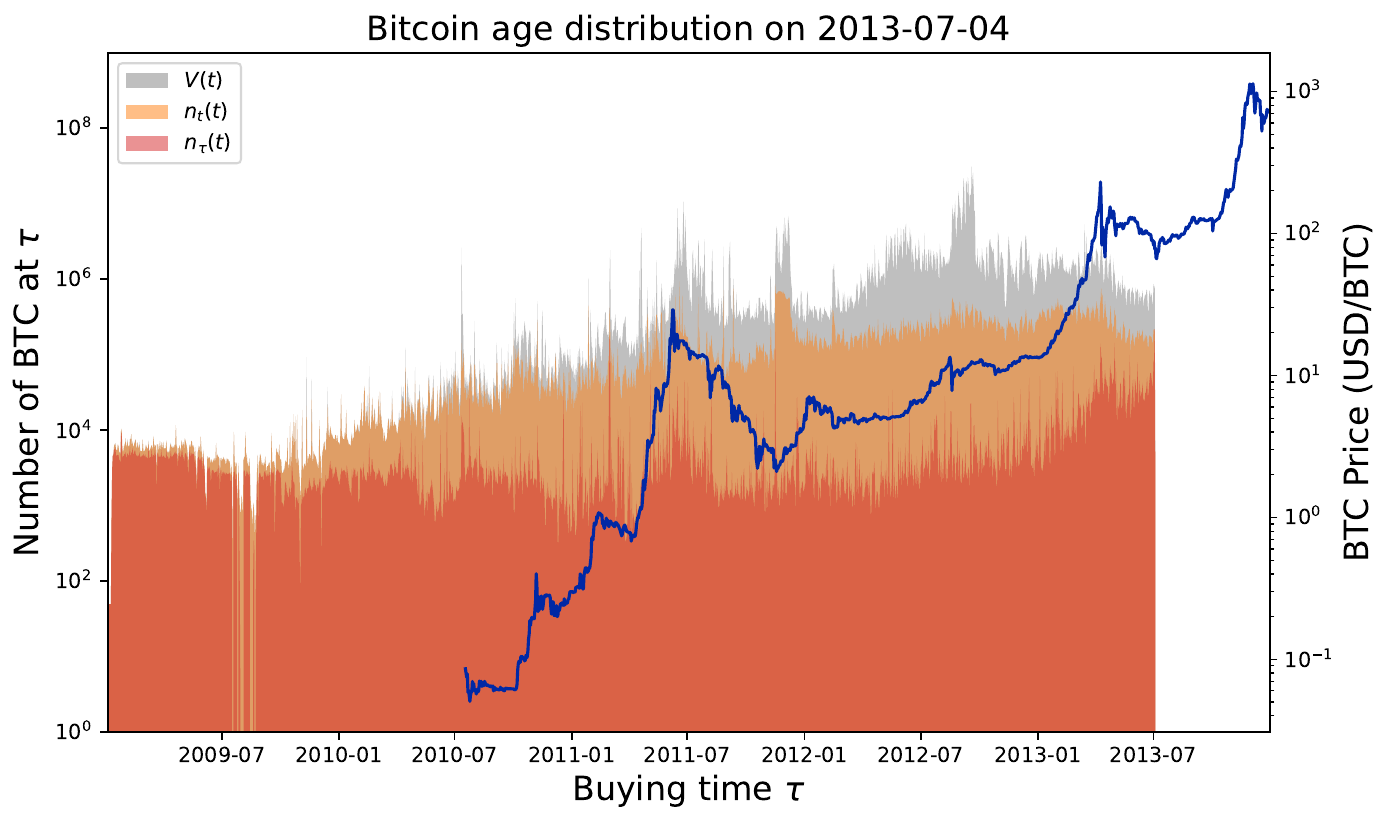}
			\label{fig:btc-age-subfig2}}\\
		
		\subfloat[Subfigure 3 list of figures text][]{
			\includegraphics[width=0.48\textwidth]{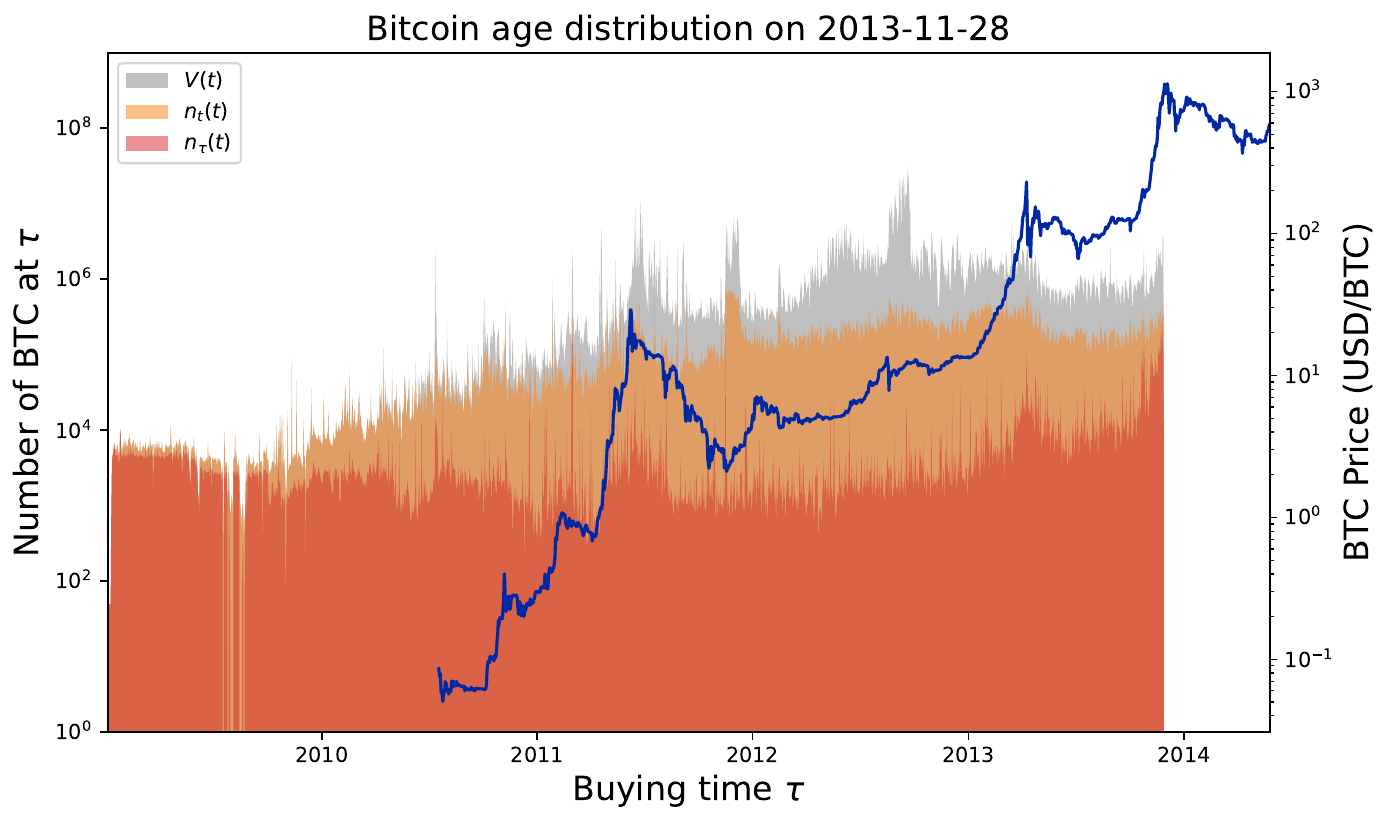}
			\label{fig:btc-age-subfig3}}
		\subfloat[Subfigure 4 list of figures text][]{
			\includegraphics[width=0.48\textwidth]{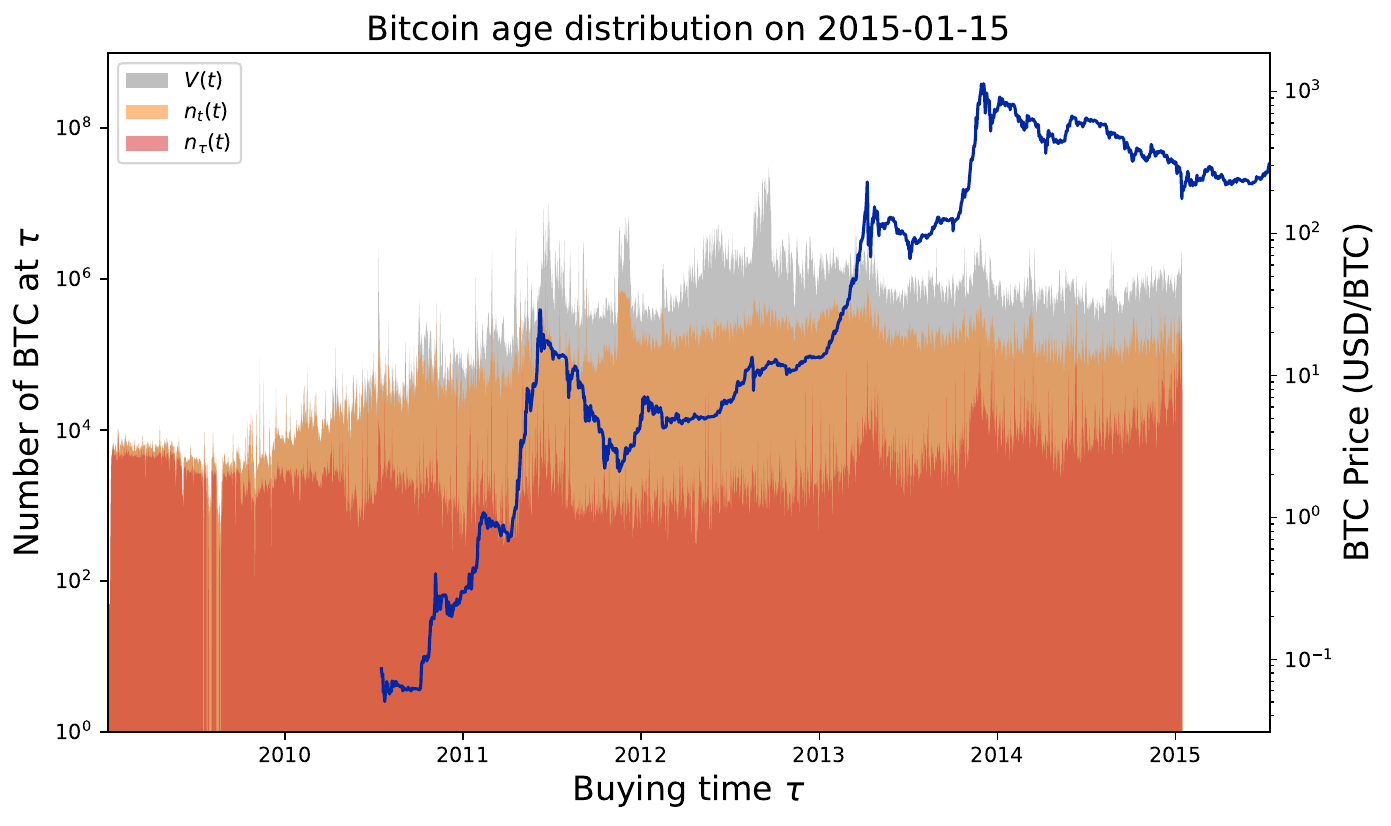}
			\label{fig:btc-age-subfig4}}\\
		
		\subfloat[Subfigure 3 list of figures text][]{
			\includegraphics[width=0.48\textwidth]{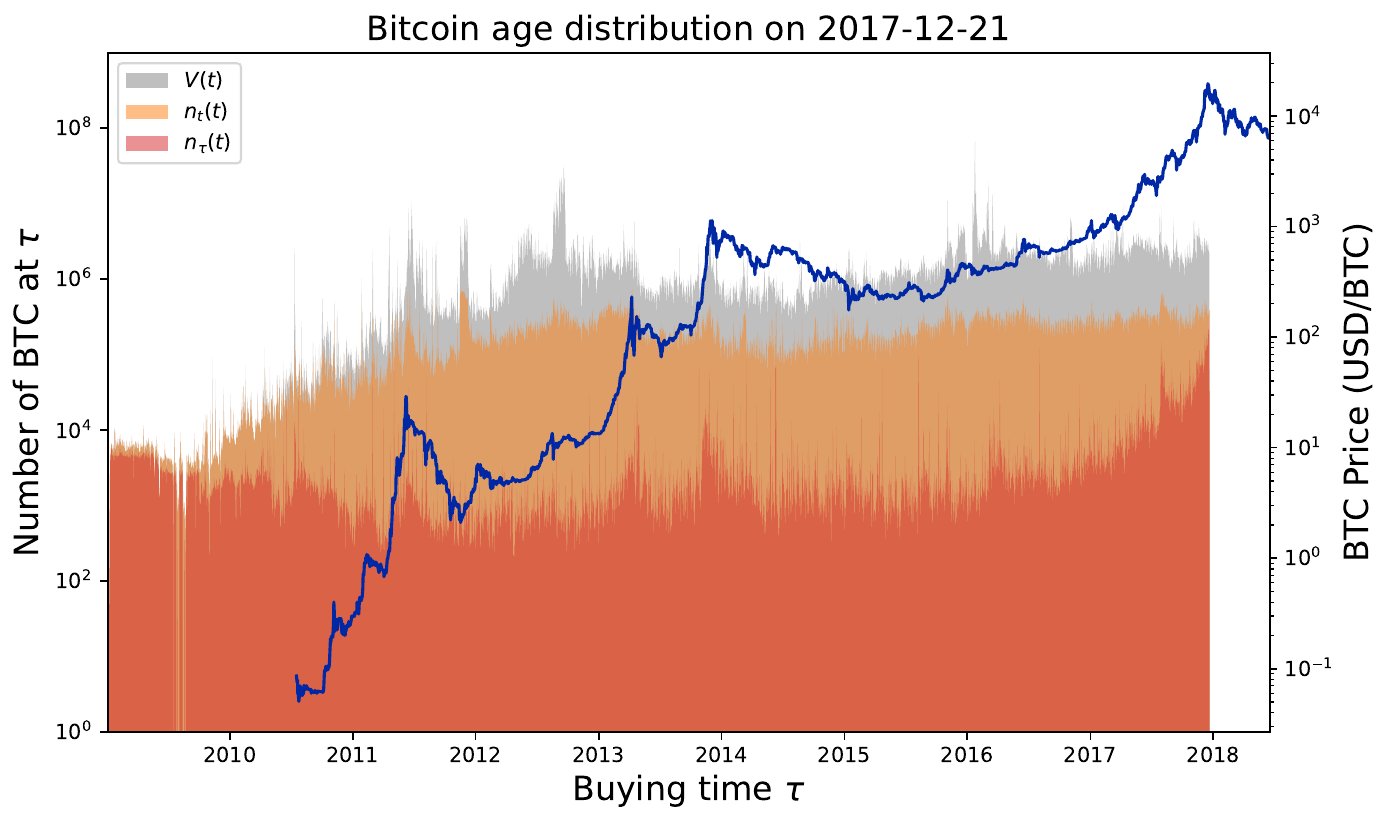}
			\label{fig:btc-age-subfig5}}
		\subfloat[Subfigure 4 list of figures text][]{
			\includegraphics[width=0.48\textwidth]{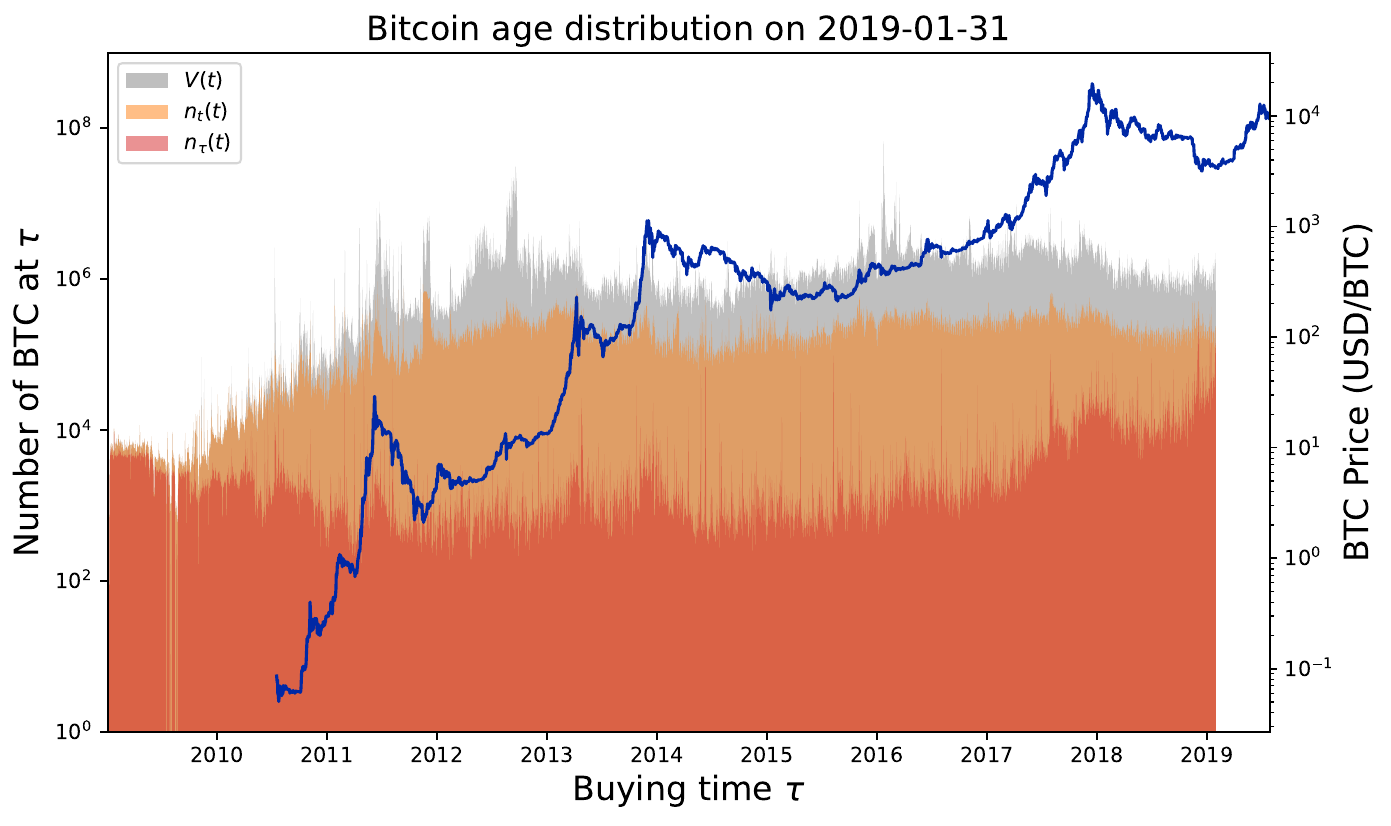}
			\label{fig:btc-age-subfig6}}
		\caption{\label{f:btc-age} The Bitcoin age / holding time distribution $n_{\tau}(t_i)\ \forall\tau, i=1, ..., 6$ as a function of $\tau$
			(left log-scale, red shaded) at the special points in time $\{t_1, ..., t_6\}$
			given in Table \ref{t:times}. The log-price (right scale, blue line) is expressed in US Dollars. In parallel, the number of coins born at $t$, $n_t(t)=V(t) + S(t)$ (yellow shaded) and the total transaction volume $V(t)$ (grey shaded) are plotted as a function of $t$. The yellow shaded area thus shows how many coins were born at each time t, and $V(t)$ shows how many coins were exchanged during the time interval $[t-\Delta t,t]$, with $\Delta t = 1$ day.}
	\end{figure}

	Figures \ref{fig:btc-age-subfig1} - \ref{fig:btc-age-subfig6} show in red the age distribution $ n_{\tau}(t_i) $ from the inception date of bitcoin to the six event times $\{t_1, ..., t_6\}$ from Table \ref{t:times}, overlaid with the log-price in USD/BTC (blue trajectory in right scale). The discrete bars of the age distribution have a width of $ \Delta t = 1$ day. The numbers of coins born at $ [\tau-\Delta t, \tau] $ (i.e. last exchanged within that interval) and still unexchanged until $t$ are summed up to obtain the corresponding value $ n_{\tau}(t) $. 
	
	Figures \ref{fig:btc-age-subfig1} - \ref{fig:btc-age-subfig6} also show
	the number of coins born and exchanged at each time $t$, $n_t(t)=V(t) + S(t)$ (yellow, equation (\ref{eq:n-t-t})) and the total transaction volume $V(t)$  (grey, equation (\ref{eq:v-t})).  
	Viewing the age distribution (in red) in relation to the number of born coins (in yellow) gives an interesting visualization of the number of coins that were bought at each previous $ \tau $ and remain unsold at time $ t $. Note that, while $n_\tau(t)$ is a distribution over $\tau$ in each panel ($t$ is fixed), $n_t(t)$ and $V(t)$ are variables given
	as a function of time $t$ running over the age time interval of the abscissa. In other words, $\tau$ becomes $t$ for the variables $n_t(t)$ and $V(t)$ 
	so that they can be compared with the number of bitcoins of different ages.
	
	Several important observations can be drawn from figures \ref{fig:btc-age-subfig1} - \ref{fig:btc-age-subfig6}. Firstly, in all plots, an initial stack of bitcoins that were bought prior to 2010 can be observed. During this transition period, bitcoin was not yet commonly traded on exchanges. Many blockchain participants might have tried out (mining) Bitcoin ``for fun" at this time, but probably did not expect for many of them a large financial gain from holding the cryptocurrency over long-term horizons. Thus, interest in the cryptocurrency might have declined for many, resulting in lost private keys and abandoned wallets that could not be accessed anymore. The initial stack of bitcoins is likely to consist of a large fraction of such ``dead coins" or ``dead wallets". 
	
	Secondly, and most importantly, there is a clear dependence of the age distribution on past times and price levels. The price-dependent memory of the age distribution is illustrated in panel \ref{fig:btc-age-subfig5}, which shows two distinct peaks of $n_\tau(t)$ around the times of the two bubble peaks in 2013, and panel \ref{fig:btc-age-subfig6} also clearly shows another peak around the bubble in the end of 2017. This indicates that investors who bought BTC in the final periods of the strongest bubble price growth seem to be reluctant to resell their coins. This is in agreement with the disposition effect \cite{shefrin1985disposition}: having bought at an expensive price, these owners are reluctant to realize their losses.  Therefore, they are holding, waiting for the price to climb back above their purchase prices.
	Indeed, if a losing position is held, there is the possibility that it will turn into a profitable position in the future. 
	Losing positions may entice investors/speculators to a longer-term stance, with supporting narratives such as ``in the long-term, it may rise again" or ``I am not a short-term speculator, but a long-run investor". The empirical evidence presented here 
	provides direct support for the existence of this behavior in markets. This will be further investigated throughout the study.
	
	\subsection{Characteristics of Short-, Medium- and Long-term Holders}
	
	As it is impractical to show the age distribution and other variables at all times, several statistics describing the distributions will be derived.
	As a first metric, we construct the fractions $f_{h_i}$ of short-, medium- and long-term holders over time defined by
	\begin{equation}\label{eq:f-holders}
		f_{h_i}:=\dfrac{\sum_{\tau\in{\mathcal{T}_i}}n_{\tau}(t)}{\sum_{\tau\leq t}n_{\tau}(t)} = \dfrac{N_{\mathcal{T}_i}(t)}{N(t)}\quad\quad i=1,2,3~,
	\end{equation}
	where the different windows $ \mathcal{T}_i $ are 
	\begin{itemize}
		\item short: \tabto{1.75cm} $ \mathcal{T}_1(t) = t-[0,30] $ 		days
		\item medium:\tabto{1.75cm} $ \mathcal{T}_2(t) = t-(30,365] $ 		days
		\item long:  \tabto{1.75cm} $ \mathcal{T}_3(t) = t-(365, \infty) $ 		days
	\end{itemize}
	
	\begin{figure}[!ht]
		\centering
		\includegraphics[draft=false,width=0.75\linewidth]{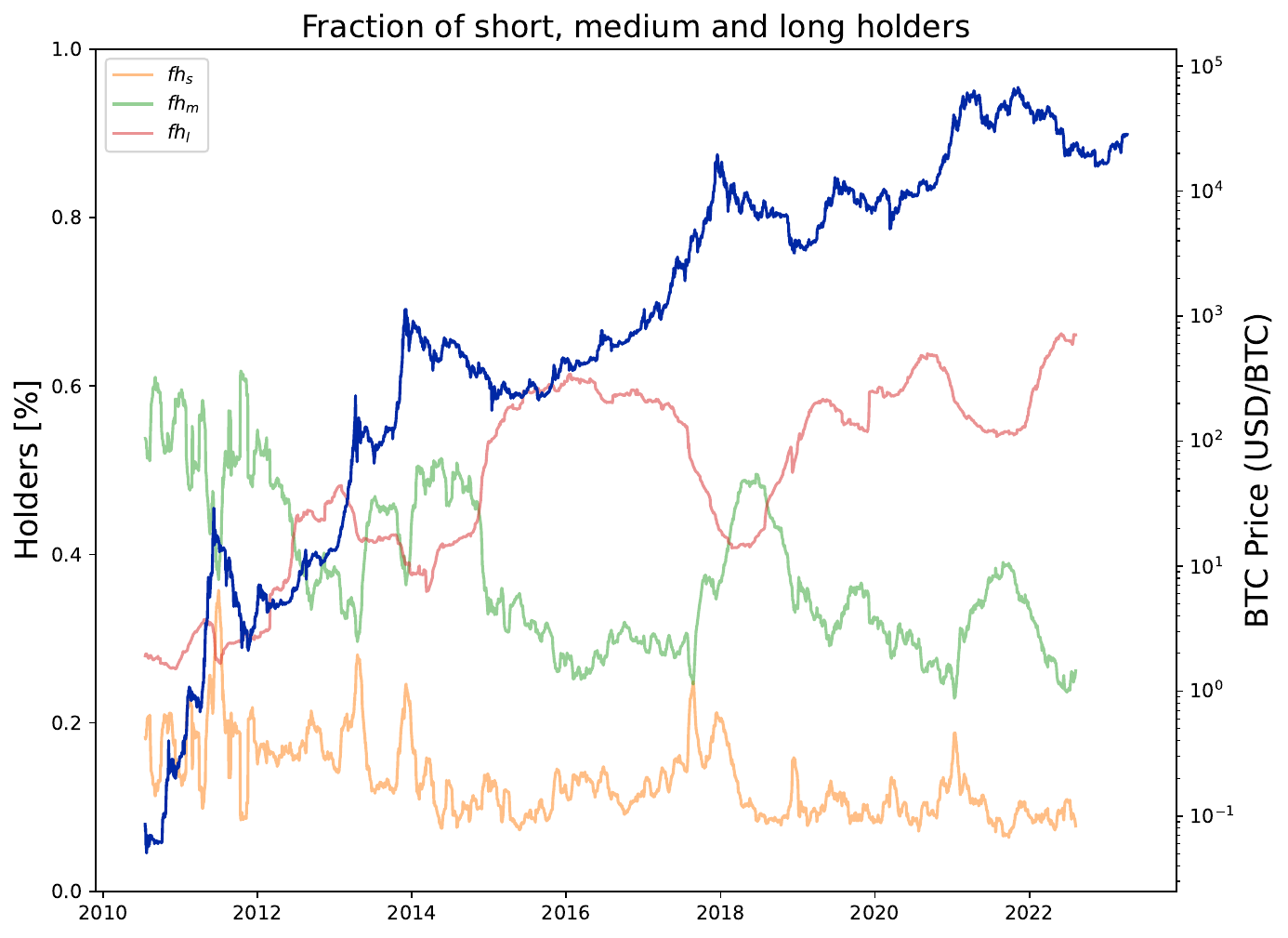}
		\caption{\label{f:fraction-holders} Fractions of bitcoin holders $fh_s$  (orange) at short- (up to 30 days), $fh_m$ (green) medium- (between one month and one year old) and $fh_l$ (red) long-term (older than one year) time scales. Bitcoin holdings in percent are defined in expression (\ref{eq:f-holders}), juxtaposed with price (right scale, blue line) over time. These fractions $fh_s,\ fh_m$ and $fh_l$ are derived from the age distribution as the percentage share of the age distribution that is located within the corresponding time intervals $[0,30],(30,365],(365, \infty)$.}
	\end{figure}
	
	Figure \ref{f:fraction-holders} shows that, for most times between 2011 and 2022, the fraction of long-term holders has been above 30\% and 
	exhibits an upward trend.  A first visual examination of the correlation between price variations and the proportion of long-term holders
	shows that, whenever the price experiences a decrease, there is a corresponding increase in the fraction of long-term holders. 
	Conversely, when the price exhibits an upward movement, the proportion of long-term holders tends to decrease especially in the later
	phases of the growth (and bubble) regimes. 
	This shows that long-term holders tend to sell their bitcoins to realize profit when the price increases. In contrast, 
	the population of long-term holders increases during down-trending markets, reflecting the tendency 
	to hold with the hope of recouping their book-to-market losses (disposition effect discussed above). 
	
	The multi-annual anti-correlation between the populations of long-term holders and 
	medium-term holders is quite striking visually, demonstrating a kind of communicating vessels phenomenon, in which the depletion of one 
	is the contribution to the increase of the population of the other and vice-versa. 
	For example, during both the 2013 bubbles and the large bubble in 2017, the fraction of long-term holders decreased while the fraction of medium-term holders increased. This indicates the tendency of short-term bitcoin buyers to hold their coins throughout the growth of the bubble and of long-term bitcoin holders to sell their bitcoins. 
	There is also a clear anti-correlation between the fraction of short-term bitcoin holders and the fraction of long-term holders. 
	The relationship between short- and medium-term holders is more complicated with positive visual co-movements
	over some long periods and opposite co-movements at times of spikes.
	Moreover, the fraction of short-term bitcoin holders can be seen to follow closely the ups and downs of the bitcoin price.
	
	\begin{table}[ht]
		\centering
		\begin{tabular}{|c|c|c|c|}
			\hline
			& $fh_s$ &$fh_m$ &$fh_l$ \\ \hline
			$fh_s$ & 1      &0.197  & -0.577   \\ \hline
			$fh_m$ & 0.197  &1      &-0.914    \\ \hline
			$fh_l$ & -0.577 &-0.914 &1 \\\hline
		\end{tabular}
		\caption{Correlation matrix among $fh_s$, $\ fh_m$ and $fh_l$.}
		\label{fsml_correlation}
	\end{table}
	
	We now put these qualitative observations on a more quantitative statistical footing.
	Because they can provide complementary information, we calculate the pairwise cross-correlations between
	$fh_s$, $\ fh_m$ and $fh_l$ and between their respective increments.
	While being a priori delicate in terms of statistical reliability and interpretation due to possible non-stationarity, we first 
	calculate the correlation matrix of $fh_s$, $\ fh_m$ and $fh_l$ shown in table \ref{fsml_correlation}. 
	The most striking result is the very strong negative correlation between $fh_m$ and $fh_l$, and the large 
	negative correlation between $fh_s$ and $fh_l$, which quantifies 
	the obvious visual observation. The smaller positive correlation found between $fh_s$ and $fh_m$ 
	is more difficult to interpret as it is likely to be in part the result of mixtures of transient positive and negative
	correlation regimes.
	
	An obvious potential problem with the results shown in table \ref{fsml_correlation} is that the time series
	of $fh_s$, $fh_m$ and $fh_l$ may be non-stationary. We thus 
	conduct the Augmented Dickey-Fuller (ADF) unit root test on the three time series $fh_s$, $fh_m$, and $fh_l$. 
	At the 95\% significance level of the ADF test, the null hypothesis (indicating the presence of a unit root in the time series) cannot be rejected 
	for both $fh_m$ and $fh_l$, regardless of whether drift and trend are considered. In contrast, the null hypothesis of the ADF test 
	is rejected for $fh_s$, implying its stationarity. The non-stationarity of $\ fh_m$ and $fh_l$ thus makes delicate the quantitative 
	interpretation of the correlation values reported in table \ref{fsml_correlation}.
	
	\begin{table}[ht]
		\centering
		\begin{tabular}{|c|c|c|c|c|}
			\hline
			& $\Delta fh_s$ &$\Delta fh_m$ &$\Delta fh_l$  & $r_t$ \\ \hline
			$\Delta fh_s$ & 1      &-0.958  & -0.160   & 0.0137 \\ \hline
			$\Delta fh_m$ & -0.958 &1       &-0.127    &-0.006 \\ \hline
			$\Delta fh_l$ & -0.160 &-0.127 &1  & -0.0276 \\\hline
			$r_t$ & 0.0137      &-0.006  & -0.0276  &1  \\ \hline
		\end{tabular}
		\caption{Correlation matrix among $\Delta fh_s$, $\Delta fh_m$ and $\Delta fh_l$. $\Delta fh_s$ is the difference of $fh_s$, namely $\Delta_t fh_s= fh_s^t -fh_s^{t-1}$. Similarly, we have $\Delta_t fh_m= fh_m^t -fh_m^{t-1}$ and $\Delta_t fh_l= fh_l^t -fh_l^{t-1}$. The bitcoin return rate $r_t=\frac{p(t)}{p(t-1)}-1$. Note that all three pairs of fraction difference variables ($\Delta fh_s$, $\Delta fh_m$ and $\Delta fh_l$) are negatively correlated.}
		\label{fsmli_correlation}
	\end{table}

	Quantifying the cross-correlation between $fh_s$, $fh_m$ and $fh_l$ 
	will be statistically more meaningful when using instead their first-order difference (or time increment) $\Delta fh_s$, $\Delta fh_m$ and $ \Delta fh_l$.
	Applying the ADF test and Ljung-Box test on $\Delta fh_s$, $\Delta fh_m$ and $ \Delta fh_l$, 
	we verify that these first-order differences are stationary and are not white noise.
	The corresponding correlation matrix of $\Delta fh_s$, $\Delta fh_m$ and $ \Delta fh_l$ 
	is shown in table \ref{fsmli_correlation}.
	The most important result is the very strong anti-correlation between the first-order differences of $fh_s$ and of $fh_m$.
	This can be interpreted in part as due to the conversion of medium-term holders into short-term holders when the former trade.
	Another contribution to this remarkably strong anti-correlation is more mundane: as $fh_l$ tends to be much smoother 
	and to vary much more slowly that $fh_s$ and $\ fh_m$ exhibiting quasi-plateaus in some regimes, then by definition 
	the variation of  $fh_s$ should be close to the negative of the variation of $fh_m$. This second contribution
	to the strong anti-correlation between $\Delta fh_s$, $\Delta fh_m$ and $ \Delta fh_l$  is thus more a testimony of the slow and weak variations of $fh_l$.
	Given this, we consider it adequate to accept the relatively weak anti-correlation between 
	$\Delta fh_s$ and $ \Delta fh_l$  and between $\Delta fh_m$ and $ \Delta fh_l$ .
	
	To explore how short-, medium- and long-term holders react to the variation of bitcoin price, we calculate the cross-correlation between 
	the bitcoin return $r_t$ and the fraction difference of short-, medium- and long-term holders. The result is shown in table \ref{fsmli_correlation}.
	This confirms the visual impression discussed above that short-term holders change their position 
	in positive relationship with the price variations of bitcoin. In contrast, the correlation between bitcoin return
	and $\Delta fh_m$ and $ \Delta fh_l$ is negative, confirming quantitatively that,
	when the bitcoin price experiences a decrease, there is a corresponding increase in the fraction of long-term and medium-term holders.

	\begin{table}[ht]
		\centering
		\begin{tabular}{|c|c|c|c|}
			\hline
			&$\Delta fh_s$ &$\Delta fh_m$ &$\Delta fh_l$ \\ 
			\hline
			AIC &(2,5)& (3,4)&(1,4)\\
			\hline
			BIC &(1,1)& (1,1)&(1,4)\\
			\hline
		\end{tabular}
		\caption{The ARMA order (p,q) of $\Delta fh_s$, $\Delta fh_m$ and $\Delta fh_l$ under AIC and BIC. }
		\label{autocorrelation_order}
	\end{table}
	
	We then calibrate the ARMA(p,q) model to $\Delta fh_s$, $\Delta fh_m$ and $\Delta fh_l$, where
	$p$ is the order of auto-regression and $q$ is the order of moving average.
	To determine the optimal orders $p$'s and $q$'s for each time series, we 
	apply both AIC and BIS criteria, and the corresponding values are given in 
	table \ref{autocorrelation_order}.
	Moreover, the calibration with the best ARMA(p,q) models shows that the coefficients of auto-regressive terms are all positive
	and significant at the 99\% confidence level. These regressions confirm that the $\Delta fh_s$, $\Delta fh_m$ and $\Delta fh_l$ time series 
	exhibit a persistence or momentum (both when decreasing and increasing).
	
	We perform a vector auto-regression (VAR) analysis for the pairs $[\Delta fh_s, \Delta fh_m]$ and $[\Delta fh_m, \Delta fh_l]$.
	For the pair $[\Delta fh_s, \Delta fh_m]$, the t-statistics confirm a highly significant momentum effect for both time series.
	In other words, the regression of $\Delta fh_s(t)$ as a function of $\Delta fh_s(t-1)$ is highly significant with a 
	positive regression coefficient of $0.30$. Similarly, the regression of $\Delta fh_m(t)$ as a function of $\Delta fh_m(t-1)$ is highly significant with a 
	positive regression coefficient of $0.12$. Moreover, the regression of $\Delta fh_s(t)$ as a function of $\Delta fh_m(t-1)$ is also highly significant with a 
	positive regression coefficient of $0.18$. In contrast, $\Delta fh_m(t)$ is not significantly influenced by $\Delta fh_s(t-1)$.
	For the pair $[\Delta fh_m, \Delta fh_l]$, the t-statistics also confirm a highly significant momentum effect for both time series, 
	with regression coefficients respectively equal to $0.12$ and $0.30$ for $\Delta fh_m(t)$ as a function of $\Delta fh_m(t-1)$
	and for  $\Delta fh_l(t)$ as a function of $\Delta fh_l(t-1)$. However, there are no statistically significant 
	cross-influences of $\Delta fh_m(t-1)$ on $\Delta fh_l(t)$ or of $\Delta fh_l(t-1)$ on $\Delta fh_m(t)$.
	
	\section{Distributions of Book-to-Market and Realised Returns}\label{sec:b2m}
	
	\subsection{Distributions of Book-to-Market Returns}
	
	The investors who last bought at time $\tau$ and still hold at time $t$ have an unrealized (positive or negative) return
	$r_{\tau\rightarrow t}$ given by Eq.\eqref{eq:log-ret}. The total number of corresponding bitcoins
	is by definition $n_{\tau}(t)$. As the holding times $t-\tau$ vary from intraday to years, for the sake of comparison, it is convenient
	to define the return per day $(\bar{r}_{\tau\rightarrow t})$ for each position
	\begin{equation}\label{log-ret-day}
		\bar{r}_{\tau\rightarrow t} = {1 \over t-\tau}~ r_{\tau\rightarrow t}~.
	\end{equation}
	
	\begin{figure}[!ht]
		\centering
		\subfloat[Sub-figure 1 list of figures text][]{
			\includegraphics[width=0.48\textwidth]{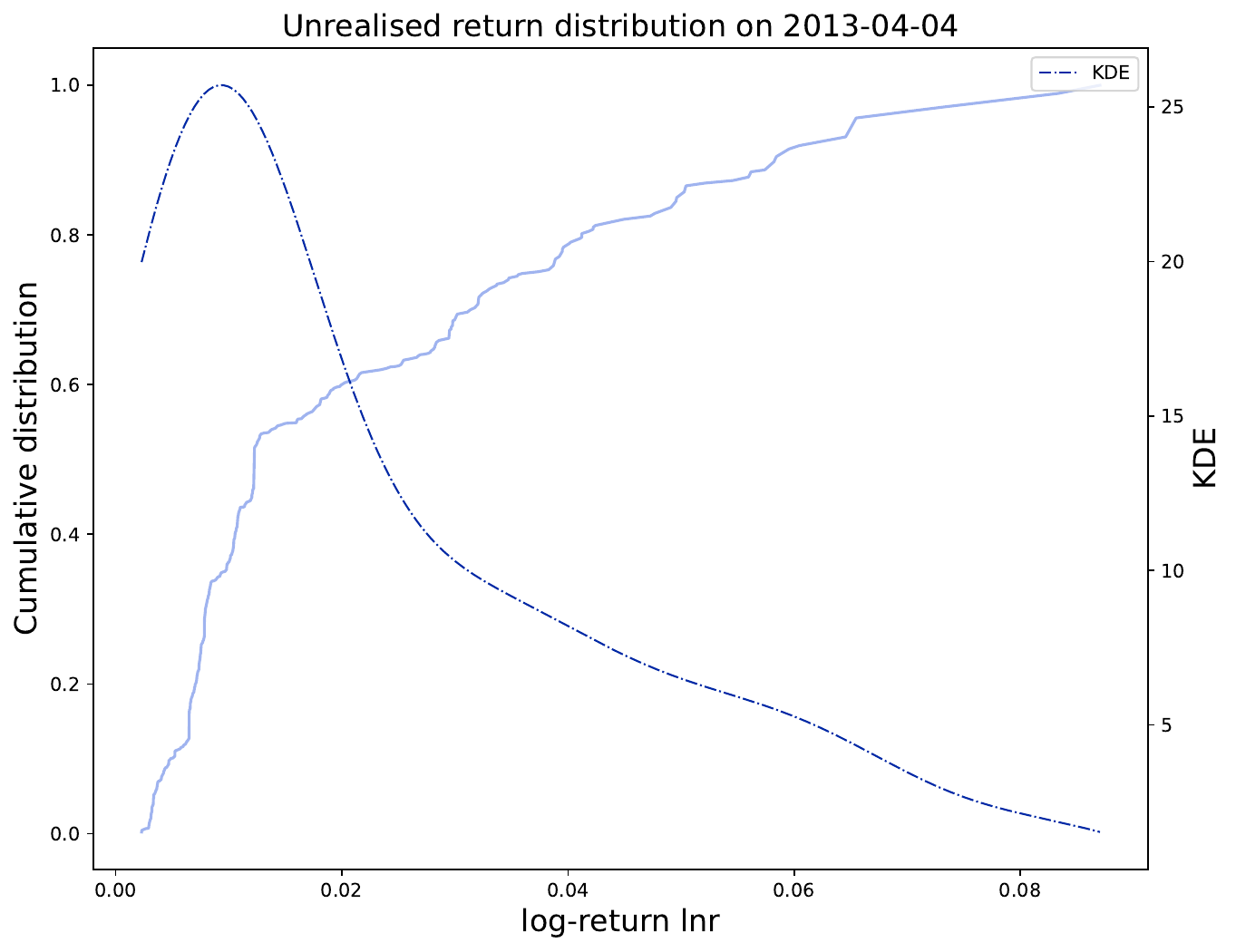}
			\label{fig:btm20130404}}
		\subfloat[Sub-figure 2 list of figures text][]{
			\includegraphics[width=0.48\textwidth]{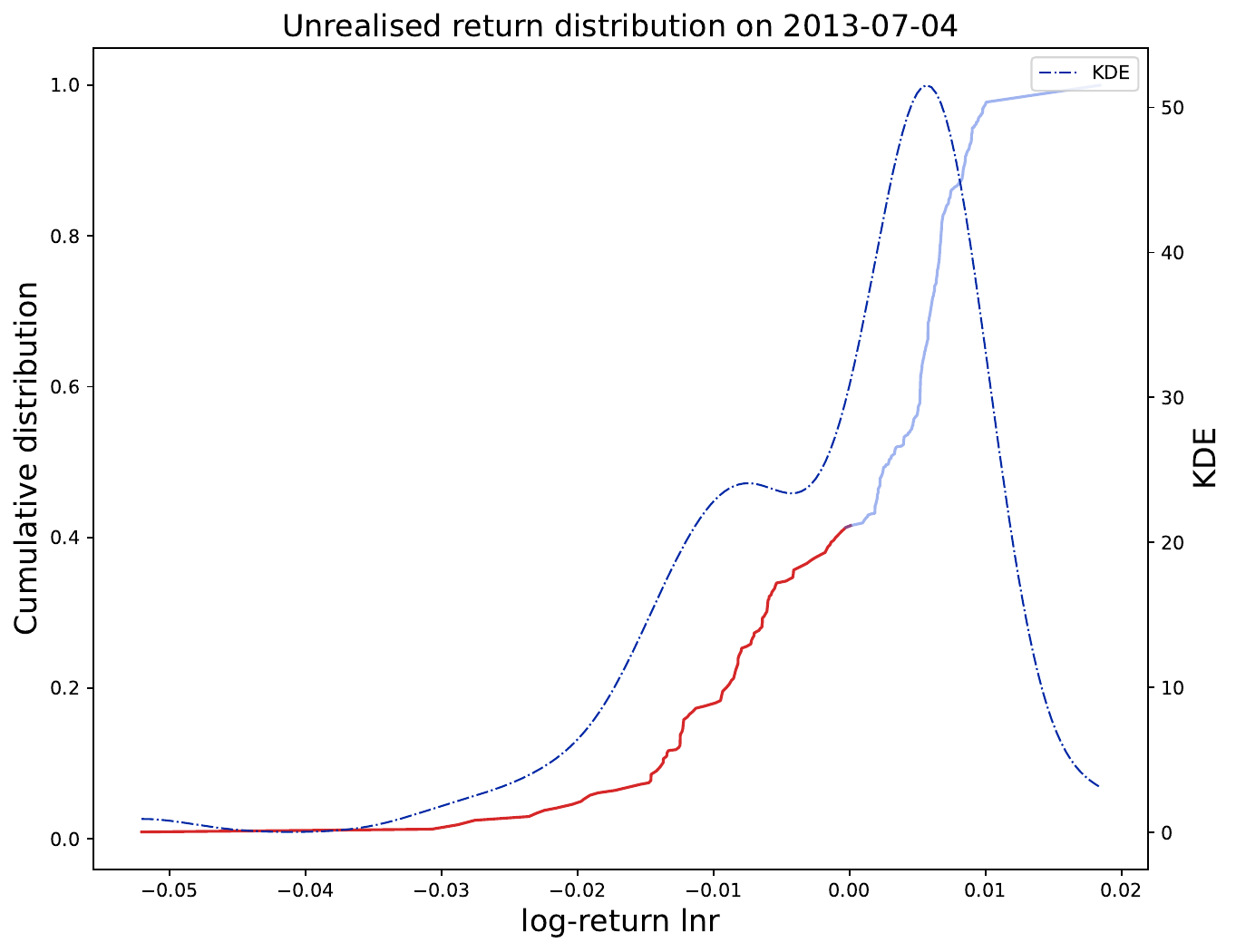}
			\label{fig:btm20130704}}\\
		
		\subfloat[Sub-figure 3 list of figures text][]{
			\includegraphics[width=0.48\textwidth]{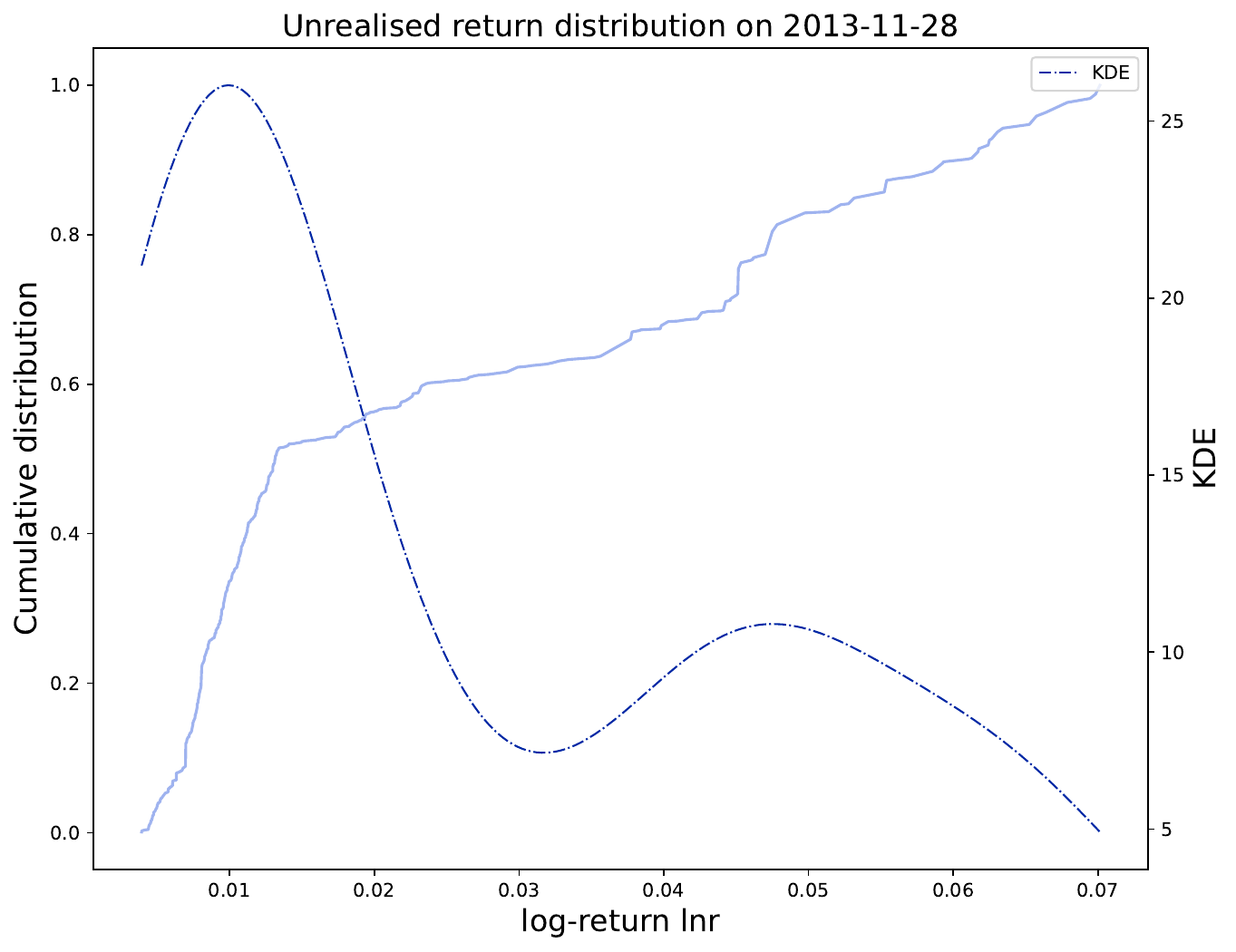}
			\label{fig:btm20131128}}
		\subfloat[Sub-figure 4 list of figures text][]{
			\includegraphics[width=0.48\textwidth]{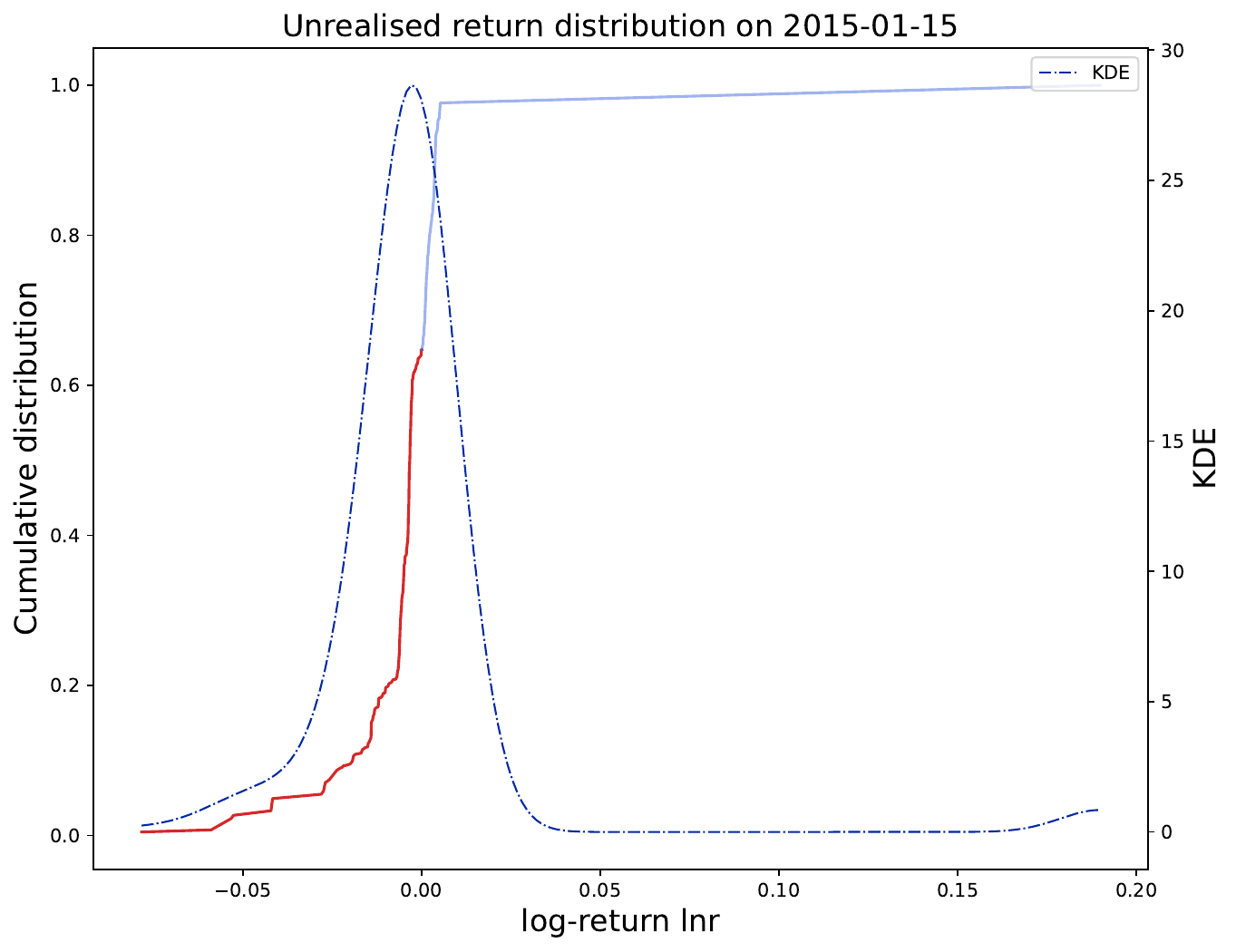}
			\label{fig:btm20150115}}\\
		
		\subfloat[Sub-figure 3 list of figures text][]{
			\includegraphics[width=0.48\textwidth]{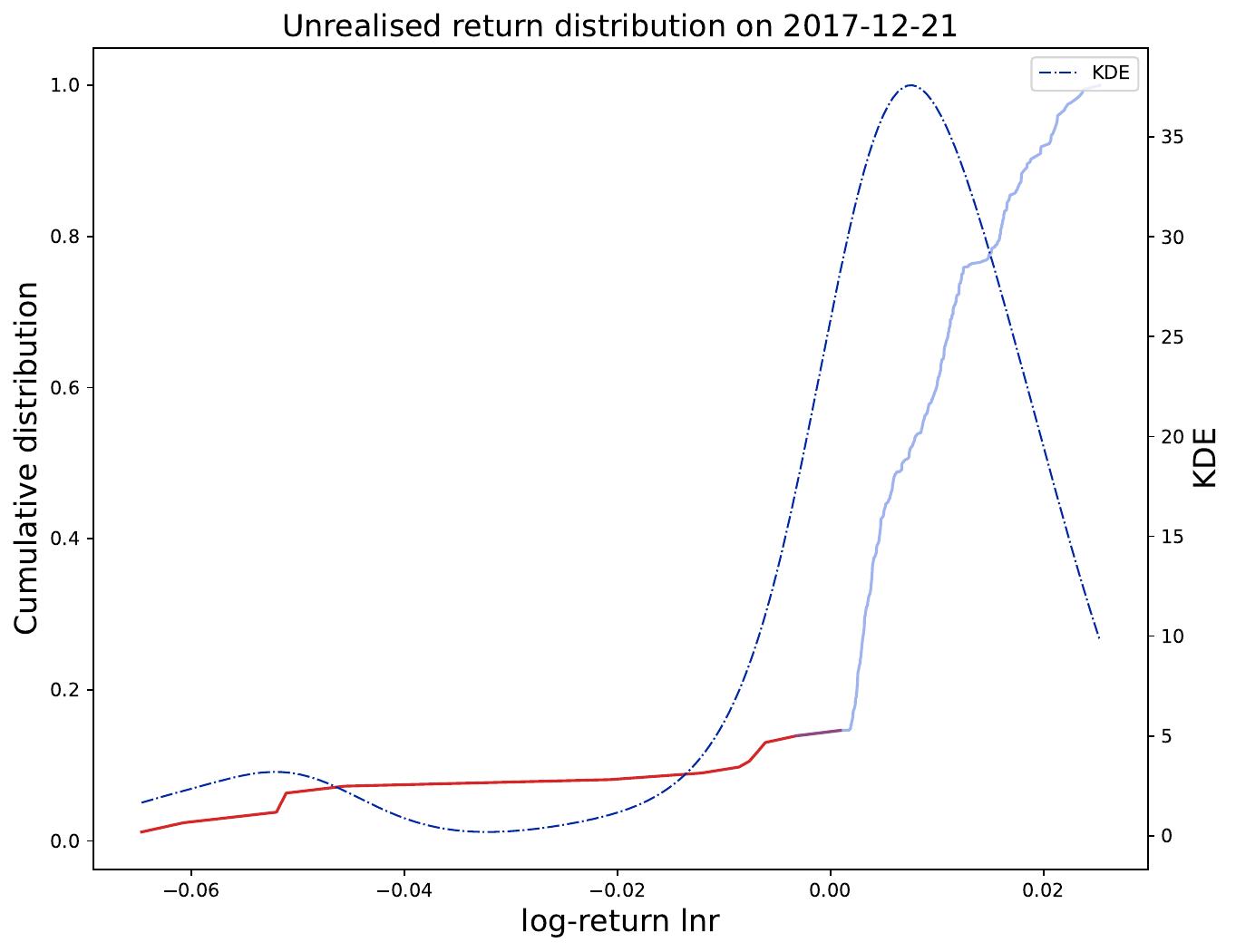}
			\label{fig:btm20171221}}
		\subfloat[Sub-figure 4 list of figures text][]{
			\includegraphics[width=0.48\textwidth]{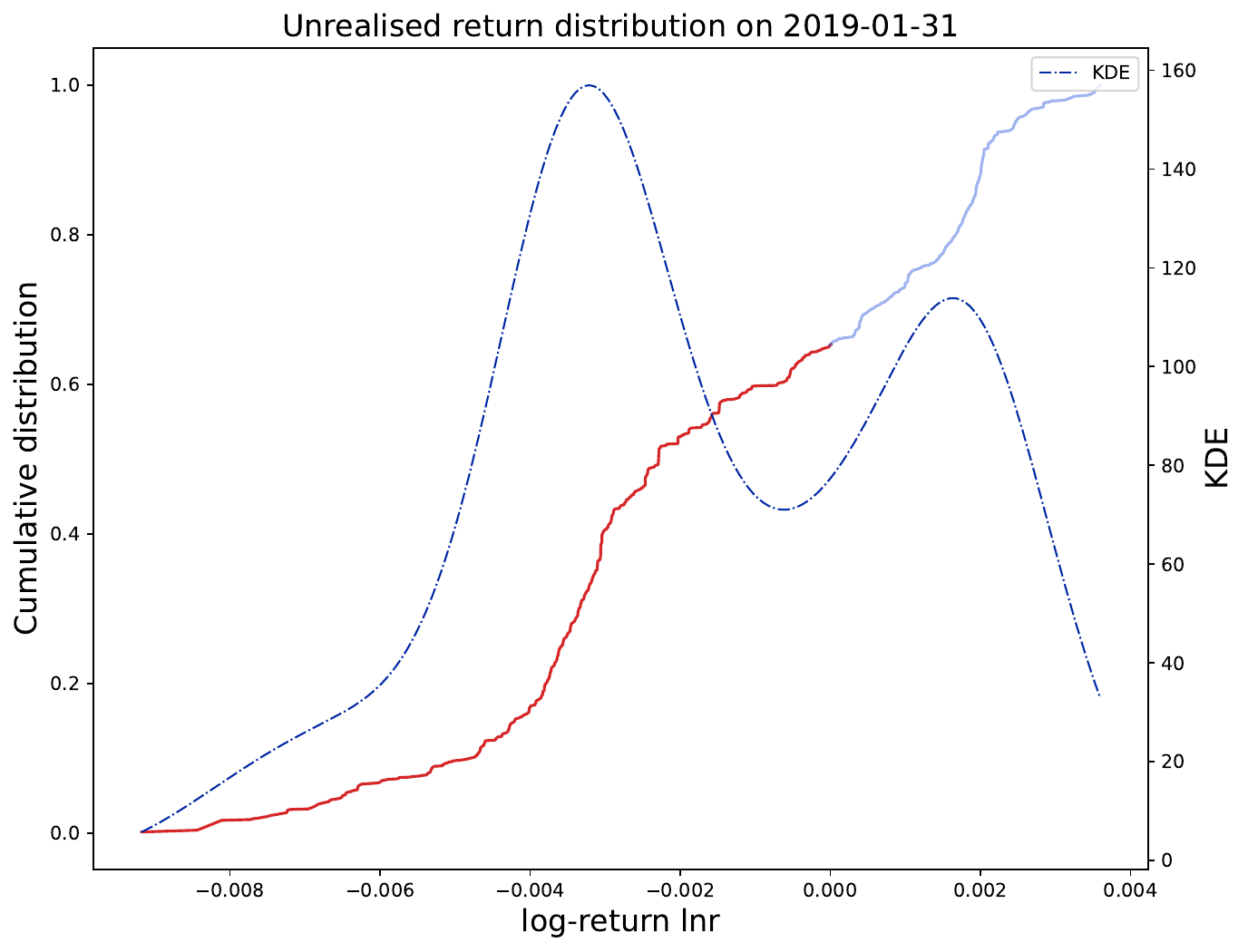}
			\label{fig:btm20190131}}
		\caption{Continuous line: empirical cumulative probability distribution functions $\text{cdf-B2M}_t(r)$ (equation (\ref{eq:b2qtgtm}))
			of (unrealized) Book-to-Market return per day of bitcoin holdings for the different analysis times given in Table \ref{t:times}.
			The range of positive (resp. negative) returns is shown in blue (resp. red). Dashed line: Smoothed probability density functions
			(smoothed derivative of the Cumulative normalized distributions shown in a continuous line).}
		\label{fig_cum_density}
	\end{figure}
	
	A given return $r$ per day is realised at time $t$ if there exist $\tau$ such that $\bar{r}_{\tau\rightarrow t} =r$.
	Summing the corresponding $n_{\tau}(t)$ over these $\tau$'s that have the same return $r$ allows us to define the distribution of Book-to-Market daily returns (B2MD) at time $t$ as
	\begin{equation}\label{eq:b2m}
		\text{B2MD}_t(r) :=  \sum_i n_{\tau_i}(t) \delta[\bar{r}_{\tau_i\rightarrow t} -r]~,
	\end{equation} 
	where $\delta(\cdot)$ is the Kronecker delta function equal to $1$ when its argument is equal to 0, and is equal to zero otherwise.
	Thus, $\text{B2MD}_t(r)$ is the sum of the $n_{\tau_i}(t)$ over all birth dates $\tau_i$ such that $\bar{r}_{\tau_i\rightarrow t} =r$.
	The corresponding cumulative quantity is defined as 
	\begin{equation}\label{eq:b2qtgtm}
		\text{cdf-B2M}_t(r) :=  \int_{-\infty}^r dx {\sum_i n_{\tau_i}(t) \delta[\bar{r}_{\tau_i\rightarrow t} -x] \over \sum_j n_{\tau_j}(t)}
		= {\sum_{i | \bar{r}_{\tau_i\rightarrow t}\leq r} n_{\tau_i}(t)] \over \sum_j n_{\tau_j}(t)}
	\end{equation} 
	This cumulative normalized B2MD can be interpreted as the empirical cumulative probability distribution function of the 
	unrealized daily return.  $\text{cdf-B2M}_t(r)$ provides a direct view of all outstanding unrealized gains and losses 
	that could be immediately realized at the given daily return ($r$) and time ($t$). 
	Figure \ref{fig_cum_density} shows $\text{cdf-B2M}_t(r)$ for the different analysis times given in Table \ref{t:times}. 
	Figure \ref{fig_cum_density} also shows the smoothed derivative of the cumulative probability distribution functions $\text{cdf-B2M}_t(r)$, 
	i.e., the corresponding probability density functions defined by ${1 \over N(t)} \text{B2MD}_t(r)$, where $N(t)$ is given by (\ref{eq:N-t-sum}).
	
	At the bubble peak times in figure \ref{fig:btm20130404}, \ref{fig:btm20131128} and \ref{fig:btm20171221}, almost 100\% of all coins are in an unrealized profit position, which reflects the fact that price peaks reached temporary all-time highs or close to all-time highs, after which the bubbles burst. 
	The other subfigures shown at times of price troughs visualize that a larger fraction of coins are, unsurprisingly, in unrealized loss positions. 
	It is interesting to note the differences in the shape of the cumulative normalized B2MD between 
	figure \ref{fig:btm20130404}, \ref{fig:btm20131128} and \ref{fig:btm20171221} on the one hand
	and figure \ref{fig:btm20130704}, \ref{fig:btm20150115} and \ref{fig:btm20190131} on the other hand. In the former cases at times close to all-time highs, the distributions tend to have thin tails on the left of the return axis, expressing the fact that there is a concentration of gains performed by investors who were attracted by the bubble rise and thus bought close to the peak. In contrast, $\text{cdf-B2M}_t(r)$ has a rather fat tail on the right of the return axis, which is associated with the bitcoin purchases of longer-term holders. In the later cases of times at price troughs, $\text{cdf-B2M}_t(r)$ exhibit fat tails on the left side of the return axis, which reflects the actions of investors who bought all along the ascent of the bubble and failed to realize their gains. These positions find themselves exhibiting the full range of losses associated with all prices above the trough during the rise of the bubbles. The part of $\text{cdf-B2M}_t(r)$ for positive returns corresponds to the medium and long-term holders who bought before the bubbles when prices were much lower.
	
	\begin{figure}[ht]
		\centering
		\includegraphics[width=\textwidth,height=\textheight,keepaspectratio]{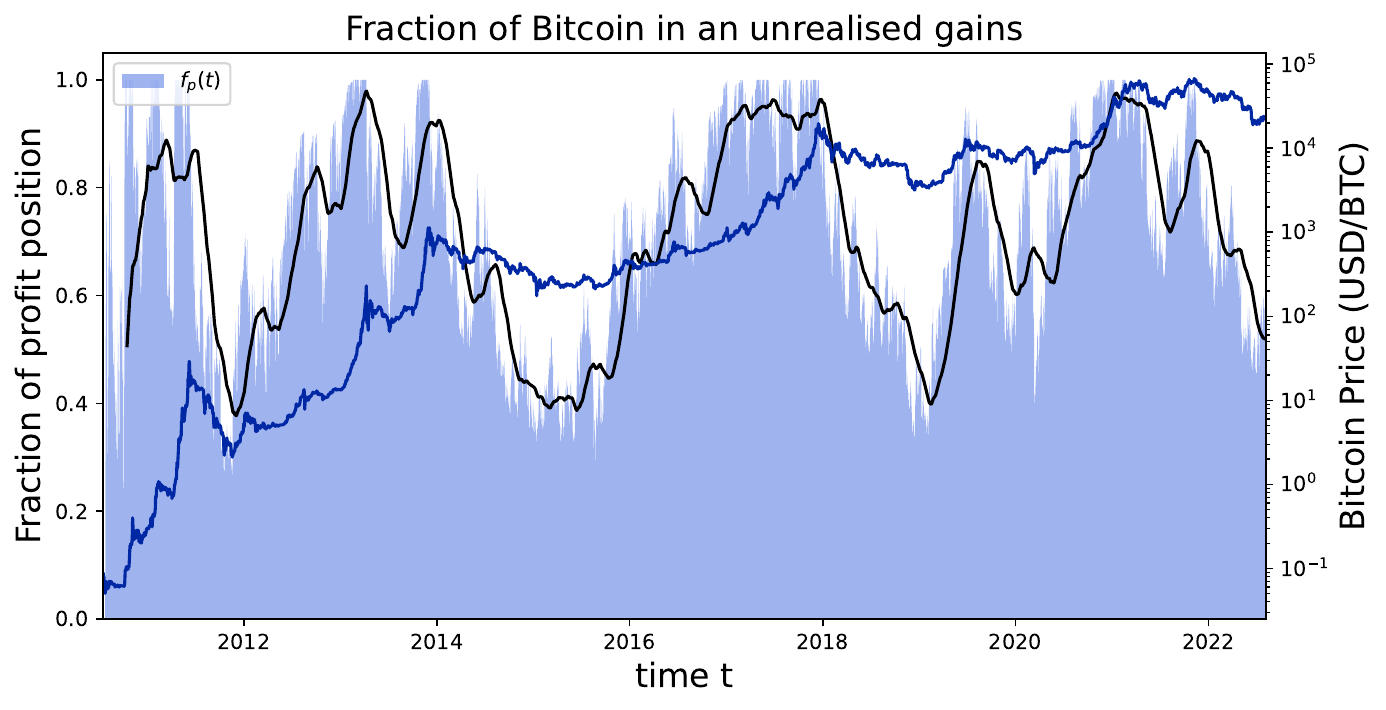}
		\caption{\label{f:unrealized-profits} Percentage $100\cdot f_{p}(t)$ defined
			by expression (\ref{eq:f-unreal-prof}) of bitcoin holdings in an unrealized profit position (blue area, left scale) 
			as a function of time. The smooth dark line is a 90-days moving average of $100\cdot f_{p}(t)$, which is offered to help interpret the 
			relationship between $f_{p}(t)$ and bitcoin price (blue line, right scale).}
	\end{figure}
	
	Figure \ref{f:unrealized-profits} shows the time dependence of the fraction of bitcoin positions in unrealized profit positions, which is defined by
	\begin{equation}\label{eq:f-unreal-prof}
		f_p(t):= \int_{r>0}  \text{pdf-B2M}_t(r) dr  ~.
	\end{equation}
	The main bubble phases during which the price repetitively marked new records can be clearly identified. 
	For all three major bubbles, the percentage of unrealized profit holdings reaches 100\% before the respective bubble peaks, followed by a sharp plunge and subsequent decay to levels of about 30\%. 
	One of the most interesting features is the presence of oscillations of $f_p(t)$ during the rise of the bubbles. 
	This phenomenon is particularly clear for the bubble starting in 2016 and ending in December 2017.
	The oscillations of $f_p(t)$ are associated with the succession of local corrections of the price during the bubble rise.
	Indeed, the local maxima of $f_p(t)$ observed during the 2016 to Dec. 2017 bubble in figure \ref{f:unrealized-profits} 
	can be mapped one-to-one onto the local maxima of the prices. This oscillatory dynamics of $f_p(t)$
	thus allows us to visualize the short-term and medium-term trading occurring during the bubble rise, which is a combination 
	of the attraction for traders to buy into and ``surf'' the bubble and the desire to realize gains (too early, which is again a signature
	of the disposition effect \cite{shefrin1985disposition}).
	
	\begin{figure}[!ht]
		\centering
		\includegraphics[width=\textwidth,height=\textheight,keepaspectratio]{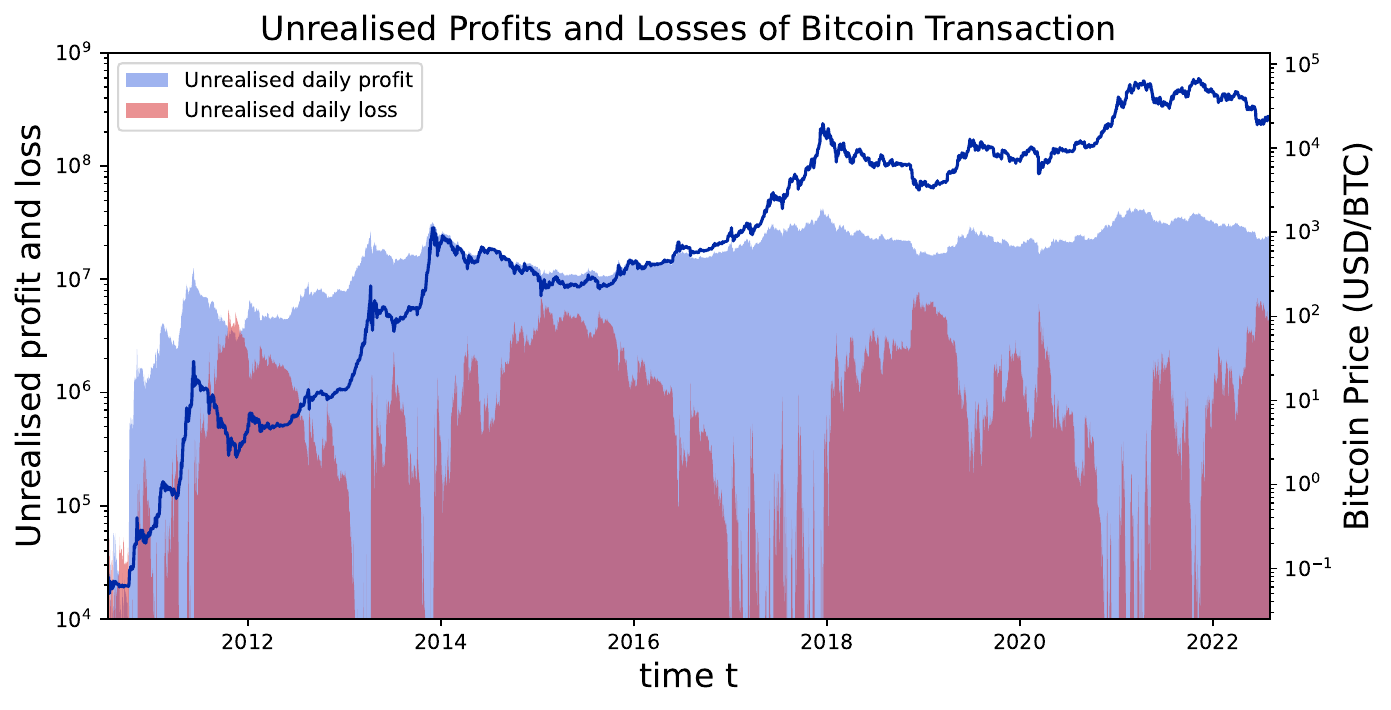}
		\caption{Total unrealized profit $P(t)$ per day
			defined by expression (\ref{eq:avg-unrel-profit}) (light blue) and total unrealised loss $L(t)$ per day defined by expression (\ref{eq:avg-unrel-loss}) (red).
			The blue line is the bitcoin price time series (right scale).}
		\label{fig:absolute_ptlt}
	\end{figure}
	
	\begin{figure}[ht]
		\centering
		\includegraphics[width=\textwidth,height=\textheight,keepaspectratio]{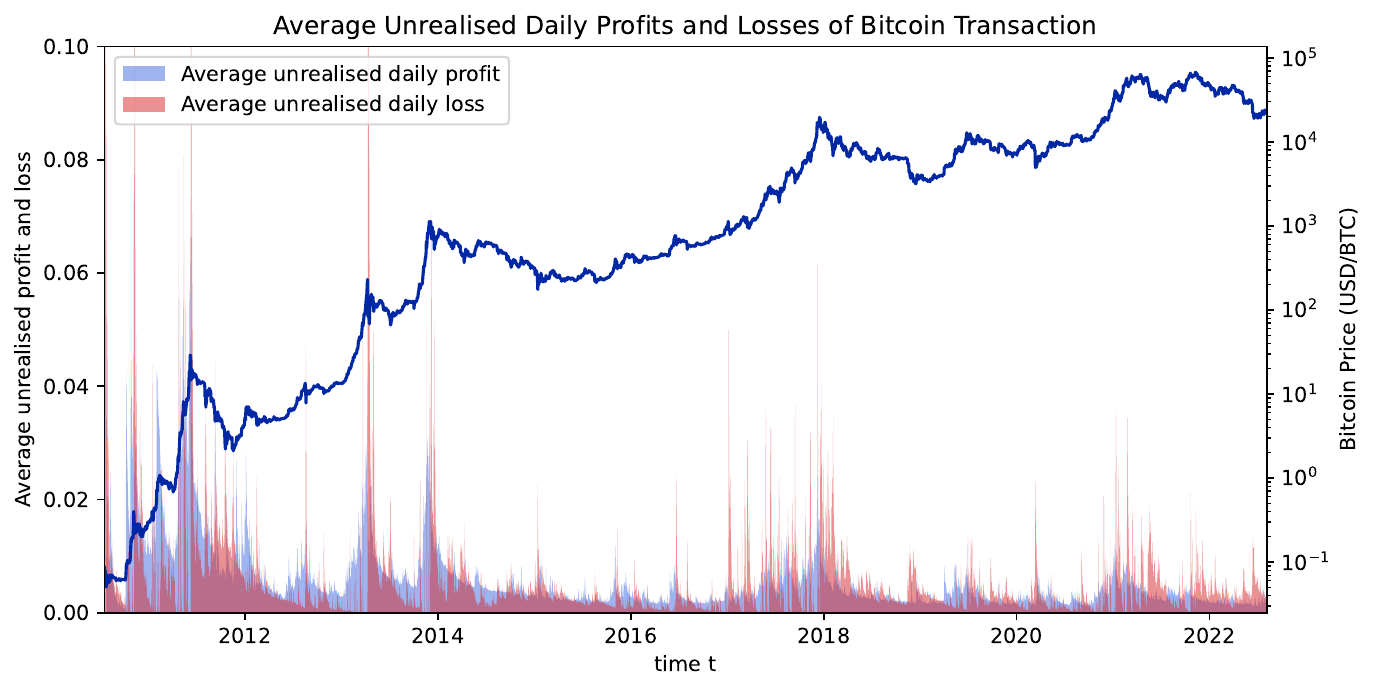}
		\caption{\label{f:avg-unrealized-gain-loss} Average unrealized profit $\bar{P}(t)$ per day
			defined by expression (\ref{eq:avg-normal-unrel-profit}) (light blue) and average loss rate $\bar{L}(t)$ per day defined by 
			expression (\ref{eq:avg-normal-unrel-loss}) (red).
			The blue line is the bitcoin price time series (right scale).}
		\label{fig:normalized_ptlt}
	\end{figure}

	Figure \ref{fig:absolute_ptlt} presents the unrealized profit $P_t$ and loss $L_t$ per day summed over all positions defined by
	\begin{equation}
		\label{eq:avg-unrel-profit}
		P_t:=\int_{r>0} r~\text{pdf-B2M}_t(r)  dr
	\end{equation}
	and 
	\begin{equation}\label{eq:avg-unrel-loss}
		L_t:=-\int_{r\leq 0} r~\text{pdf-B2M}_t(r)   dr~,
	\end{equation}
	respectively.
	
	Figure \ref{fig:normalized_ptlt} shows the normalised profit $\bar{P}(t)$ and loss $\bar{L}(t)$ defined by . 
	\begin{equation}
		\label{eq:avg-normal-unrel-profit}
		\bar{P}(t):=\frac{\int_{r>0} r~\text{pdf-B2M}_t(r)dr}{\int_{r>0} \text{pdf-B2M}_t(r)dr}
	\end{equation}
	and 
	\begin{equation}\label{eq:avg-normal-unrel-loss}
		\bar{L}(t):=-\frac{\int_{r\leq 0} r~\text{pdf-B2M}_t(r)dr}{\int_{r\leq 0} \text{pdf-B2M}_t(r)dr}~,
	\end{equation}
	
	As shown in figure \ref{fig:absolute_ptlt}, in general, the average unrealized profits exceed the amplitude of the average loss most of the time, which can be expected in an overall rising price dynamics.
	Prior to the peaks of the three major bubbles, the average unrealized loss is zero and there are distinct peaks in the average unrealized profit. At times of crashes, the average loss sharply increases. The sharp increases are more visible 
	in figure \ref{fig:normalized_ptlt}, which uses a linear vertical scale compared with the logarithmic scale in figure \ref{fig:absolute_ptlt}.
	Moreover, this figure exemplifies that the average daily profit return increases fast toward the end of bubbles,
	while the average daily loss return jumps sharply just after each bubble peak.
	Then, both the average daily profit return and daily loss return tend to decay following bubble peaks.
	
	During the growth of the bubble that ends in Dec. 2017 \cite{gerlach2019dissection}, the local peaks of the average profit 
	can be associated with the succession of local corrections of the price during the bubble rise, as previously discussed
	for the dynamics of $f_p(t)$ shown in figure \ref{f:unrealized-profits}.
	These characteristic bubble price patterns are well-fitted by the Log-periodic Power Law Singularity (LPPLS) model of financial bubbles, 
	as for instance documented in \cite{wheatley2019bitcoin}.
	
	\pagebreak
	\clearpage
	
	\subsection{Distributions of Realized Returns}     
	A Book-to-Market return at some time $t$ is a theoretical value that an investor would enjoy if he was to sell at the precise time $t$, 
	at that specific quoted price, neglecting transaction costs and slippage. 
	When an investor owns bitcoins, he can enjoy the feeling of obtaining virtual gains or suffer from the feeling of virtual losses.
	Book-to-Market financial data allows for an instantaneous characterization of existing positions and thus of the risk 
	exposition as well as potential returns. But this is the realm of expectation and hope.
	Then, for various reasons including the implementation of rational trading strategies, but also the influence of less rational rumours, herding influences, FOMO (fear of 
	missing out), need for cash and so on, the investor will sell. The sale transforms a virtual gain or loss into a real one.
	Realized returns include an additional key element to the statistical properties of financial market prices, namely the decision of the investor to make a transaction.
	
	Using a loose analogy with quantum mechanics and in particular with the de Broglie-Bohm pilot wave theory \cite{Bohm52-a,Bohm52-b},
	the Book-to-Market prices are like pilot trajectories guiding the probability of obtained realised prices. 
	Realized prices are obtained under the decision and action of the investor that transforms a potential gain or loss into 
	a real gain or loss, often close to the potential one but sometimes different due to lack of market liquidity.
	Book-to-Market returns are the domain of what is potential while
	the realized returns are hard facts that impact directly the wealth and welfare of the investor. Realized returns
	are obtained through market or limit orders \cite{book-limit-order16} that can also impact the price of the market, when large.
	As the disposition effect illustrates, Book-to-Market returns are susceptible to the fallacy of delaying selling losing 
	assets in order to avoid ``taking the loss'' while hoping that the market will recover. 
	Given the fact that going from a virtual (Book-to-Market) return to a realized one requires the decisive step of 
	making a decision to sell and go through the concrete steps of implementing the decision, one can surmise that 
	the distribution of Book-to-Market returns described in the previous subsection and the distribution of realized returns may have
	different properties while sharing others.
	
	Equation (\ref{eq:b2m}) has defined the Book-to-Market return distribution $\text{B2MD}_t(r)$.
	In contrast,  the number of coins traded within the interval $ [t-\Delta t, t] $ that were bought at time $ \tau $ is $ n_{\tau}(t-1) - n_{\tau}(t) $. The corresponding total realized return resulting from the transaction of these bitcoins is $ \ln(p(t)/p(\tau)) $. 
	To make possible the comparison over different holding times $t-\tau$, as before, we use the average return per day (\ref{log-ret-day}).
	The density distribution of realized returns (profit and loss) is thus given by
	\begin{equation}\label{eq:pnld}
		\text{PnL}(r) :=  {\sum_{i \in {\cal S}_r} \left(n_{\tau_i}(t-\Delta t) - n_{\tau_i}(t)\right)  \over  N(t-\Delta t) - N(t)} ,  
	\end{equation} 
	where ${\cal S}_r$ is the set of $i$'s such that $\bar{r}_{\tau_i\rightarrow t} =r$, where $\bar{r}_{\tau_i\rightarrow t}$ is defined by Eq. (\ref{log-ret-day}).
	Remarkably, the Bitcoin ledger allows us to reconstruct this distribution $\text{PnL}(r)$ explicitly, which is otherwise
	unobservable in standard financial markets.
	
	\begin{figure}[ht]
		\centering
		
		\subfloat[Subfigure 1 list of figures text][]{
			\includegraphics[draft=false,width=0.48\textwidth]{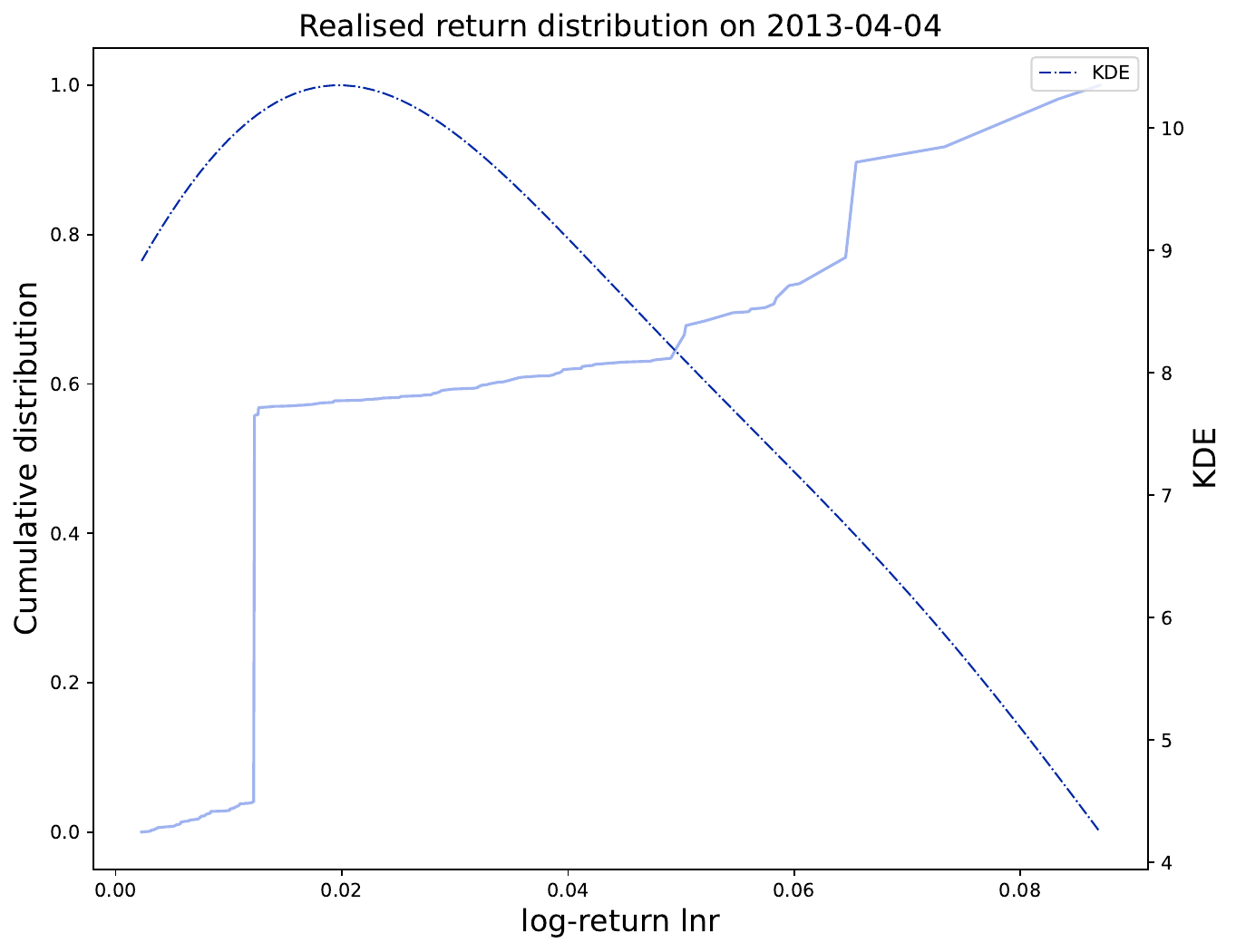}
			\label{f:rel-ret-subfig1}}
		\subfloat[Subfigure 2 list of figures text][]{
			\includegraphics[draft=false,width=0.48\textwidth]{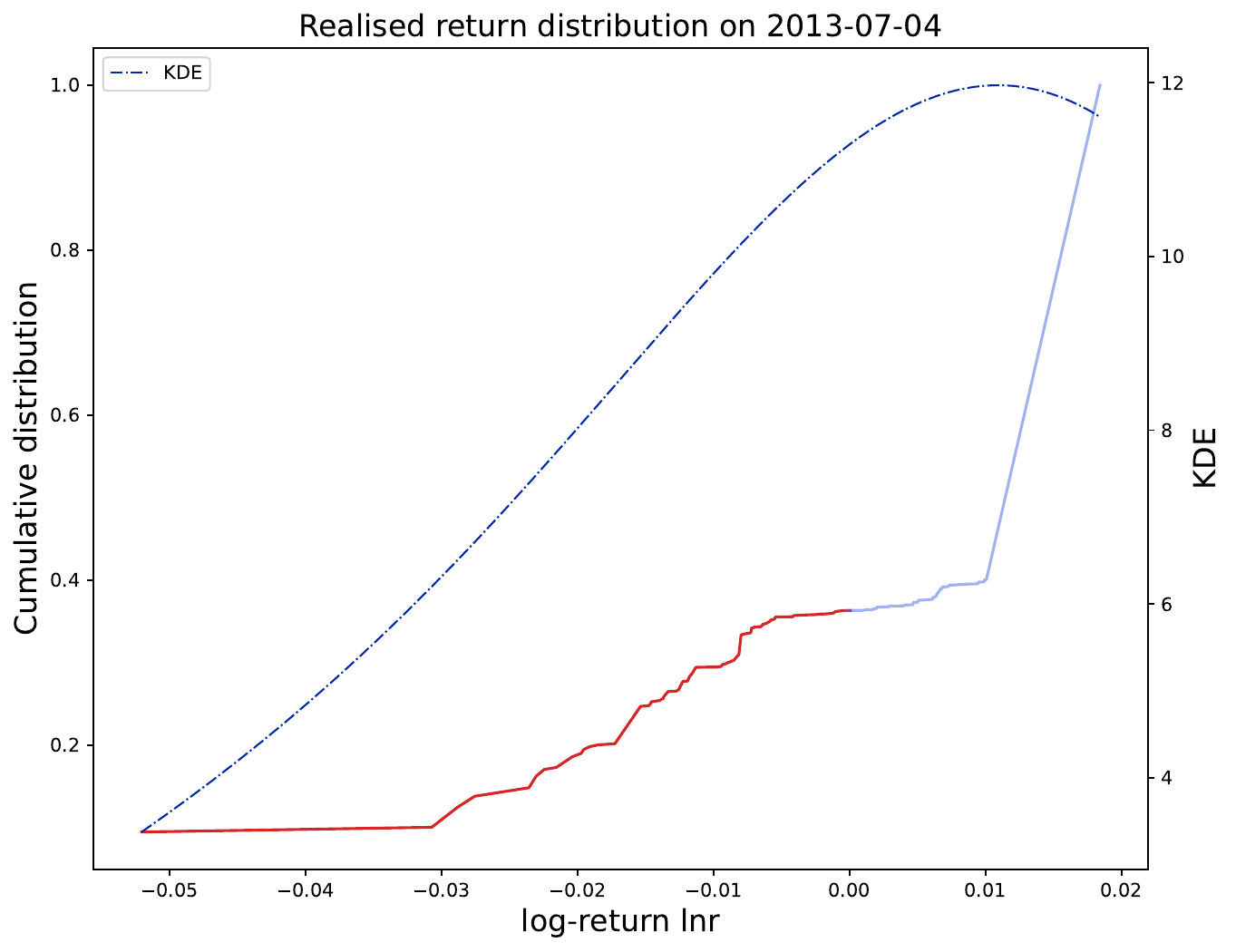}
			\label{f:rel-ret-subfig2}}\\
		
		\subfloat[Subfigure 3 list of figures text][]{
			\includegraphics[draft=false,width=0.48\textwidth]{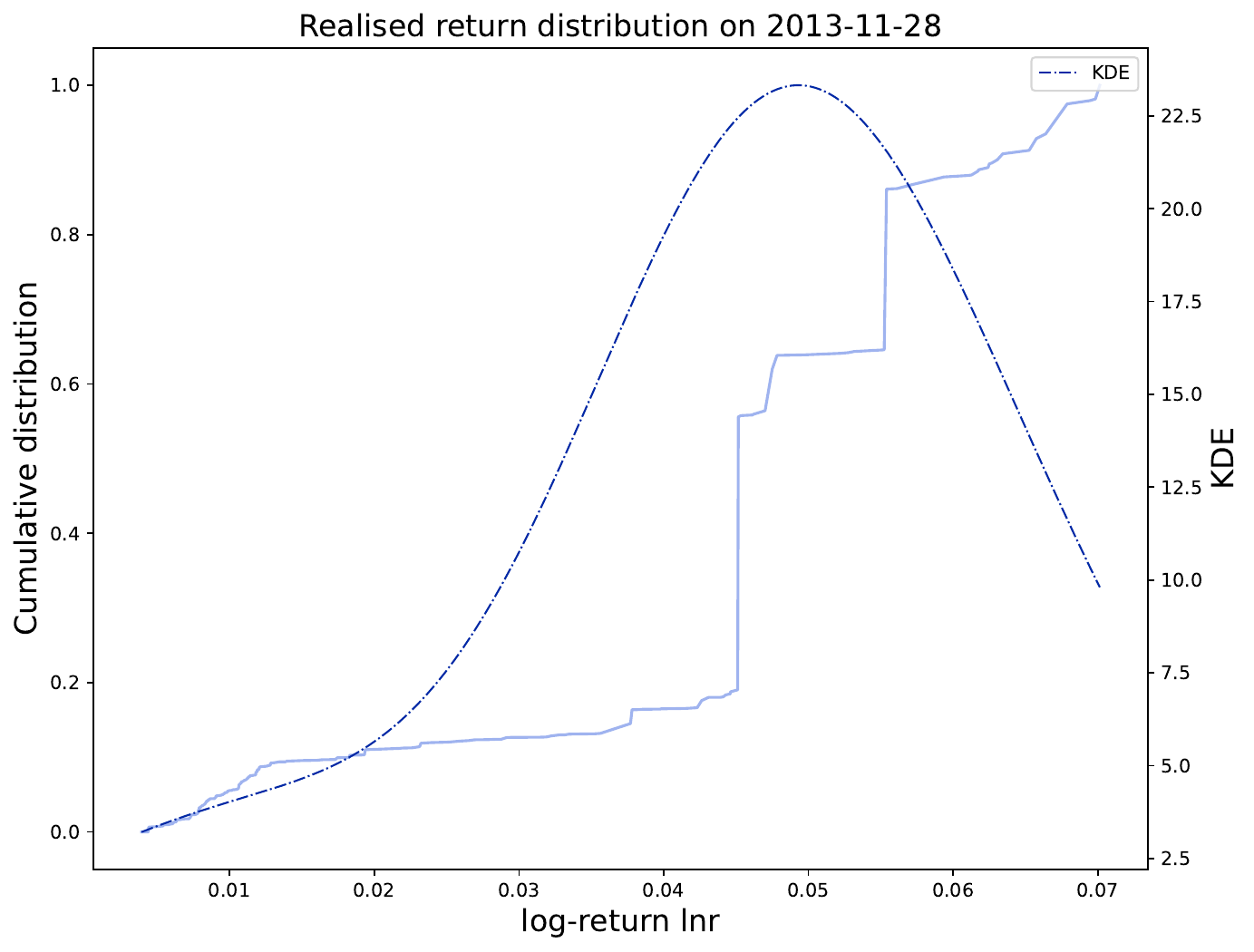}
			\label{f:rel-ret-subfig3}}
		\subfloat[Subfigure 4 list of figures text][]{
			\includegraphics[draft=false,width=0.48\textwidth]{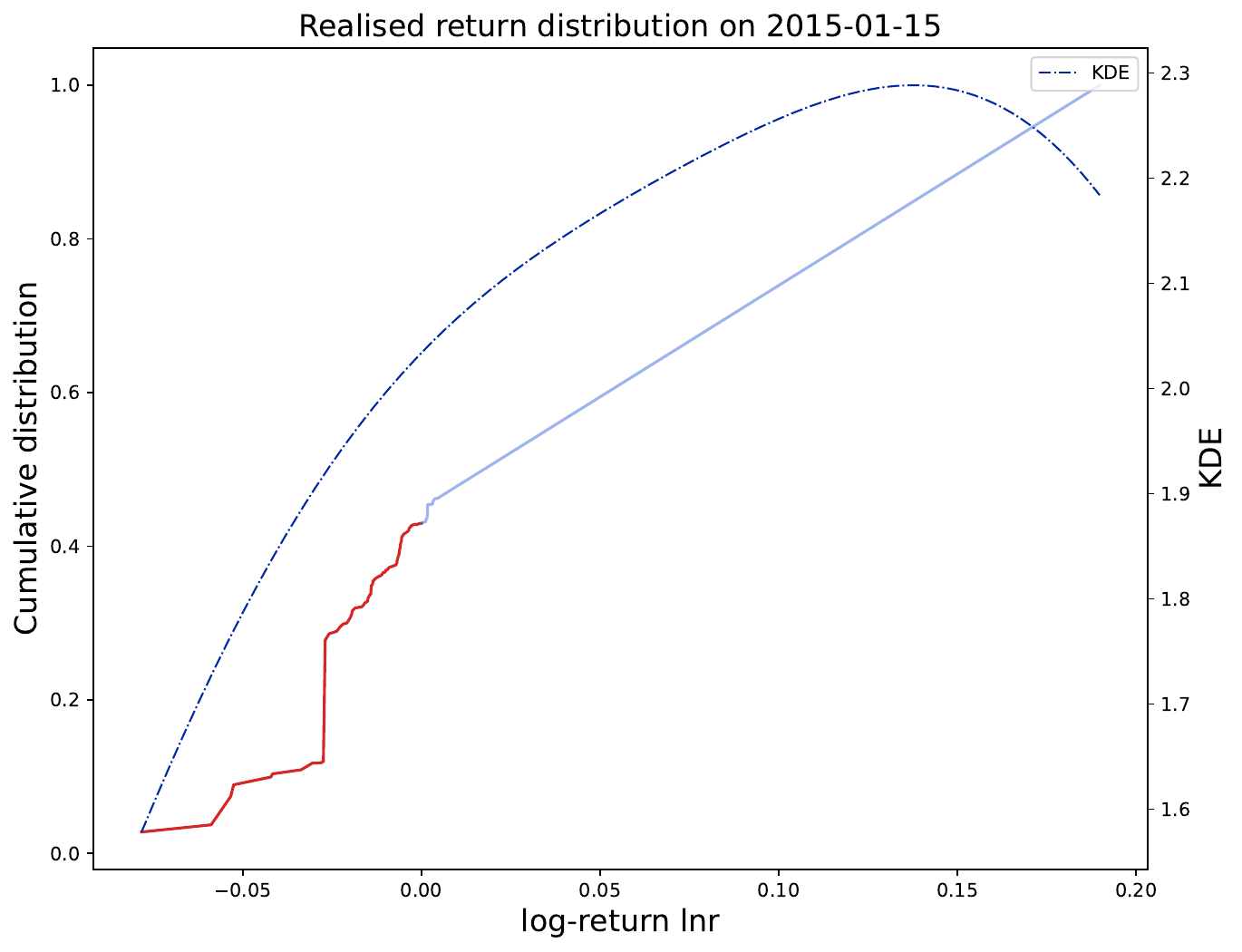}
			\label{f:rel-ret-subfig4}}\\
		
		\subfloat[Subfigure 3 list of figures text][]{
			\includegraphics[draft=false,width=0.48\textwidth]{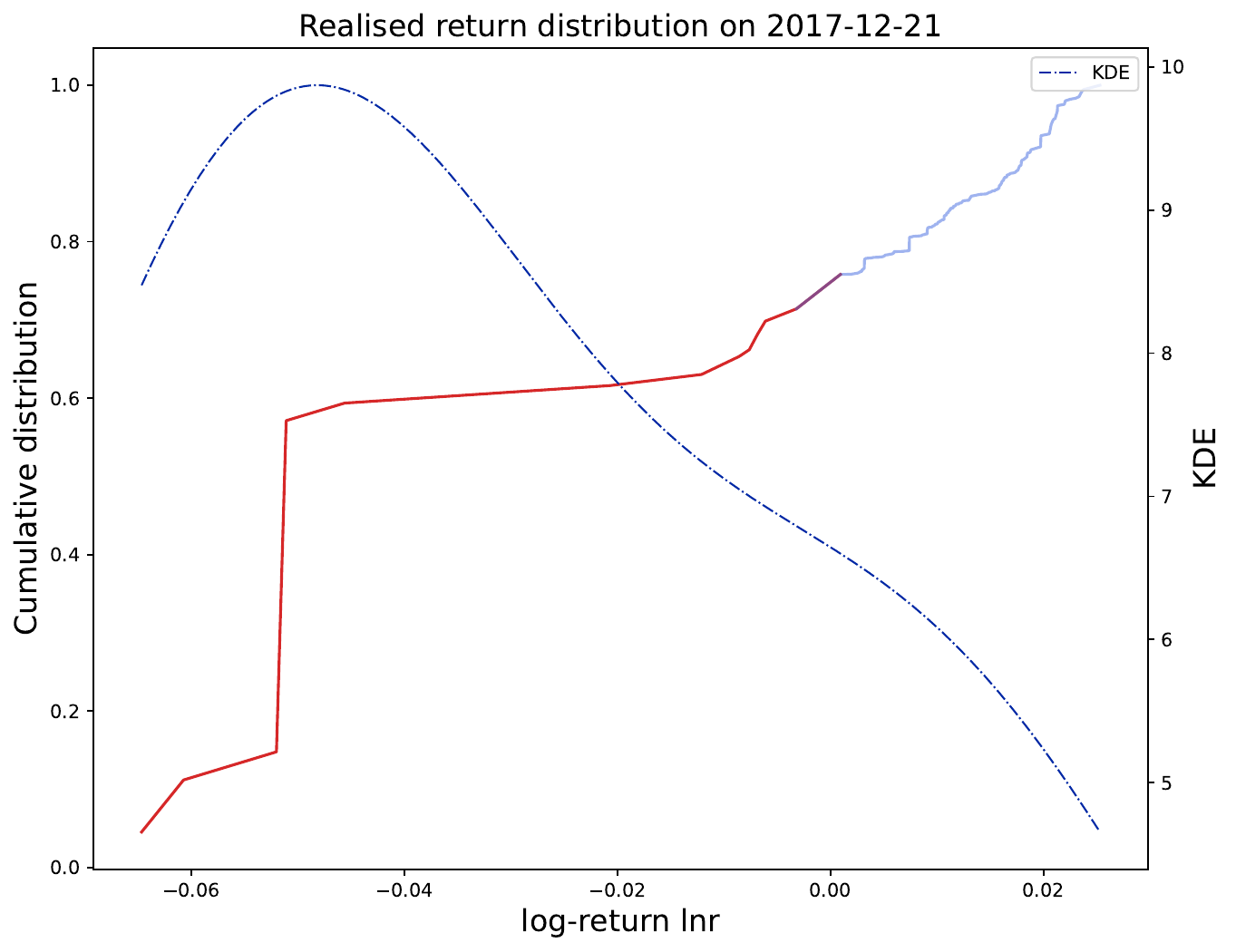}
			\label{f:rel-ret-subfig5}}
		\subfloat[Subfigure 4 list of figures text][]{
			\includegraphics[draft=false,width=0.48\textwidth]{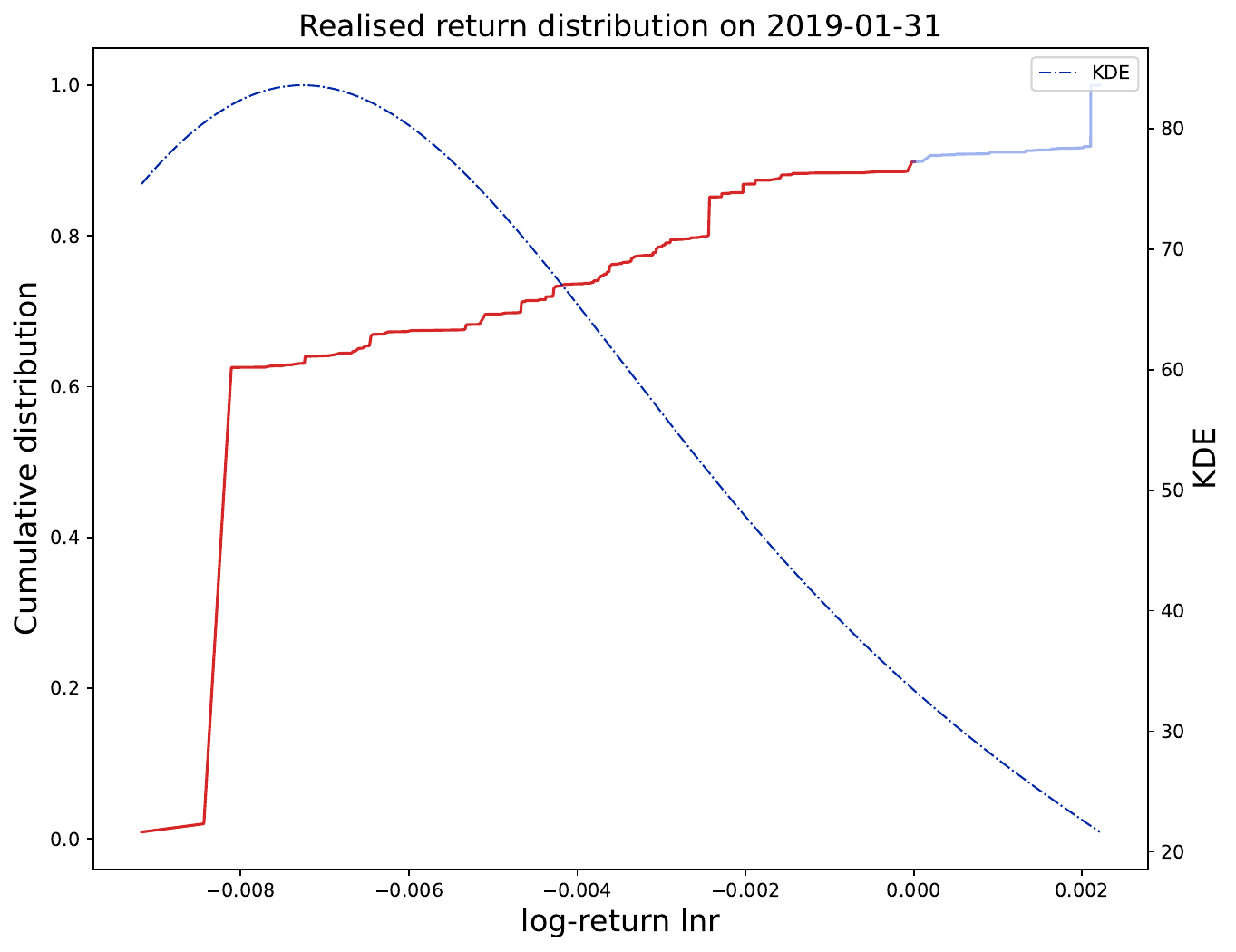}
			\label{f:rel-ret-subfig6}}
		
		\caption{\label{f:rel-ret} Cumulative normalized distribution $\int_{-\infty}^r \text{PnL}(r') dr'$, where ${PnL}(r)$
			is defined by expression (\ref{eq:pnld}), of realized returns per day of bitcoins traded within the interval $ [t-\Delta t, t]$, 
			at the six different analysis times $\{t_i, i=1, ..., 6\}$ given in Table \ref{t:times}. 
			The range of positive (resp. negative) returns is shown in blue (resp. red).
			The panels should be compared with the cumulative normalized distribution of (unrealized) Book-to-Market returns shown in figure. \ref{fig_cum_density}.
		}
		\label{fig_realised_cum_distr}
	\end{figure}
	
	Figure \ref{f:rel-ret} shows the cumulative distribution $\int_{-\infty}^r \text{PnL}(r') dr'$ at the six different analysis times $\{t_i, i=1, ..., 6\}$ given in Table \ref{t:times}. One can observe several striking differences compared with the cumulative normalized distribution of (unrealized) Book-to-Market returns shown in figure \ref{fig_cum_density}.
	For the peak times (panels (a), (c), and (e)), the realized returns are still mostly positive but they are significantly smaller 
	overall than the Book-to-Market returns. This is explained by looking at the transition probability function (see section \ref{rwyhnbrg2}),
	whose power law structure shows that the major fraction of the transaction volume occurs short term over
	the first few time intervals from the current time.  
	Further, panels (a), (b), and (c) of figure \ref{f:rel-ret} show a strong uni-modal distribution, i.e., the returns are concentrated 
	on specific value, which can be interpreted as ``taking profit'' decisions of investors who have bought at a specific time in the past.
	
	For the trough times (panels (b), (d) and (f)), a much larger percentage (close to or larger than $40\% $, even $80\%$) of realized returns are negative compared with those in figure \ref{fig_cum_density}. This puts in evidence a tendency for investors to have exhausted
	their patience and sell finally, thus taking their losses. 
	
\begin{figure}[ht]
		\centering
		\includegraphics[draft=false,width=0.75 \linewidth]{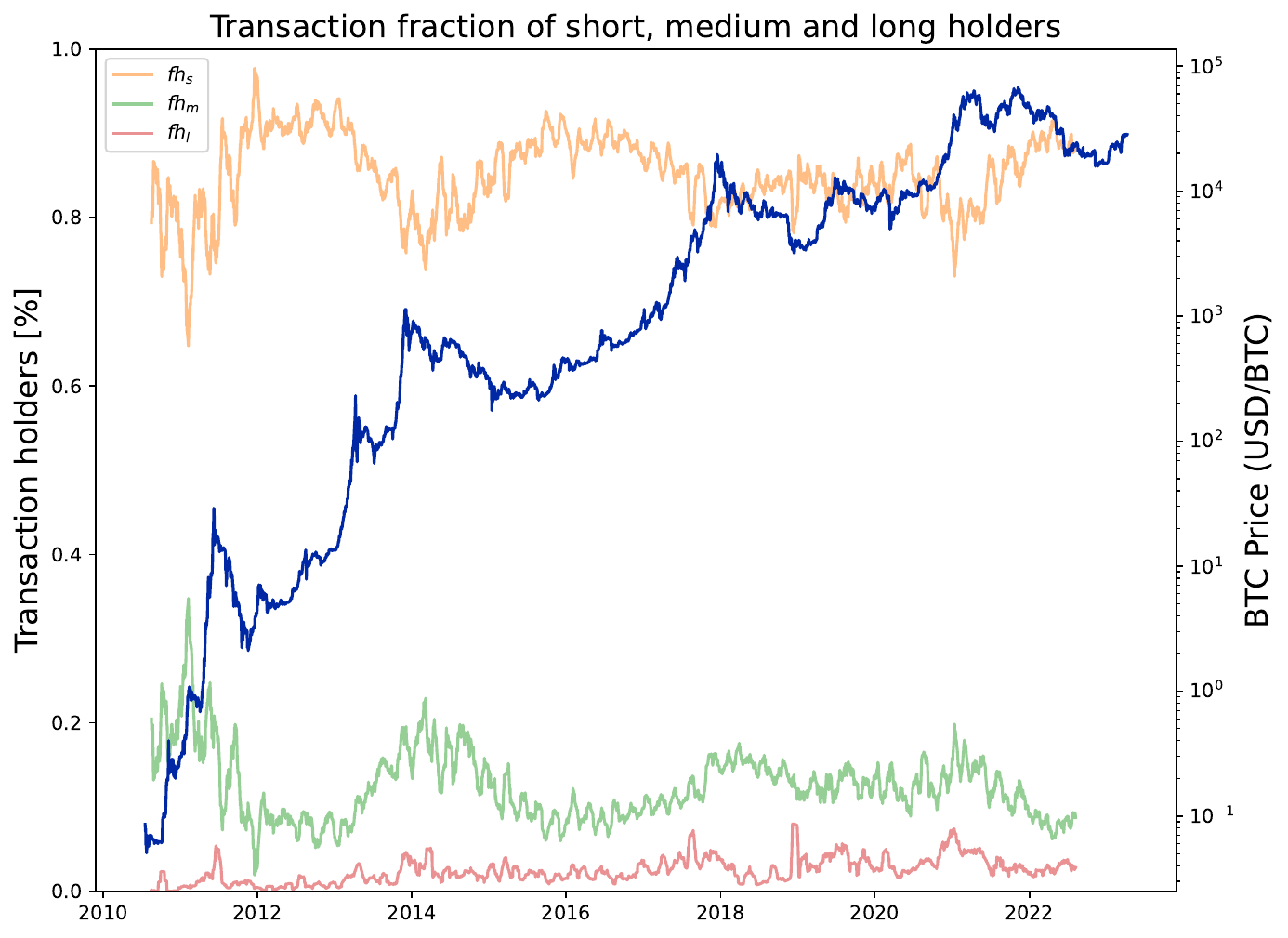}
		\caption{\label{f:fraction-volume} Analogously to figure \ref{f:fraction-holders}, 
			this figure shows the fractions of short-term (up to 30 days), medium-term (between one month and one year old) and long-term (older than one year) bitcoin volume contribution to the total transaction volume $V(t)$ at each point in time.  The log price of bitcoin (right scale, blue line) is also shown.
			The orange (green) line indicates the centered 90-day moving median of the boundaries between the short-term and medium-term fraction time series, which provide smoothed estimations of the fraction of long-term sand of medium-term holders.
		}
		\label{re_sml}
	\end{figure}

	Another interesting question to investigate is how the short-, medium- and long-term holders transacted with price and time. As a  metric, we construct the fractions $V_{f_{h_i}}(t)$ of short-, medium- and long-term holders who transacted over time defined by
	\begin{equation}\label{eq:vf-holders}
		V_{f_{h_i}}(t) :=\dfrac{\sum_{\tau\in{\mathcal{T}_i}}n_{\tau}(t)\pi_{\tau\rightarrow t}(t)}{\sum_{\tau\leq t}n_{\tau}(t)\pi_{\tau\rightarrow t}(t)} = \dfrac{V_{\mathcal{T}_i}(t)}{V(t)}\quad\quad i=1,2,3~,
	\end{equation}
	where the different windows $ \mathcal{T}_i $ are defined as the same as in equation \ref{eq:f-holders}; $V(t)$ denotes the total transaction volume at time $t$, $V_{\mathcal{T}_i}(t)$ denotes the total number of transacted bitcoins whose latest transaction time is in window $ \mathcal{T}_i $. 
	Analogously to figure \ref{f:fraction-holders}, figure \ref{f:fraction-volume} shows the fractions of short- (up to 30 days), medium- (between one month and one year old) and long-term (older than one year) bitcoin volume contribution to the total transaction volume $V(t)$ at each point in time.  
	The temporal structures are similar to those observed in figure \ref{f:fraction-holders}, with a large difference in an inversion of relative importance.
	Here, the short-term holders dominate the transaction volume by far.
	Figure \ref{re_sml} demonstrates the very interesting phenomenon that short-term holders are more active when the price decreases. When the price increases, short-term holders tend to keep bitcoins and are less active. In contrast, the activity of medium-term holders is approximately synchronized with the ups and downs of bitcoin price.
	
	\begin{figure}[ht]
		\centering
		\includegraphics[draft=false,width=1\linewidth]{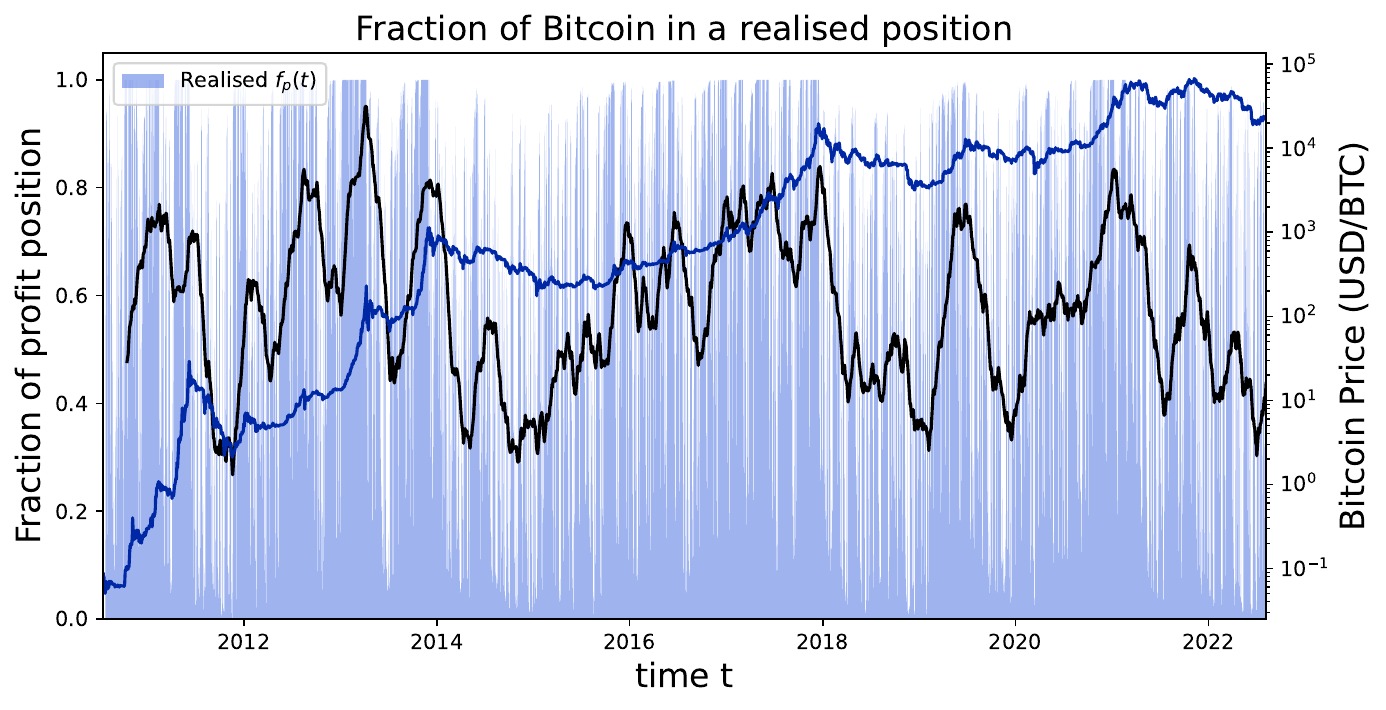}
		\caption{\label{f:realized-profits} Percentage $100 \cdot f_{p}(t)$ of bitcoin holdings sold at a realized profit (blue area, left scale) over time. It is computed as the value of the complementary cumulative PnL distribution expressed at zero return. As the curve is quite erratic, we also show the centered 90-day moving average of $f_{p}(t)$ in dark.}
	\end{figure}
	
	\begin{figure}
		\centering
		\includegraphics[width=1 \linewidth]{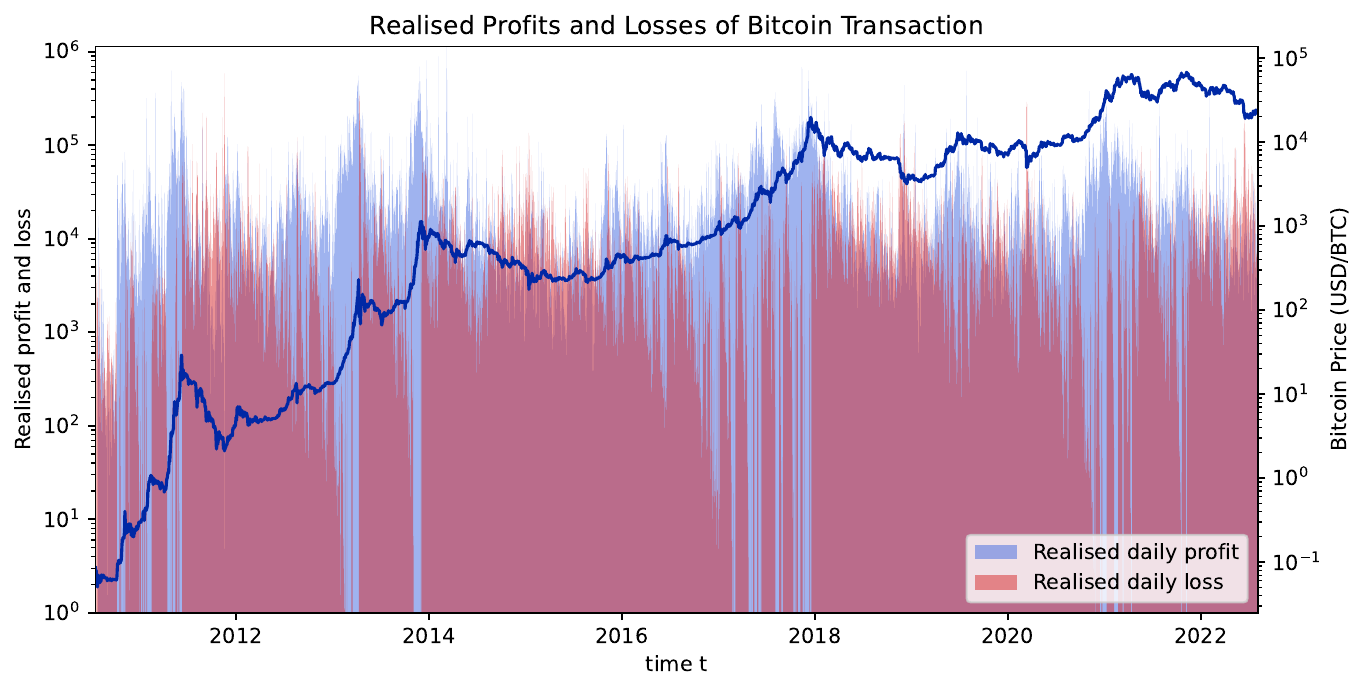}
		\caption{Analogously to figure \ref{fig:absolute_ptlt},
			this figure shows the absolute realized gains (blue) and losses (red).
			The blue line is the bitcoin price time series (right scale).}
		\label{fig:my_label}
	\end{figure}
	
	\begin{figure}[ht]
		\centering
		\includegraphics[draft=false,width=1\linewidth]{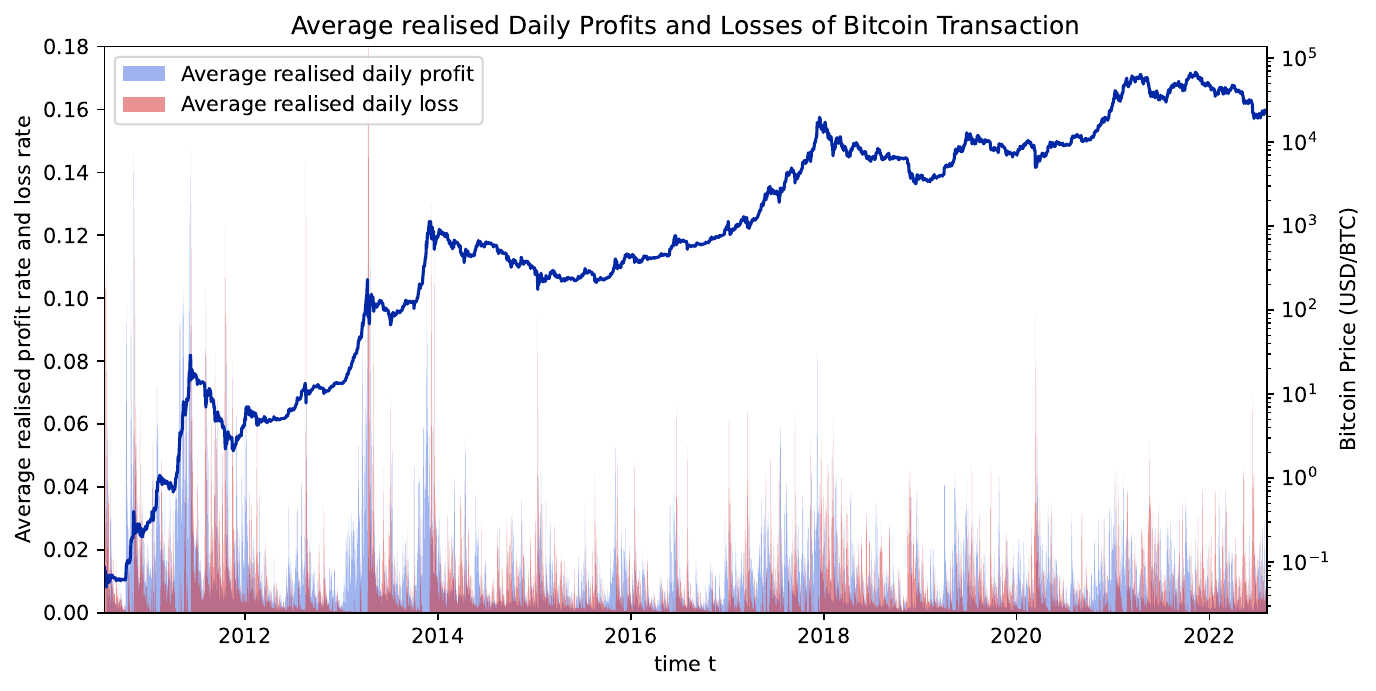}
		\caption{\label{f:avg-realized-gain-loss} Analogously to figure \ref{f:avg-unrealized-gain-loss},
			this figure shows the average realized gains $\bar{P}(t)$ (blue) and losses $\bar{L}(t)$ (red).
			The blue line is the bitcoin price time series (right scale).
		}
	\end{figure}
	
	Analogously to Equations \eqref{eq:f-unreal-prof} and Equations (\ref{eq:avg-unrel-profit}-\ref{eq:avg-unrel-loss}), the fraction of profitable transactions, as well as the average realized profit and loss, are shown in figures \ref{f:realized-profits} and \ref{fig:my_label}. 
	Qualitatively, both plots look like chopped versions of their counterparts for the Book-to-Market returns. 
	The percentage of profitable transactions shown in figure \ref{f:realized-profits}, though showing similar trends as in figure \ref{f:unrealized-profits}, strongly fluctuates between values far above and below 50\%. At no time does it remain steady, or evolve slowly, except 
	over a few months before the three major bubble peaks, for which the upward price acceleration makes a short-term trading loss almost impossible.
	This chopped structure reveals the existence of very active short-term trading.
	Throughout the growth periods of the three bubbles, the fraction $f_{p}(t)$ of bitcoin holdings sold at a realized profit 
	shoots up to levels above 95\%, sometimes even touching 100\%. 
	Unsurprisingly as the price reached a new height, during bubble regimes, the majority of investors made a profit when selling their bitcoins, irrespective of the buying time or other decisions. The post-bubble peaks are exhibiting large drawdowns in which $f_{p}(t)$ drops
	transiently below 10\%, indicating almost all trades are made at a loss for the sellers.
	
	Analogously to figure \ref{f:avg-unrealized-gain-loss},
	figure \ref{f:avg-realized-gain-loss} shows the average realized gain rate $\bar{P}(t)$ (blue) and loss rate $\bar{L}(t)$ (red).
	The message is similar to that of figure \ref{f:realized-profits}, when comparing figure \ref{f:avg-unrealized-gain-loss} and figure 
	\ref{f:avg-realized-gain-loss}: the realized gains and losses are much more chopped, allowing us 
	to diagnose the prevalence of short-term trades. And one can visualize the phases of strong price growth
	that are associated with positive average returns (gain domain in blue). When gains are significant and persistent,
	this is likely to generate a general sentiment that ``one can only win in this market'' and ``fear of missing out'' 
	the great gains that others are accumulating.
	
	These data provide insights into the kind of mentality that is likely to prevail during the formation of an asset bubble.
	Given the extraordinary rate of profitable transactions that are observed during the major bubble regimes but also 
	at other times that could qualify as small bubbles (see e.g. \cite{gerlach2019dissection}), the probability that an investor meets another investor who did not recently benefit from trading BTC becomes so small that overwhelming positive experiences and news are exchanged between investors. This dynamic creates and fosters an enthusiastic market sentiment such that the demand increases to ever higher levels. Then, 
	as the price peaks and bitcoin price starts to fall, almost all new short-term trades become losses, triggering 
	negative sentiments that may percolate through the network of investors.
	This is visible for instance after the peak of the 2017 bubble when the fraction of profitable transactions drops sharply immediately after the peak, with some temporary rebounds afterwards. Because the major contributions to transaction volume stem from short-term trades, this shows that, already during the post-peak drawdown, most investors re-bought, in the hope of buying at a cheaper price, while supposedly expecting
	the price ascent to continue, as it has done for the preceding months. As the hope is not realized, many short-term traders
	exit their positions at a loss, as quantified in figure \ref{f:realized-profits}. 
	The persistence of generally low values of the fraction $f_{p}(t)$ of bitcoin holdings sold at a realized profit 
	during the whole drawdown duration, interspersed by short-lived peaks, 
	suggests that investors remain hopeful for months to years that the bearish trend will reverse. 
	Notice that the short-lived periods when $f_{p}(t)$ passes over 50\% during the long-lived drawdowns are associated with 
	short-lived price rebounds. 
	
	\pagebreak
	\clearpage
	
	\section{Dependence of Transaction Flows on Holding Times}
	
	\subsection{Power-law Distribution of Bitcoin Holding Times \label{rwyhnbrg2}}
	
	Equation (\ref{eq:pi-tau-t}) allows one to determine the transition probability $\pi_{\tau\rightarrow t}(t)$ of bitcoins. It is by definition the fraction of exchanged bitcoins born at $\tau$ to those born at $\tau$ but still held at $t-\Delta t$. Thus, for a bitcoin born at $\tau$ and next exchanged at $t$, $t-\tau$ is the holding time.
	$\pi_{\tau\rightarrow t}(t)$ is thus a function
	of both the birth date $\tau$ and ``present'' time $t$. Keeping in mind the analogy between $n_{\tau}(t)$
	and an age distribution (see section \ref{trhj3yrbgq}), it is natural to reinterpret $\pi_{\tau\rightarrow t}(t)$
	as a function of age (or holding time) $z :=t-\tau$ and present time $t$. It is thus convenient to change the notation to $\pi_{z}(t)$. 
	
	\begin{figure}[ht]
		\centering
		\subfloat[Subfigure 1 list of figures text][]{
			\includegraphics[width=0.48\textwidth]{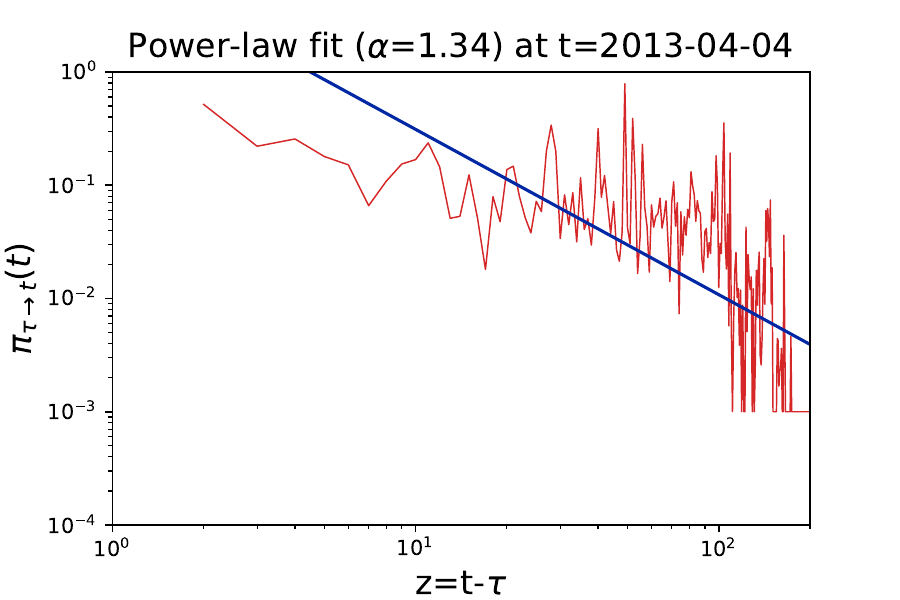}
			\label{f:stp-pl-subfig1}}
		\subfloat[Subfigure 2 list of figures text][]{
			\includegraphics[width=0.48\textwidth]{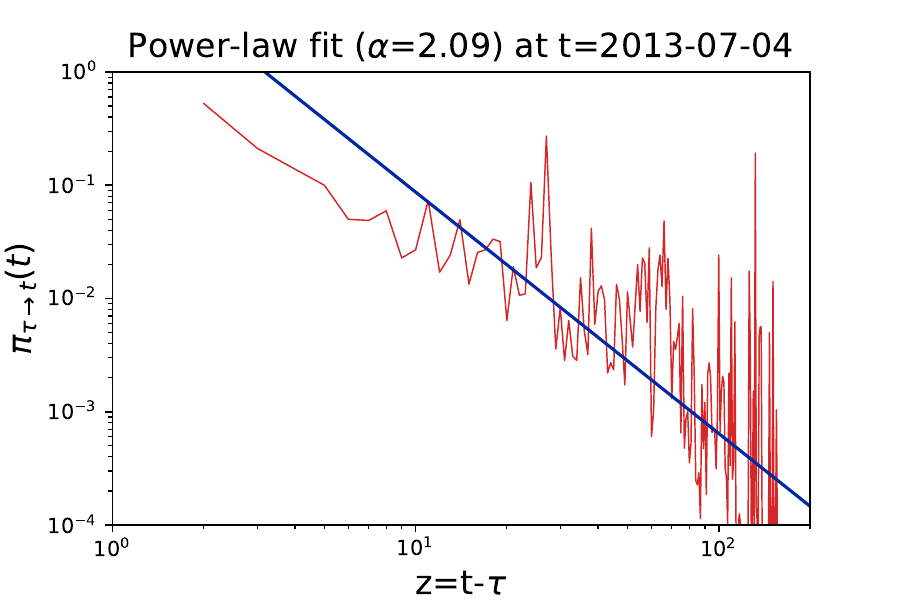}
			\label{f:stp-pl-subfig2}}\\
		
		\subfloat[Subfigure 3 list of figures text][]{
			\includegraphics[width=0.48\textwidth]{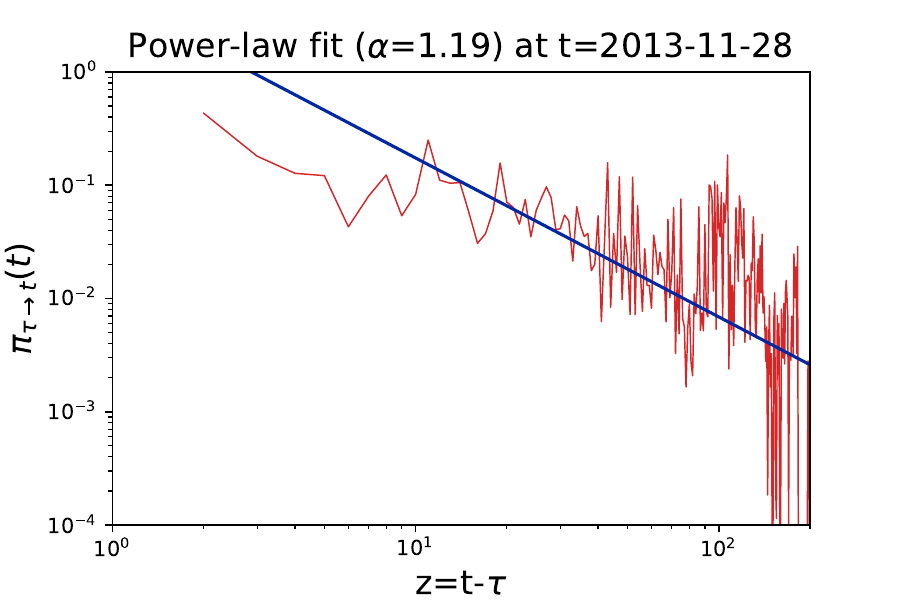}
			\label{f:stp-pl-subfig3}}
		\subfloat[Subfigure 4 list of figures text][]{
			\includegraphics[width=0.48\textwidth]{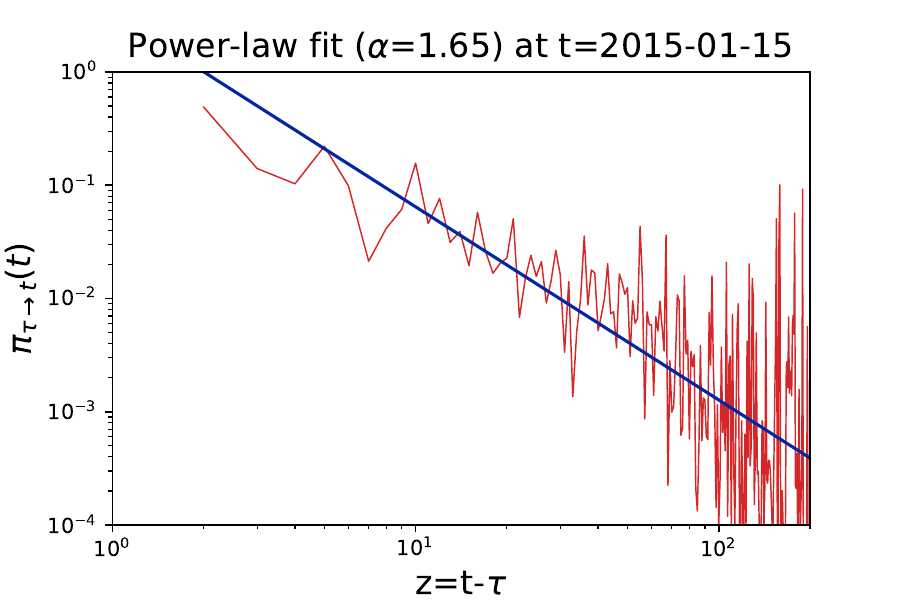}
			\label{f:stp-pl-subfig4}}\\
		
		\subfloat[Subfigure 3 list of figures text][]{
			\includegraphics[width=0.48\textwidth]{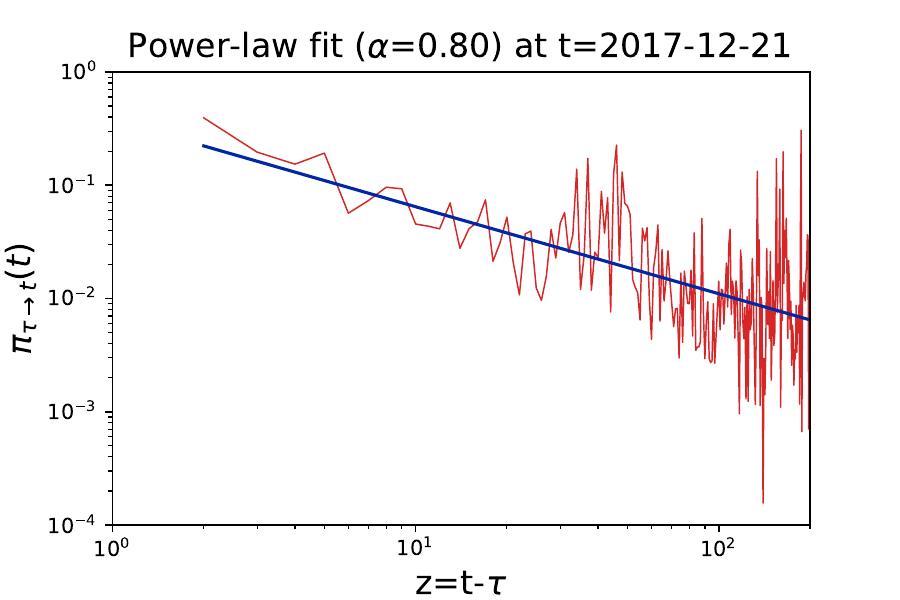}
			\label{f:stp-pl-subfig5}}
		\subfloat[Subfigure 4 list of figures text][]{
			\includegraphics[width=0.48\textwidth]{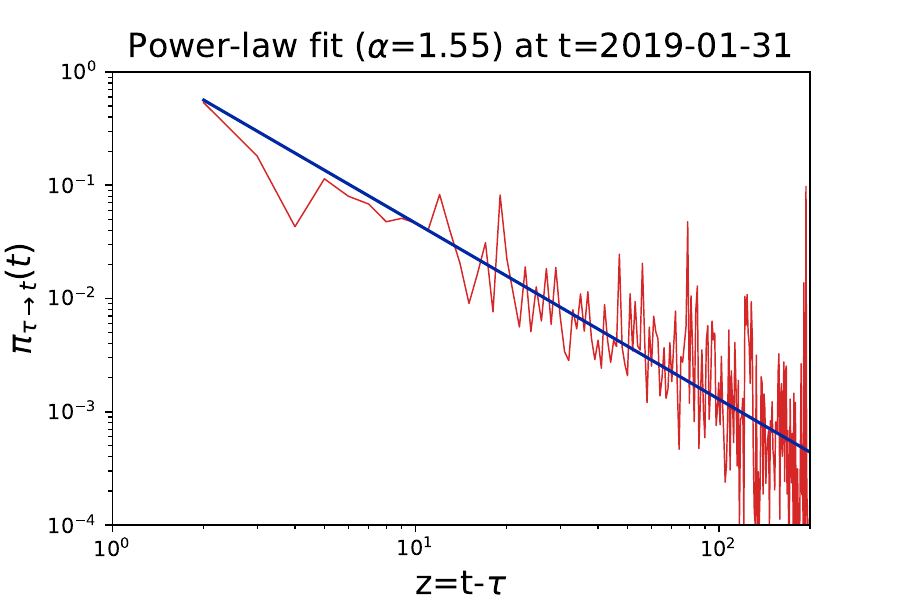}
			\label{f:stp-pl-subfig6}}
		\caption{\label{f:stp-pl} Double logarithmic representation of the fraction $\pi_{z}(t_i)$ of 
			bitcoins that have been exchanged at the six different analysis times $\{t_i, i=1, ..., 6\}$ given in Table \ref{t:times} as a function of age $z := t-\tau$ from 1 week to 200 weeks.
			The blue straight lines represent the best calibration of the red curve by a power law $\ln [\pi_{z}(t_i)] = -\alpha_i \ln(z) + \beta_i$, where the exponents $\alpha_i$ are given above each plot. In order to assess the quality of the power law fit, we 
			perform the calibration over different windows $[1,...,z_{max}]$ where the upper boundary $z_{max}$ is swept in integer steps from $125\Delta t$ to  $200\Delta t$, where $\Delta t =1$ week, amounting to 76 nested calibrations. The reported exponent $\alpha$ is the mean over these 76 calibrations.}
	\end{figure}
	
	Figure \ref{f:stp-pl} shows $\pi_{z}(t_i)$ as a function of age $z$, where $z$ spans from 1 week to 200 weeks 
	for the six different analysis times $\{t_i, i=1, ..., 6\}$ given in Table \ref{t:times}. The first observation is that the double-logarithmic scale
	suggests that $\pi_{z}(t)$ decays approximately as a power law function of age $z$, roughly as $1/z^\alpha$.
	This power law hypothesis is made quantitative by calibrating a power law $\ln [\pi_{z}(t_i)] = -\alpha_i \ln(z) + \beta_i$ to the empirical transition probabilities where the exponents $\alpha_i$ are given above each plot. Because of the large fluctuations, the fits are not perfect but indicative of the possibilities that $\pi_{z}(t)$ might be suitably represented by a power law. We observe that the exponent $\alpha_t$ of the power law fluctuates across the six analysis times. This may reflect
	(i) statistical fluctuations due to insufficient data and/or noisy data, (ii) genuine variations as a function of market regimes 
	or (iii) a more subtle multifractal structure (which is investigated in the next section).
	
\begin{figure}[ht]
		\centering
			\includegraphics[width=0.75\textwidth]{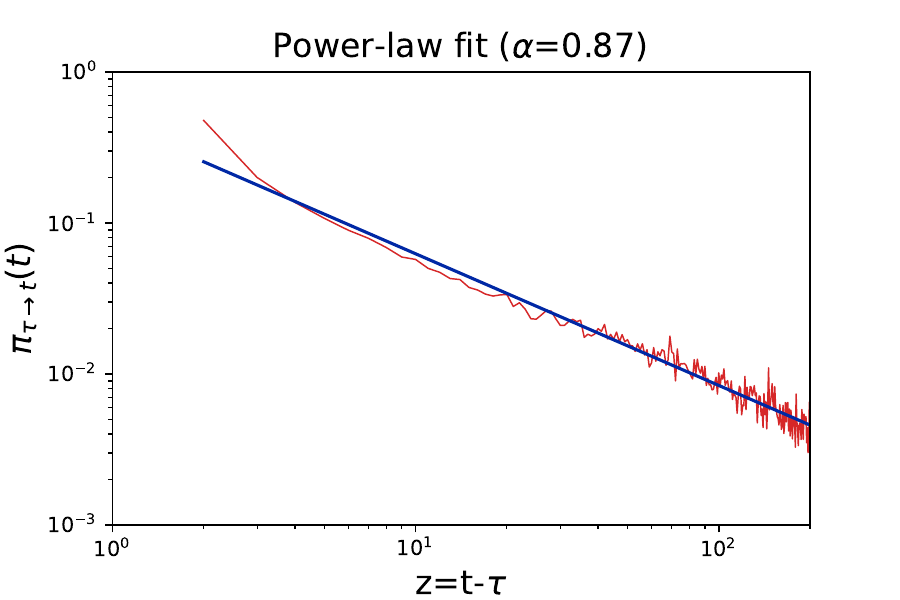}
			\label{f:exp-stp-subfig1}
			\caption{\label{f:exp-stp} Time average transition probability $\pi(z)$ defined by expression (\ref{eq:estpd})
				as a function of age $z$, by sampling with the resolution $\Delta t=1$ week all times $t \in [2013.1,2022.8]$. 
			}
		\end{figure}

	While ensemble averaging of noisy power law functions is a delicate procedure (see e.g. the discussion in Chapter 14 in \cite{sornette2006critical}), it is nevertheless interesting to ask whether a well-defined transition probability as a function of age can emerge, which could be approximately independent of time. In other words, is there a stationary transition probability $\pi_{z}$ independent of $t$? If it exists, it can be obtained by taking the average 
	\begin{equation}\label{eq:estpd}
		\pi(z):=E_t[\pi_z(t)]~.
	\end{equation}
	We perform this expectation by sampling with the resolution $\Delta t=1$ week all times $t \in [2013.1,2022.8]$. Figure \ref{f:exp-stp} shows the time average transition probability $\pi(z)$, which is visually well described by a pure power law function $\pi(z) \simeq {1 \over z^\alpha}$ with $\alpha \approx 0.87$. In order to check the validity of the power law, we performed a calibration of the logarithm of the empirically determined $\pi(z)$ by the function $\ln [\pi_{z}] = -\alpha \ln(z) + \beta$ in intervals $z \in [1,z_{max}]$, where $z_{max}$ was varied from $125$ to $200$. In this set of 76 calibrations, the exponent $\alpha$ is found to be centered on $0.855$ with a maximal value $0.873$ and a minimum value $0.842$, representing less than a $0.9\%$ variability around the mean value.
	
		The fact that the transition probability $\pi(z) \simeq {1 \over z^\alpha}$ is found to have an exponent $\alpha \approx 0.87$ smaller than $1$ implies that we observe here an intermediate asymptotic, which has to cross over to a faster decay with an exponent larger than $1$ or some other functional form in order to ensure normalization. Connecting with the response function formalism of complex social systems to exogenous or endogenous shocks 
		\cite{sornette2003endogenous,sornette2004endogenous,crane2008robust,saichev2010generation}, this result $\alpha \approx 0.87$ can be related to the so-called exo-critical regime where a bare response function $~\simeq {1 \over t^{1+\theta}}$, with $\theta >0$, to an external shock is renormalized into $~\simeq {1 \over t^{1-\theta}}$ by the cascades of social interactions up to some time scale $t^*$ beyond which the decay resumes in the form $~\simeq {1 \over t^{1+\theta}}$. For time scales smaller than $t^*$, the system behaves as in a critical state. Our result $\alpha \approx 0.87$ can thus be interpreted as the dynamics of bitcoin trading being in a critical state
		up at least to time scales of 200 weeks. We note that the value $\theta \approx 0.13$ found here is smaller than the values $\approx 0.3-0.4$ reported in previous studies \cite{sornette2003endogenous,sornette2004endogenous,crane2008robust,sornette2005multifractal}.
		
		In summary of this section, while there are significant sample-to-sample variations of the distribution of holding times, 
		the empirical evidence supports the hypothesis that the distribution of holding times has a heavy-tailed structure in the
		form of a power law with typical exponent $\alpha$ less than $1$. Taken at face value, this would imply
		an extremely wild regime, where the distribution is non-normalizable. In practice, there is obviously a natural
		holding time cut-off, which cannot be longer than the lifetime of Bitcoin's existence. Another possibility, as mentioned above,
		is the transition to another decay regime beyond the time scales investigated here.
		The fact that the power law exponent is so small indicates that bitcoin holding times are very broadly distributed
		with a significant fraction of investors holding for very long times. This behavior is common knowledge 
		in the specialized literature. One standard explanation for this behavior is the bet to profit from the growing scarcity of 
		new bitcoins, which are hard-coded in the bitcoin mining protocol.
		
		\pagebreak
		\clearpage
		
		\subsection{Transaction Flows Conditional on Holding Time}
		
		We introduce two variables $D_{\tau}^{+}(t)$ and $D_{T}^{-}(t)$ obtained from the Bitcoin blockchain data
		that inform on the contribution of different age cohorts to the transactions occurring at a given time.
		
		Recall that $n_{\tau}(t-1) - n_{\tau}(t)$ is the number of bitcoins born at time $\tau$ that are traded
		in the last time interval ending at $t$ and $S(t)$ is the number of bitcoins mined at $t$. 
		We define the fraction of the total transaction flow occurring at time $t$ contributed by the
		cohort of bitcoins with age equal to $t-\tau$ as
		\begin{equation}\label{eq:supply}
			D_{\tau}^{+}(t)  := \dfrac{n_{\tau}(t-1) - n_{\tau}(t)}{V(t) + S(t)}		\quad\quad\quad\forall~\tau < t
		\end{equation} 
		We refer to $D_{\tau}^{+}(t)$ as the ``$\tau \to t$ transaction flow fraction''.
		The $D_{\tau}^{+}(t)$'s obey the following identity 
		\begin{equation}\label{eq:d+sum}
			\sum_{\tau< t}D_{\tau}^{+}(t) +  {S(t) \over V(t) + S(t)}  = 1~.
		\end{equation}
		To derive Eq.\eqref{eq:d+sum}, we use identities \eqref{eq:n-t-t}, \eqref{eq:N-t-rec} and \eqref{eq:N-t-sum} and write
		\begin{eqnarray}\label{eq:dwrb22emand}
			S(t) + \big(V(t)+S(t)\big)\sum_{\tau< t}D_{\tau}^{+}(t) & =&  S(t) + \sum_{\tau< t}n_{\tau}(t-1) - \sum_{\tau< t}n_{\tau}(t)  \nonumber  \\
			& =&  S(t) + N(t-1)  -  (N(t) - n_t(t))  \nonumber \\
			&=&  S(t) + (N(t-1)  -  N(t)+ n_t(t)~.
		\end{eqnarray}
		Using \eqref{eq:n-t-t} and \eqref{eq:N-t-rec}, this yields
		\begin{equation}
			S(t) + \big(V(t)+S(t)\big)\sum_{\tau< t}D_{\tau}^{+}(t) =  S(t) - S(t) + V(t) + S(t) = V(t) + S(t)
		\end{equation}
		which obtains Eq.\eqref{eq:d+sum}.

		\begin{figure}[ht]
			\centering
			\subfloat[Subfigure 1 list of figures text][]{\includegraphics[draft=false,width=0.48\textwidth]{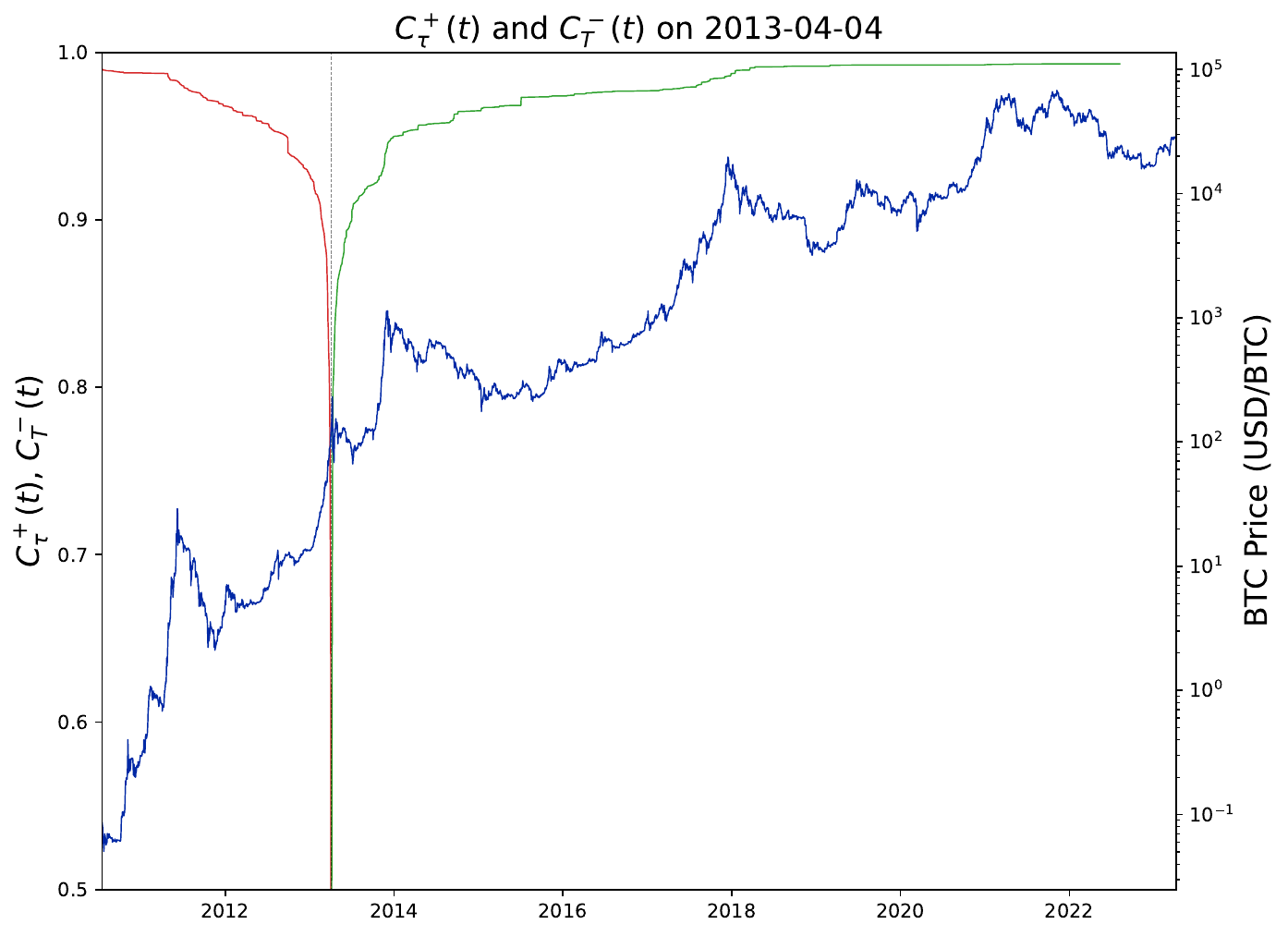}
				\label{f:dep-dep-subfig1}}
			\subfloat[Subfigure 2 list of figures text][]{
				\includegraphics[draft=false,width=0.48\textwidth]{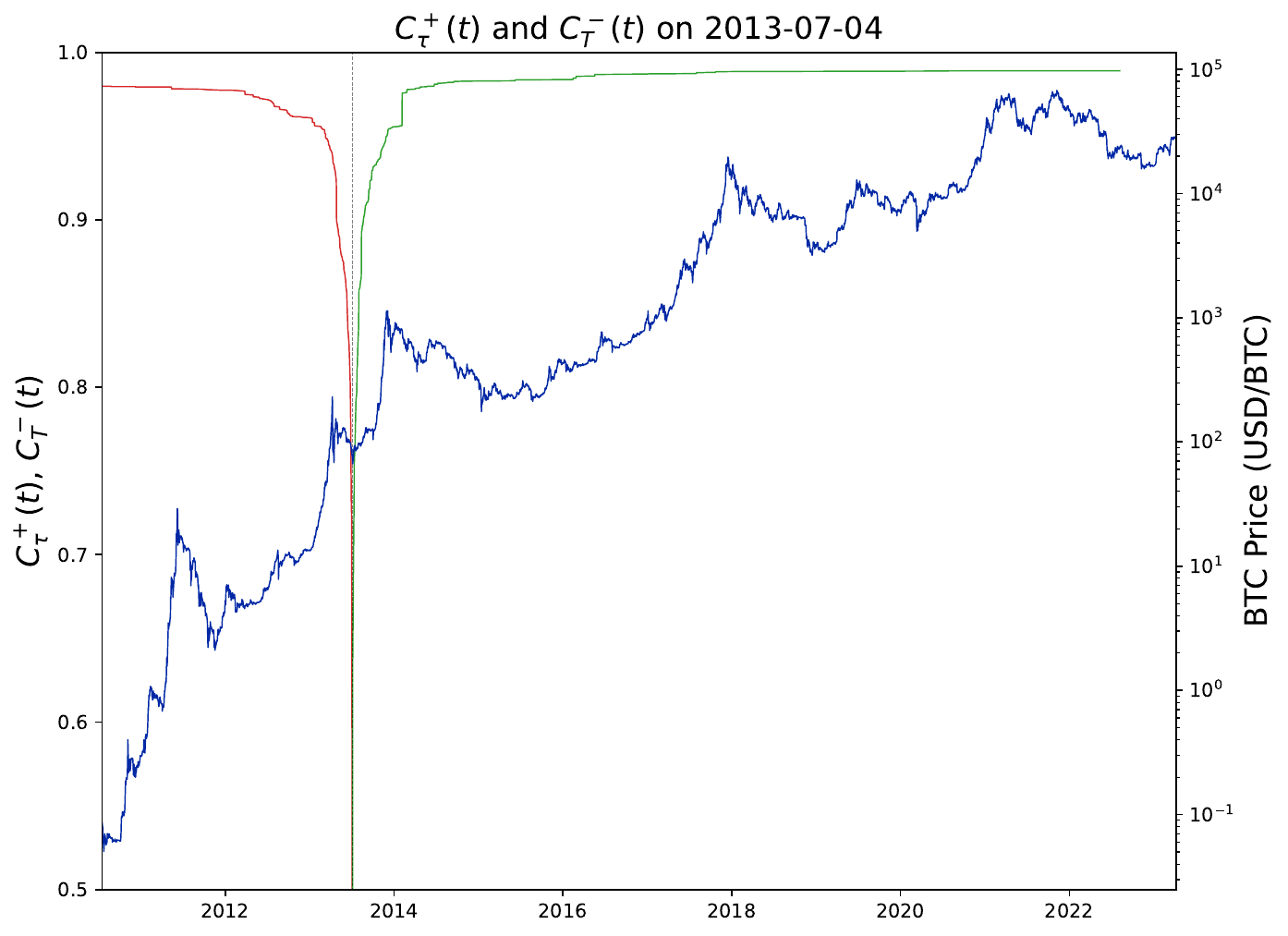}
				\label{f:dep-dep-subfig2}}\\
			\subfloat[Subfigure 3 list of figures text][]{\includegraphics[draft=false,width=0.48\textwidth]{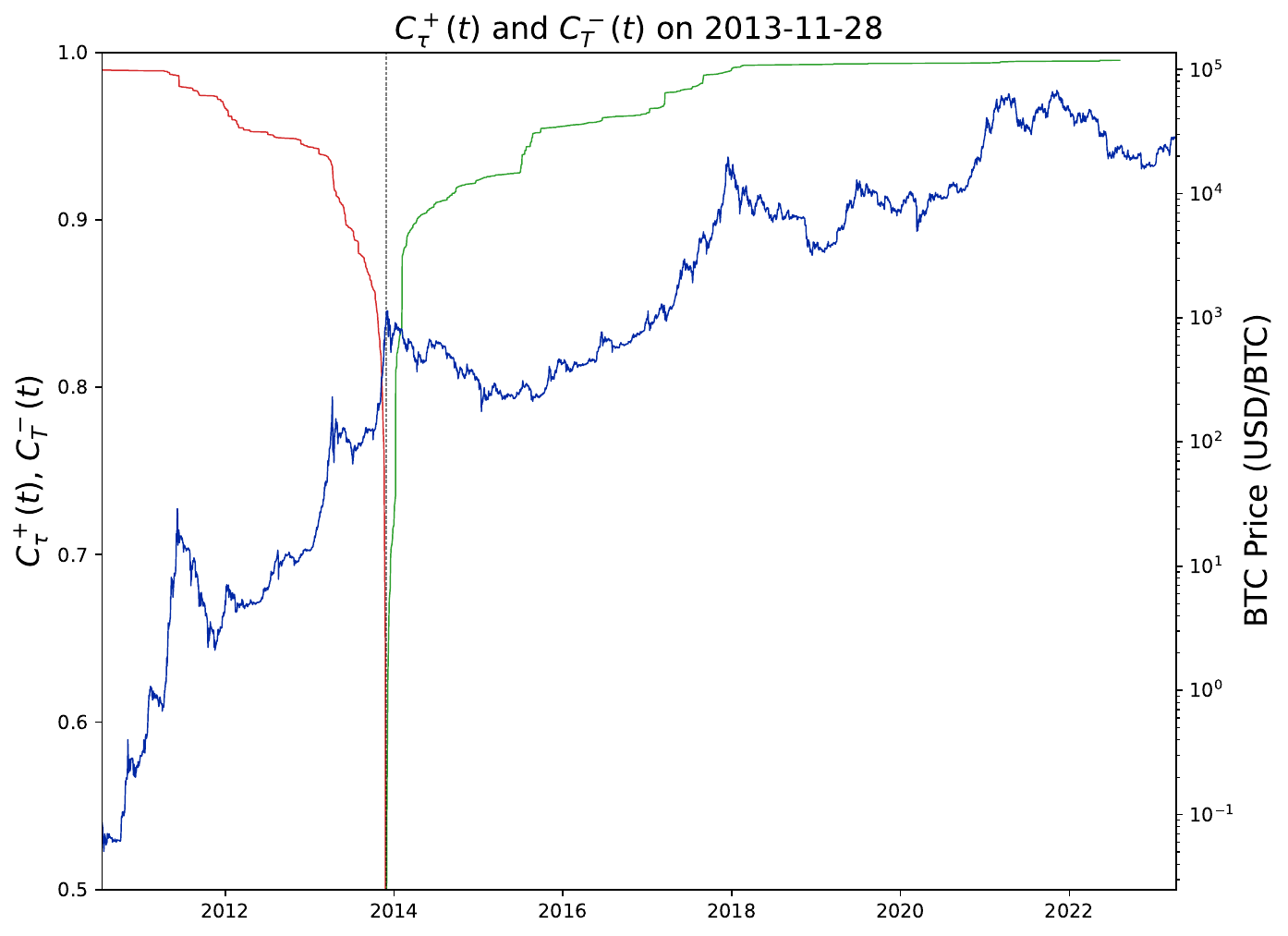}
				\label{f:dep-dep-subfig3}}
			\subfloat[Subfigure 4 list of figures text][]{
				\includegraphics[draft=false,width=0.48\textwidth]{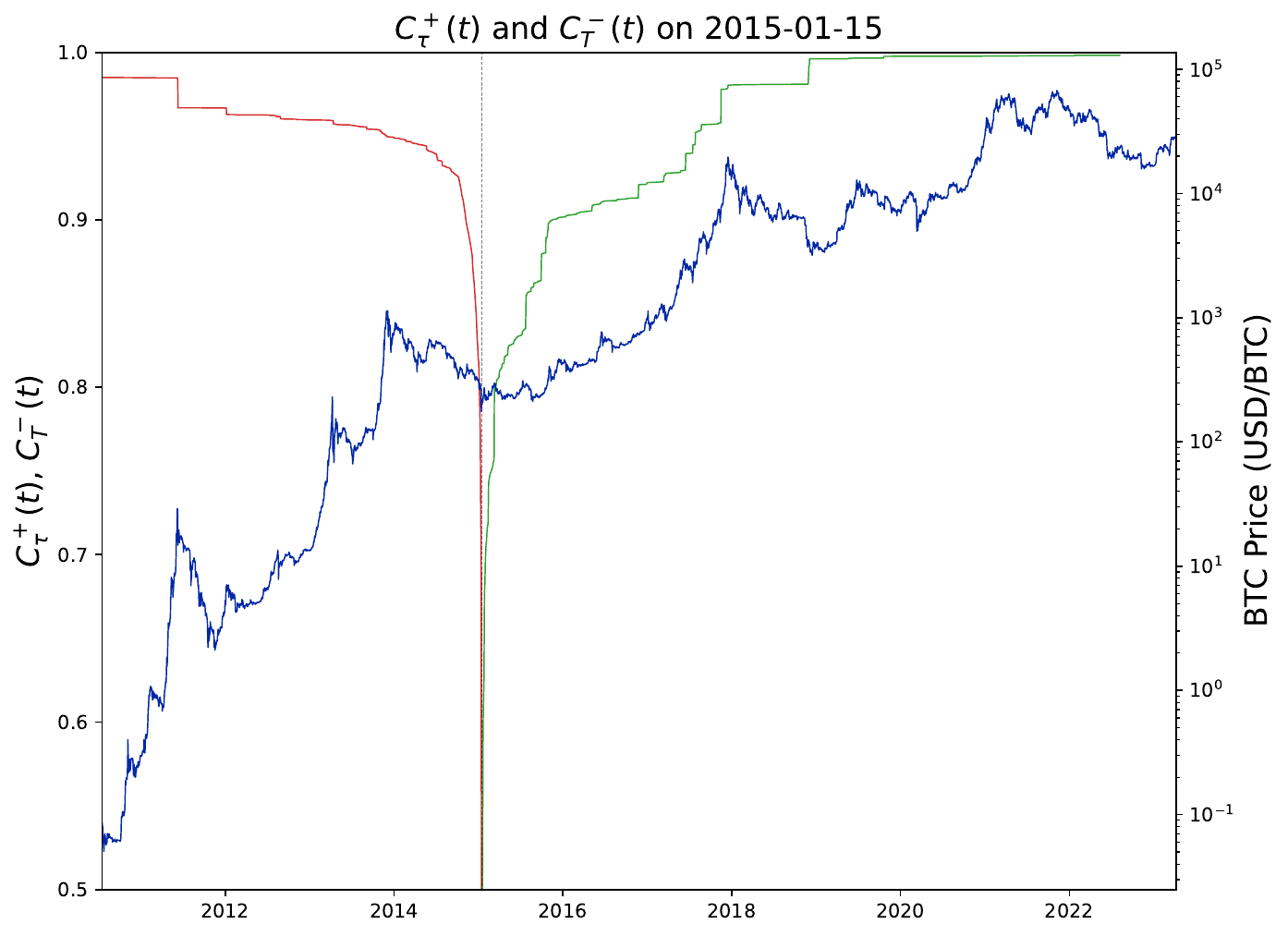}
				\label{f:dep-dep-subfig4}}\\
			
			\subfloat[Subfigure 3 list of figures text][]{
				\includegraphics[draft=false,width=0.48\textwidth]{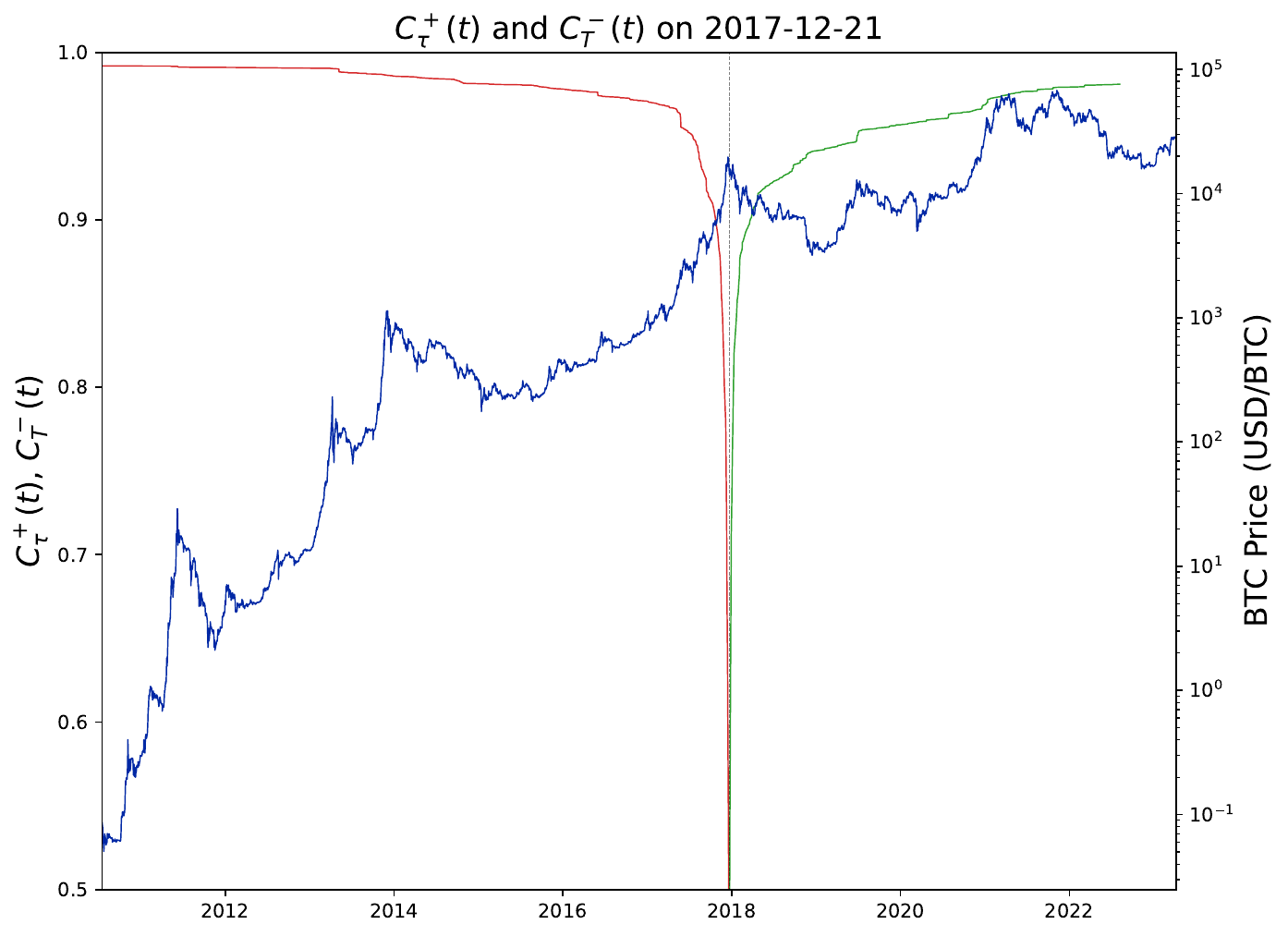}
				\label{f:dep-dep-subfig5}}
			\subfloat[Subfigure 4 list of figures text][]{
				\includegraphics[draft=false,width=0.48\textwidth]{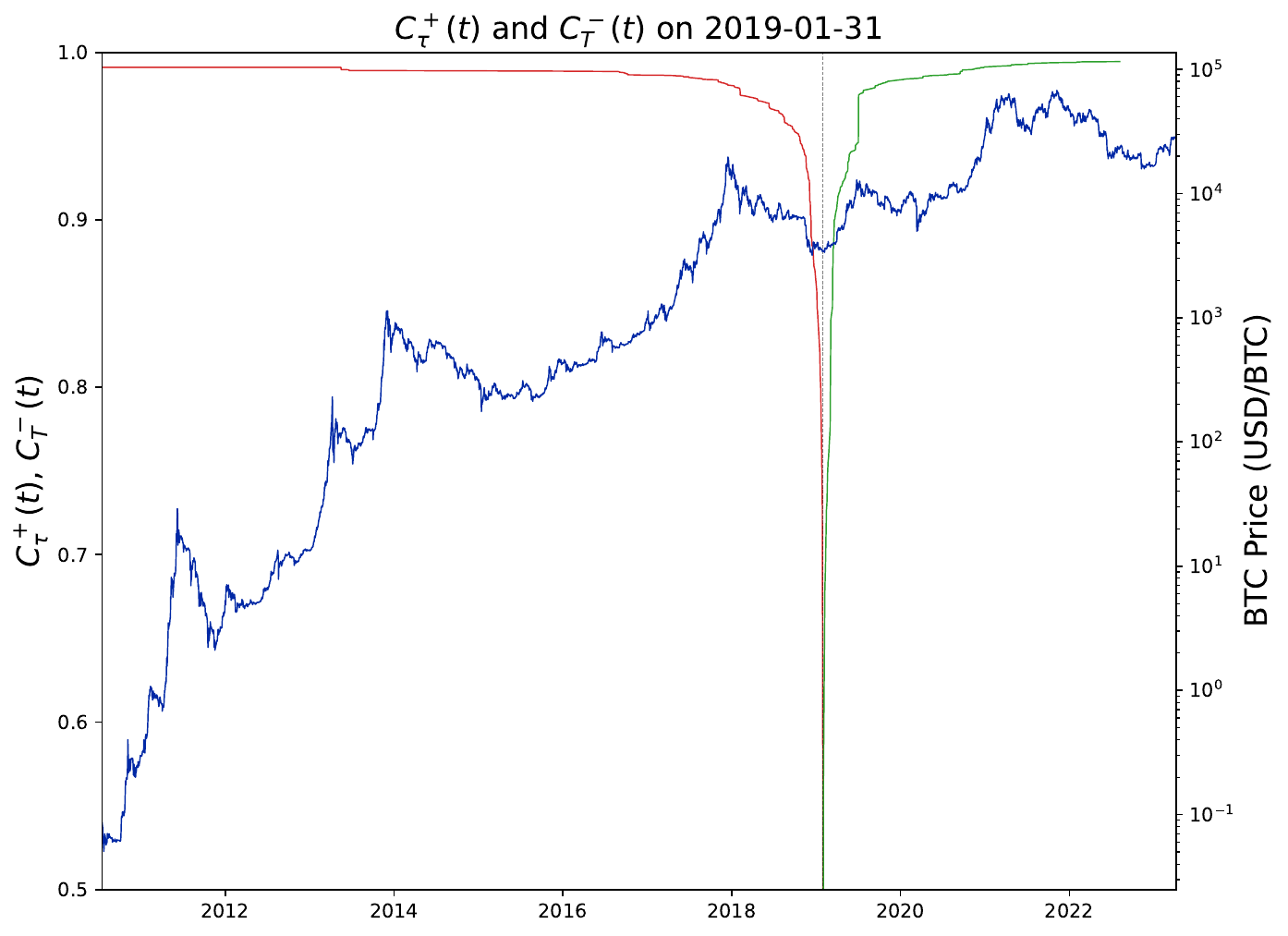}
				\label{f:dep-dep-subfig6}}
			\caption{\label{f:dep-dep} Transaction in-flow $C_{\tau}^{+}(t)$ (red, left scale) defined by expression (\ref{hrwtgqa54}) as a function of $\tau$ and  
				transaction out-flow $C_{T}^{-}(t)$ (green, left scale)
				defined by expression (\ref{h8uiqa54}) as a function of $T$, for the six different analysis times $t \in \{t_i, i=1, ..., 6\}$ given in Table \ref{t:times}. 
				These six times are each shown as the vertical dashed black lines separating $C_{\tau}^{+}(t)$ and $C_{T}^{-}(t)$.
				The blue line is the bitcoin price (right scale).
			}
		\end{figure}
		
		We now introduce the ``$t \to T$ transaction flow fraction'' $D_{T}^{-}(t)$, which mirrors
		the ``$\tau \to t$ transaction flow fraction'' $D_{\tau}^{+}(t)$, in the sense of being the fraction of
		bitcoins born at $t$ and exchanged at some later time $T > t$:
		\begin{equation}\label{eq:demand}
			D_{T}^{-}(t) := \dfrac{n_{t}(T-1)-n_{t}(T)}{V(t) + S(t)}	\quad\quad\quad\quad \forall~ T > t~.
		\end{equation}
		Note that, since $T>t$, $D_{T}^{-}(t)$ is not knowable causally at time $t$, in contrast to $D_{\tau}^{+}(t)$.
		It obeys the normalization condition
		\begin{equation}\label{eq:d-sum}
			\sum_{T > t}D_{T}^{-}(t) = 1 ~.
		\end{equation}
		To derive (\ref{eq:d-sum}), we write 
		\begin{eqnarray}
			\left(V(t)+S(t)\right) \sum_{T > t}D_{T}^{-}(t) & =& \sum_{T > t} \left( n_{t}(T-1)-n_{t}(T)\right)  \nonumber \\ 
			& =&  n_{t}(t) - \lim_{T\rightarrow\infty}n_{t}(T)~.
		\end{eqnarray}
		Assuming that all bitcoins born at $t$ eventually ``die'', i.e. are finally exchanged in a transaction at a future time $T$
		however distant in the future, this implies $\lim_{T\rightarrow\infty}n_{t}(T)  =  0$, and using \eqref{eq:n-t-t}, 
		this obtains identity (\ref{eq:d-sum}).
		This assumption can be relaxed by accounting for the number of bicoins that are stuck inside wallets forever and will not be exchanged ever again. Calling this residual number of bitcoins bought at time $ t $ as 
		\begin{equation}
			\lim_{t\rightarrow\infty}n_{t}(\tau) = n^*_{t},
		\end{equation}
		equation \eqref{eq:d-sum} becomes
		\begin{equation}\label{eq:d-sumwrtbt}
			\sum_{T > t}D_{T}^{-}(t) = \dfrac{n^*_{t}}{V(t)+S(t)} := D_{t}^{*-} ~.
		\end{equation}
		The two identities (\ref{eq:d+sum} and (\ref{eq:d-sum}) simply express the conservation of bitcoins,
		when neglecting lost wallets and forgotten passwords.
		
		Given the definition of the ``$\tau \to t$ transaction flow fraction'' $D_{\tau}^{+}(t)$ defined by expression (\ref{eq:supply})
		and of the ``$t \to T$ transaction flow fraction'' $D_{T}^{-}(t)$ defined by expression (\ref{eq:demand}), it is convenient
		to introduce the corresponding running cumulative transaction flow fractions as respectively
		\begin{equation}
			C_{\tau}^{+}(t) : = \int_{\tau}^{t}  D_{x}^{+}(t) dx~,
			\label{hrwtgqa54}
		\end{equation}
		\begin{equation}
			C_{T}^{-}(t) : = \int_{t}^{T}  D_{y}^{-}(t)dy~,
			\label{h8uiqa54}
		\end{equation}
		$C_{\tau}^{+}(t)$ is the fraction of the total transaction flow occurring at time $t$ contributed by the
		cohort of bitcoins with ages between $t-\tau$ and $t-\Delta t$.
		$C_{T}^{-}(t)$ is the fraction of the total transaction flow occurring between time $t$ and time $T$ contributed by the
		cohort of bitcoins that were born a time $t$. We can picture $C_{\tau}^{+}(t)$ as the total in-flow of transactions of bitcoins no older than $t-\tau$
		to the point time $t$ while $C_{T}^{-}(t)$ is the out-flow of transactions out of point time $t$ up to time $T$.
		
		Figure \ref{f:dep-dep} shows $C_{\tau}^{+}(t)$ defined by expression (\ref{hrwtgqa54})
		as a function of $\tau$ and  $C_{T}^{-}(t)$ 
		defined by expression (\ref{h8uiqa54}) as a function of $T$, 
		for the six different analysis times $t \in \{t_i, i=1, ..., 6\}$ given in Table \ref{t:times}.
		The most striking observation is the singular-like shape of the in-flow $C_{\tau}^{+}(t)$ and out-flow $C_{T}^{-}(t)$ 
		around the analysis times $t$, expressing the fact that most of the trading volume originates from short-term transactions.
		One can observe the presence of sharp jumps, corresponding to a large amount of bitcoin transactions at specific times. 
		These jumps correspond to important events. In Panel \ref{f:dep-dep-subfig4} for instance, there is a large sell-off 
		of approximately 4\% of all coins bought at $ t= $15 Jan. 2015 around the peak of a bubble on 21 Dec. 2017. 
		In panels \ref{f:dep-dep-subfig1} and \ref{f:dep-dep-subfig2}, one can infer that there remains about  
		$1\%$ of bitcoins born at the time $t$ of analysis of the corresponding panel that are still held and have not been transacted after five years.
		
		\begin{figure}[ht]
			\centering
			\subfloat[Subfigure 1 list of figures text][Time-averaged $D_{\tau}^{+}(t)$  ]{
				\includegraphics[draft=false,width=0.48\textwidth]{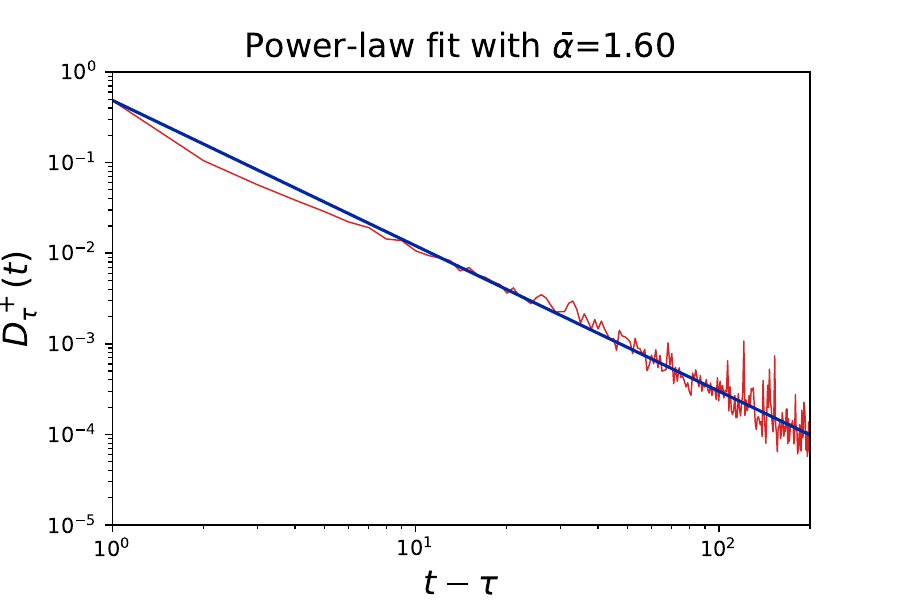}
				\label{fig:subfig1}}
			\subfloat[Subfigure 2 list of figures text][Time-averaged $D_{T}^{-}(t)$]{
				\includegraphics[draft=false,width=0.48\textwidth]{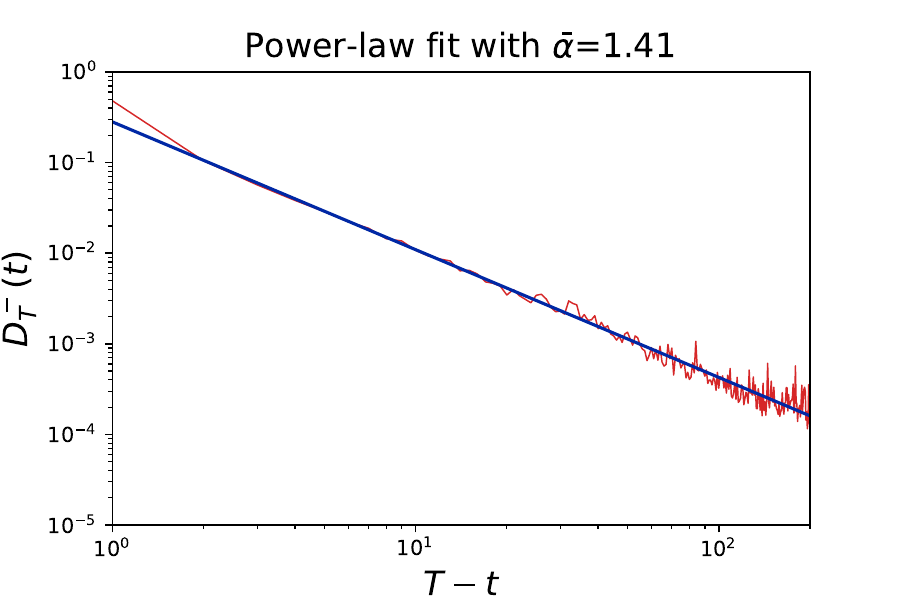}
				\label{fig:subfig2}}
			\caption{\label{fig:exp-dep-dep-pl} The time-averaged  ``$\tau \to t$ transaction flow fraction'' $D_{\tau}^{+}(t)$ (left panel) and time-averaged 
				``$t \to T$ transaction flow fraction'' $D_{T}^{-}(t)$ (right panel)
				are constructed analogously to the procedure described in figure \ref{f:exp-stp} by calculating $D_{\tau}^{+}(t)$ and $D_{T}^{-}(t)$ 
				for many different $t$'s, then translating them to the same origin of time and then averaging. 
				The averaging step is performed by sampling with the resolution $\Delta t=1$ week all times $t \in [2013.1,2022.8]$. Both variables are well-fitted by 
				power laws, qualified as straight lines in these log-log plots, with exponents respectively equal to $1.60$ and $1.41$.}
		\end{figure}
		
		The shapes of the in-flow $C_{\tau}^{+}(t)$ and out-flow $C_{T}^{-}(t)$ depend on the bitcoin price, its price growth rate and its volatility at $t$.
		For instance, in both Panel \ref{f:dep-dep-subfig1} and \ref{f:dep-dep-subfig3}, a large part of the bitcoins bought at the earlier peak 
		could still be profitably sold and are thus 
		exchanged fast throughout the growth of the bitcoin price. However, the bitcoins bought at the late 2013 peak are exchanged at a much slower rate over time. 
		While about 70\% of all bitcoins bought at the peak were already sold at the beginning of 2014 at realized losses, some other investors have been holding their losing positions. Presumably, from a psychological point of view, they do so in order not to realize these losses, a behavior that we have already referred to as the disposition effect.
		Panel \ref{f:dep-dep-subfig3} provides another interesting observation, namely the tendency for some investors to take profits too early. 
		This can be seen from the shoulder or near flat plateau in the dependence of the in-flow $C_{\tau}^{+}(t)$ as a function of $\tau$.
		There is a large jump of about 20\% in the out-flow $C_{T}^{-}(t)$, occurring in the early beginning of 2017. In other words, one jump occurred about two-thirds through the price growth of the major 2017 bubble as an early ``take profit'' trade.
		The other jump corresponds to the final price crash in the drawdown that ended the 2015 bubble. 
		Around the late stage of the 2013 bubble, both jumps correspond to sell-offs of bitcoins leading to losses. 
		These bitcoins are either sold too early before the bubble peaked or too late, that is, already in the drawdown after the bubble peak. 
		Another example of this behavior is shown in Panel \ref{f:dep-dep-subfig4}: bitcoins are bought around a local minimum of the price at the start of 2015; As a few minor bubble peaks occur throughout 2015-2017, close to 95\% of all these ``cheaply" bought bitcoins have already been sold at the start of 2017, just when the actual strongest phase of the bubble growth had just begun. These investors thus missed the opportunity to make the biggest gain.
		This illustrates again the other side of the disposition effect, namely the tendency to sell winning assets too early.
		
		Another interesting quantity is the percentage of bitcoins that ``survive" from one major bubble event to the next 
		major bubble event. For instance, only about 2-3\% of bitcoins bought at the peak of the late 2013 bubble (Panel \ref{f:dep-dep-subfig3}) survive until the peak of the 2017 bubble. Thus, only a very small fraction of investors who bought at the high price of the late 2013 peak managed to withstand the pressure of selling their coins throughout 2017 to finally end with a profit. Comparing the lifetime of these coins to the ones in Panel \ref{f:dep-dep-subfig4} which were bought around the 2015 low, this shows that only about 2-3\% of these coins survived until the bubble peak in 2017. This reflects the fact that the corresponding bitcoins were bought at a much better price. Thus, they were in a profit position much earlier, and more than 90\% of them were already sold in mid 2016. 
		
		Another example of poor trading performance can be observed in Panel \ref{f:dep-dep-subfig6} by looking at the in-flow $C_{\tau}^{+}(t)$  (red). 
		There is a large growth rate of the in-flow $C_{\tau}^{+}(t)$ around the first strong crash ending the 2017 bubble. 
		This means that a fraction of the investors who bought into the ``falling knife" after the bubble peak finally sold their coins exactly 
		around the price bottom in 2019. In other words, they sold at the worst possible time when their realized loss was maximized. 
		
		From the six panels of figure \ref{f:dep-dep}, one can almost feel unraveling the greed, enthusiasm, fear, and pain connected to the ups and downs of different price regimes and the corresponding trading decisions. The trading activity on the blockchain is clearly dominated by short-term activity but 
		is dependent on the history of the transaction entry prices. Also, there is the ubiquitous presence of some 
		traders who are badly performing as they buy for instance at a local peak, sell too early in the gain domain, and realize their profits at the wrong moment or too late. 
		
		Underlying the quirks and twists of investor trades revealed by the irregular shapes of the in-flow $C_{\tau}^{+}(t)$ and out-flow $C_{T}^{-}(t)$,
		a universal power law dependence is revealed in figure \ref{fig:exp-dep-dep-pl}. Averaging
		the  ``$\tau \to t$ transaction flow fraction'' $D_{\tau}^{+}(t)$ and the ``$t \to T$ transaction flow fraction'' $D_{T}^{-}(t)$
		over the set of times $t \in [2013.1,2022.8]$ with sampling step $\Delta t=1$ week, figure \ref{fig:exp-dep-dep-pl} shows 
		power law fits of good quality, qualifying the following dependence:
		\begin{equation}
			\label{eq:demqreret1gand}
			D_{\tau}^{+}(t) :=  {1 \over {\bar \alpha}_+ +1} \cdot  {\Delta t^{{\bar \alpha}_+ -1} \over (t-\tau)^{{\bar \alpha}_+}}~~~~~{\rm for}~\tau \leq t-\Delta t
		\end{equation}
		such that 
		\begin{equation}
			C_{\tau}^{+}(t) : = \int_{\tau}^{t} dx  {1 \over {\bar \alpha}_+ +1} \cdot  {\Delta t^{{\bar \alpha}_+ -1} 
				\over x^{{\bar \alpha}_+}} = 1 - {1 \over (t - \tau)^{{\bar \alpha}_+ -1}},
			\label{hrwth221a54}
		\end{equation}
		and
		\begin{equation}
			\label{eq:87jand}
			D_{T}^{-}(t) :=  {1 \over {\bar \alpha}_- +1} \cdot  {\Delta t^{{\bar \alpha}_- -1} \over (T-t)^{{\bar \alpha}_-}}~~~~~{\rm for}~ T \geq t-\Delta t
		\end{equation}
		such that 
		\begin{equation}
			C_{T}^{-}(t) : = \int_{t}^{T} dy {1 \over {\bar \alpha}_- +1} \cdot  {\Delta t^{{\bar \alpha}_- -1} \over y^{{\bar \alpha}_-}} = 1 - {1 \over (T - t)^{{\bar \alpha}_- -1}}~.
			\label{hy2h221a54}
		\end{equation}
		The power law fits shown in figure \ref{fig:exp-dep-dep-pl} gives ${\bar \alpha}_+ = 1.6$ and  ${\bar \alpha}_- = 1.4$.
		Being smaller than $2$, the exponents imply that the mean holding time of bitcoins involved in transactions is theoretically infinite.
		The values of these exponents are close to $1.5$, which is the expected value from priority queuing theory \cite{Grinstein1,Grinstein2,saichevsor09}.
		The corresponding intuitive interpretation is that the transaction flow fractions are proportional to distributions of 
		waiting times from the last transaction to the next. In priority queuing theory, the waiting times occur as a 
		competition between a flow of incoming tasks (taken in a general sense of something to be done, either some work to perform,
		an action to take and so on) with different priorities or sense of urgency and a rate of performing these tasks.
		In general, we tend to function in a regime where the rate of performing tasks is no larger than the rate of incoming tasks, 
		so that we tend to never be idle. In this regime, the distribution of waiting time exhibits an intermediate asymptotic power law
		with exponent $1.5$  \cite{Grinstein1,Grinstein2,saichevsor09}.

		\pagebreak
		\clearpage
		
		\section{Multifractality and Multiscaling Analysis}
		
		\subsection{Multiscaling in the Distribution of Bitcoin Holding Times \label{wrtnthbgrvqf}}
		
\begin{figure}[ht]
			\centering
			\includegraphics[draft=false,width=1\linewidth]{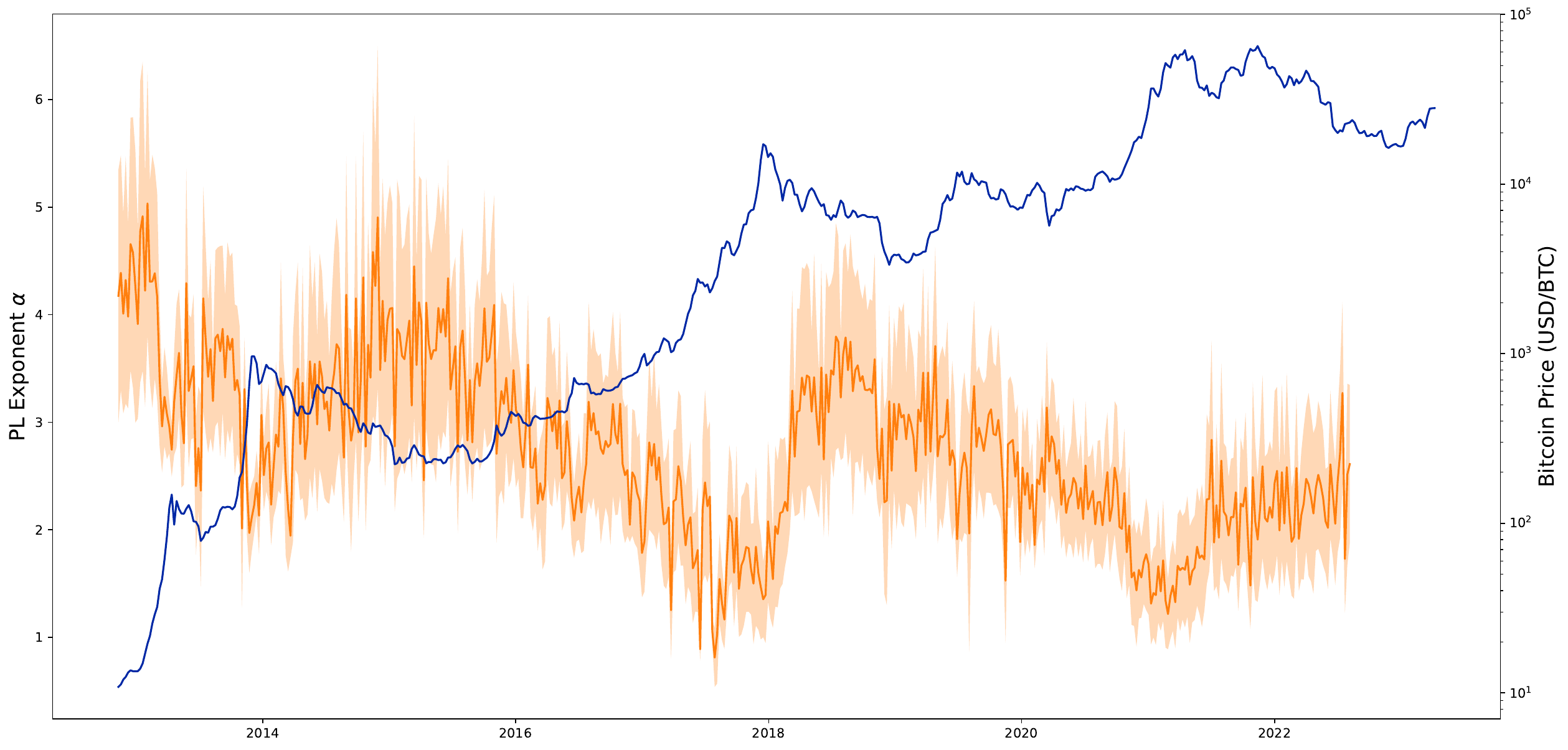}
			\caption{Time dependence of the power law exponent $\alpha_t$ for $t \in [2013.1,2022.8]$.
				The mean value of $\alpha_t$ is shown with the orange curve and its $\pm 1$ standard deviation bounds 
				are shown with the red shades. The standard deviation of $\alpha_t$ is obtained from the set of 
				exponents obtained in different intervals $[1,...,z_{max}]$ of the abscissa of the log-log plot at each time $t$, 
				where the upper boundary $z_{max}$ is swept
				as before in integer steps from $125\Delta$ to  $200\Delta t$, where $\Delta t =1$ week, amounting to 76 nested calibrations.}
			\label{alpha_series}
		\end{figure}

		Averaging over all time periods in our database provided evidence for a power law distribution
		of holding times shown in figure \ref{f:exp-stp}. 
		However, we have also observed in  figure \ref{f:stp-pl} significant variations of the value of the power law exponent $\alpha$
		for the six special times  $\{t_i, i=1, ..., 6\}$ given in Table \ref{t:times}.
		We now explore further the time dependence of the exponents $\alpha$ of the power law dependence of
		the fraction $\pi_{z}(t_i)$ of bitcoins as a function of age $z$.
		We extend the analysis given in figure \ref{f:stp-pl} to 
		all times  $t \in [2013.1,2022.8]$ and show in figure \ref{alpha_series} the dependence of $\alpha_t$
		as a function of time $t$. An estimation of the error bars on the $\alpha_t$'s is obtained from the set of 
		exponents obtained in different intervals $[1,...,z_{max}]$ of the abscissa of the log-log plot at each time $t$, 
		where the upper boundary $z_{max}$ is swept
		as before in integer steps from $125\Delta t$ to  $200\Delta t$, where $\Delta t =1$ week, amounting to 76 nested calibrations.
		
		The most important observation in figure \ref{alpha_series} is that $\alpha_t$ varies 
		significantly from low values as small as $0.3$ to high values as large as $6$.
		The existence of seemingly robust plateaus of $\alpha_t$ at very different levels seems to rule out the 
		hypothesis that $\alpha_t$ is constant. Thus, it is likely that the power law shown in figure \ref{f:exp-stp},
		obtained as the result of averaging over different power laws at different times, is insufficient to account
		for the rich structure of fluctuations of the holding time distributions.
		
\begin{figure}[ht]
			\centering
			\includegraphics[width=0.8\linewidth]{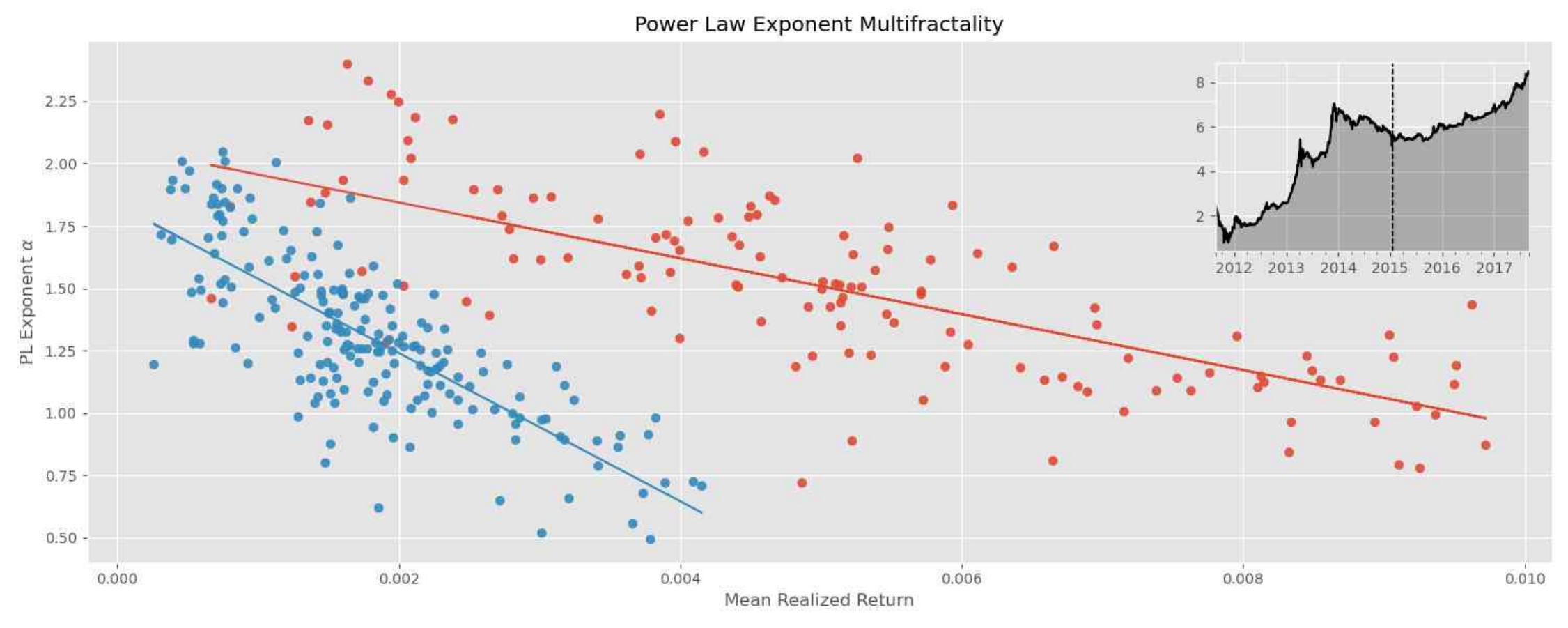}
			\caption{\label{f:alpha-multifractal} Scatter plot of $\alpha_t$ as a function of the mean 
				realized return defined in expression (\ref{log-ret-day}).  The mean realized return is the upper mean of the 
				Profit-and-Loss distribution shown in figure \ref{f:rel-ret}. 
				The plot shows two clusters of data that are each approximately described by a linear fit.  
				The red (blue) cluster represents all data before (after) 14$^{th}$ Jan 2015 (date of Bitcoin's lowest value between 2014 and 2018).
				The inset shows the bitcoin price and the date 14$^{th}$ Jan 2015  separating the two regimes is indicated by the vertical dashed line.
				The correlation coefficient corresponding to the linear regression of $\alpha_t$ as a function of realized return 
				over the first period before 14$^{th}$ Jan 2015 (blue dots) is $-0.6$.
				For the second period after 14$^{th}$ Jan 2015, the correlation coefficient is $-0.25$.
			}
		\end{figure}

		The second qualitative observation is that large values of $\alpha_t$ are found around troughs and 
		in the time periods following them, while small values of $\alpha_t$ tend to be associated with price peaks.
		The existence of a relationship between $\alpha_t$ and realized return is demonstrated quantitatively in 
		figure \ref{f:alpha-multifractal}, where $\alpha_t$ is plotted as a function of the mean 
		realized return defined in expression (\ref{log-ret-day}). The data is clearly divided into two clusters,
		the blue (red) cluster corresponding to data before (after) 14$^{th}$ Jan 2015 (date of Bitcoin's lowest value between 2014 and 2018).
		One can observe a negative correlation between $\alpha_t$ and realized return,
		with a cross-correlation of  $-0.6$ for the data before 14$^{th}$ Jan 2015 and $-0.25$ after 14$^{th}$ Jan 2015.
		Intuitively, these observations are logically coherent, as the larger the realized return, the smaller the exponent $\alpha_t$ of the power law distribution 
		of holding times, and thus the broader is the range of holding times. In other words, larger returns catalyze a
		broader range of behaviors, exemplified by a fat-tailed distribution of holding times. In contrast, for small or negative realized returns,
		the distribution of holding times is much narrower, indicating a more homogenous investment behavior.
		Identifying the specific mechanism and developing a theoretical model accounting quantitatively for these observations remain a challenge for the future.
		
		\pagebreak
		\clearpage

		\subsection{Multifractality in the Transaction Volume Domain}
		
		The previous subsection \ref{wrtnthbgrvqf} has documented the existence of a kind of multiscaling 
		in the dynamics of bitcoin transactions in the fact that the power law exponent $\alpha_t$ of the distribution of holding times is
		changing with time and is correlated with the realized bitcoin return. While being reminiscent
		of multifractality in the time domain \cite{MRW-time03,ouillon2005magnitude, sornette2005multifractal, ouillon2009multifractal, tsai2012new,nandan2021global},
		the presented behavior of $\alpha_t$ is not a definite proof of multifractality, which would require to construct
		the multifractal moment function and corresponding spectrum (see e.g. chapter 5 in \cite{sornette2006critical}, \cite{jiang2019multifractal}). 
		
\begin{figure}
			\centering
			\includegraphics[width=1\textwidth]{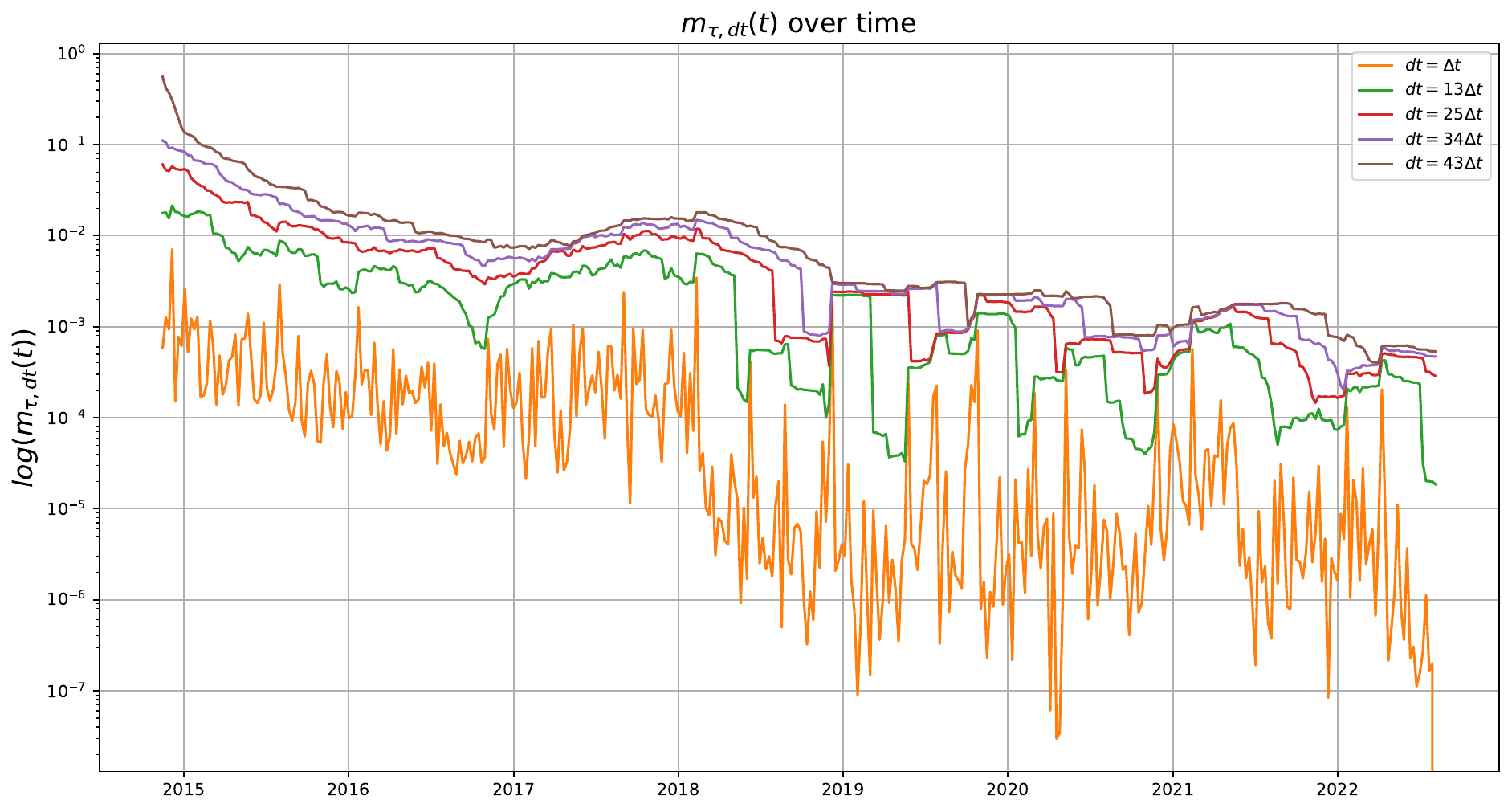}
			\caption{\label{f:m-dt}: Time dependence of the local measure $m_{\tau,dt}(t)$ defined by expression (\ref{eq:m-dt-4}) with (\ref{eq:vfd})
				for different values of the time interval $dt$ indicated in the inset. Here, $\tau$=2014-1-11 and $\Delta t = 1$ week, $dt$ plays the role of the size of the ``box'' in the standard box-counting approach of multifractality.
			}
			\label{fig:subfig_m_dt}
		\end{figure}

		We thus explore the possible existence of multifractality in the bitcoin dynamics by searching
		for the existence of fractal measures of local fractal sets.
		Here, we propose a natural measure defined as the normalized number of bitcoins exchanged at a given time
		\begin{equation}\label{eq:vfd}
			v_\tau(t):=\frac{n_\tau(t-\Delta t)-n_\tau(t)}{V(t)}~.
		\end{equation}
		The numerator in Eq.(\ref{eq:vfd}) describes the number of coins exchanged at time $t$ that last exchanged hands at time $\tau$. 
		The denominator $V(t)$ is the total number of bitcoins that have been transferred at time $t$, as defined by expression (\ref{eq:v-t}).
		By construction of $v_\tau(t)$, we have 
		\begin{equation}\label{eq:vfd-sum}
			\sum_{\tau<t}v_\tau(t)=1\quad\forall t
		\end{equation}
		For a fixed $\tau$, we introduce the local measure $m_{\tau,dt}(t)$ given by
		\begin{equation}\label{eq:m-dt-4}
			m_{\tau,dt}(t):=\int_{t-dt}^{t}v_\tau(t')dt'~,
		\end{equation}
		which represents the fraction of all exchanged bitcoins from $t-dt$ to $t$ that were last exchanged at time $\tau$ in the past. 
		The time interval $dt$ plays here the role of the size of the ``box'' in the standard box-counting approach of multifractality.
		Figure \ref{fig:subfig_m_dt} shows the time series of $m_{\tau,dt}(t)$ for $dt / \Delta t = 1, 13, 25, 34$ and $43$, where
		$\tau$=2014-1-11 and $\Delta t = 1$ week. One can observe approximate stationary time series (in particular after 2018) with large fluctuations in particular 
		on the downside.
		
\begin{figure}[ht]
			\centering
			\includegraphics[width=0.9\textwidth]{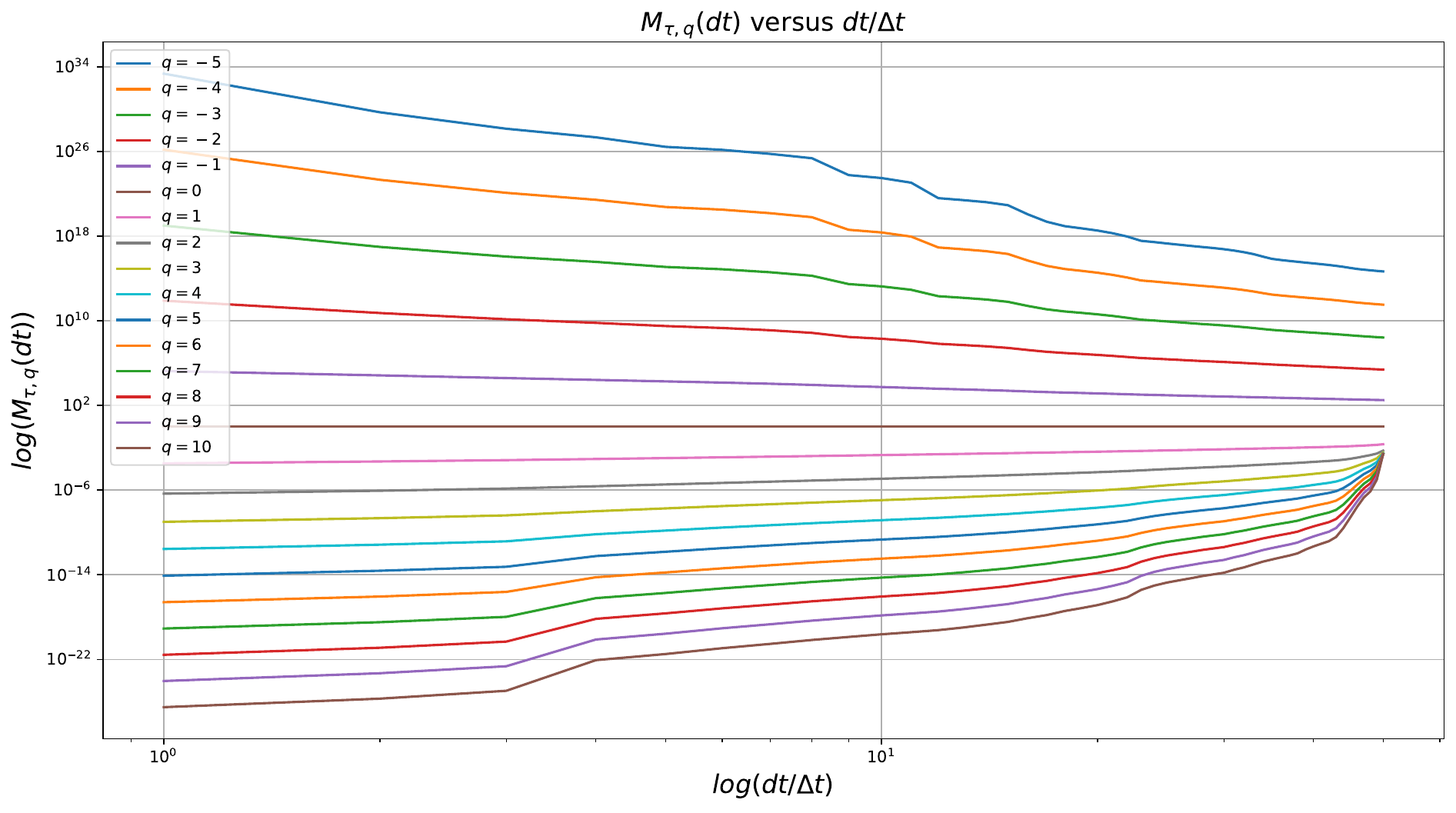}
			\label{fig:subfig6}
			\caption{\label{f:M-q} Log-log plot of $M_{\tau,q}(dt)$ defined by expression (\ref{eq:M-q}) 
				as a function of $dt$ for $dt / \Delta t =1$ to $50$, for different values of $q$ indicated in the inset.}
		\end{figure}

\begin{figure}[ht]
			\centering	
			\includegraphics[width=0.9\textwidth]{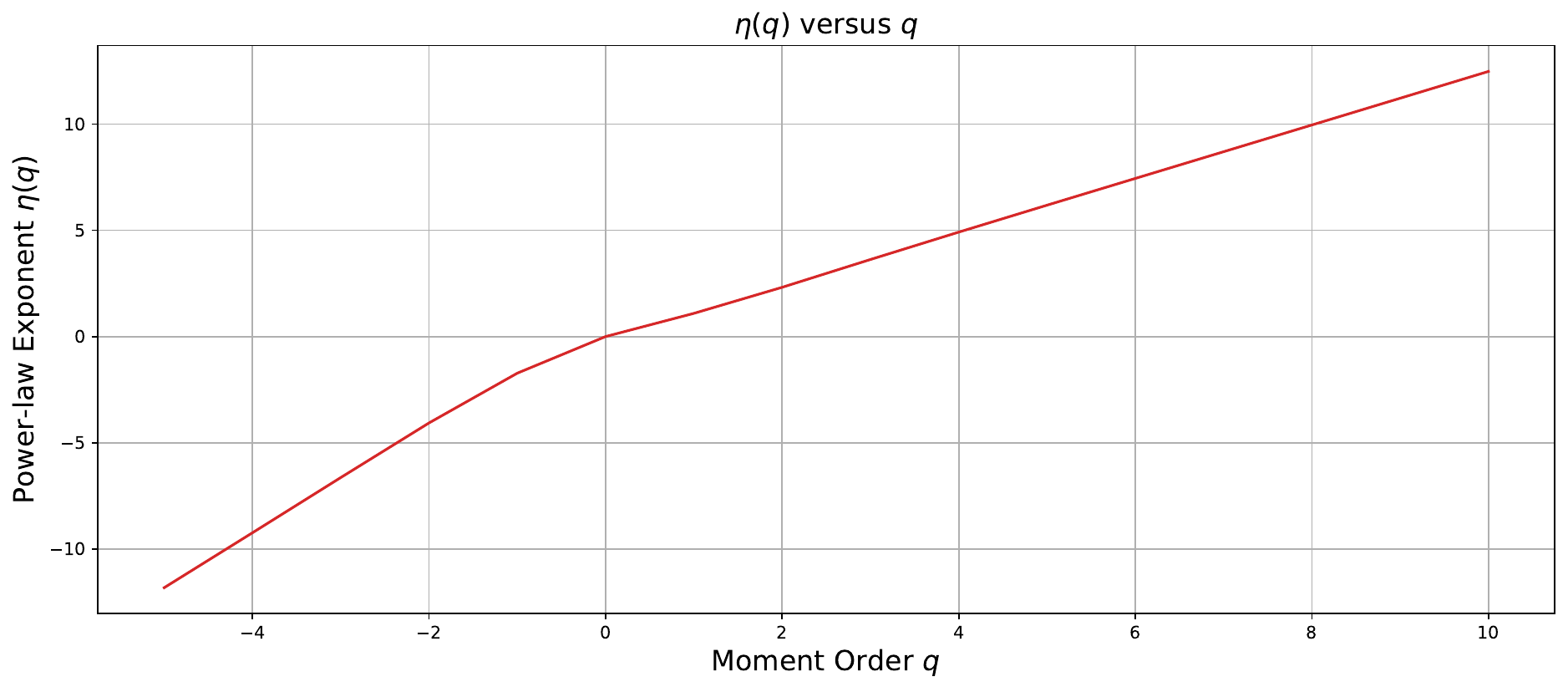}
			\label{fig:subfig6}
			\caption{\label{f:taus} Power law exponent $\eta(q)$ defined in expression (\ref{trh3tgrv2qf} as 
				a function of the order $q$ of the moment $M_{\tau,q}(dt)$.) 
			}
		\end{figure}

		We then follow the standard procedure to test for multifractality (see e.g. chapter 5 in \cite{sornette2006critical}, \cite{jiang2019multifractal}). 
		For each $\tau$, we estimate the $q$-th moment 
		\begin{equation}\label{eq:M-q}
			M_{\tau,q}(dt):=E_t[(m_{\tau,dt}(t))^q]~,
		\end{equation}
		where the expectation is performed over all times $t$ in the interval $[2013.1, 2022.8]$.
		We keep $\tau$=2014-1-11 fixed and estimate $M_{\tau,q}(dt)$ for $q\in[-5,-4,...,10]$.
		Figure \ref{f:M-q} shows a log-log plot of $M_{\tau,q}(dt)$ as a function of $dt$ for $dt / \Delta t =1$ to $50$.
		One can observe an approximate power law scaling visualised by straight-line behaviors representing the dependence
		\begin{equation}
			M_{\tau,q}(dt) \propto dt^{\eta(q)}
			\label{trh3tgrv2qf}
		\end{equation}
		where $\eta(q)$ is a function of the moment-order $q$ describing the dependence of the scaling of the moment $M_{\tau,q}(dt)$ as a function of the time interval $dt$ on which 
		the measure $m_{\tau,dt}(t)$ is estimated.
		Multifractality is qualified by a nonlinear concave dependence of the exponent $\eta(q)$ as a function of $q$, which is shown in figure \ref{f:taus}.
		
		This quantification of the existence of multifractality is useful as a tool for characterizing the fluctuating dynamics of bitcoin transactions.
		However, the underlying mechanism remains to be deciphered. Here, multifractality is just another quantification tool to help classify the dynamics.

		\pagebreak
		\clearpage

		\section{Conclusion}
		
		We have presented a quantitative analysis of the holding and transaction properties of bitcoins, building on the unique
		availability of the full record of transactions recorded on the public ledger of Bitcoin.
		The record of time stamps and identities of individual bitcoins offers an unparalleled opportunity to study and analyze investors' behavior, 
		which are otherwise hidden in standard financial markets. The available data has allowed us to 
		document how bitcoin investors' behaviors change with bitcoin's price and with their holding time of bitcoins.  
		
		After providing a suitable mathematical formulation of the flow of bitcoin transactions, we have explained how to use the blockchain structure
		to extract unique information on holding times and flow of transactions.
		We have constructed the age distribution of bitcoins, defined at the time from the last transaction to the present.
		We have compared age distributions with price dynamics, in particular for six specific times corresponding to six
		important events in the life of Bitcoin (three peaks and three troughs of price). We have documented a clear
		dependence of the age distribution on past times and price levels.
		By classifying bitcoin holders into three categories (short-, medium- and long-term holders), 
		we find that different holders have different behaviors in reaction to price changes. 
		For example, if the bitcoin price increases, the long-term holders tend to sell their bitcoins, while the short-term holders tend to hold their bitcoins. 
		The short- and medium-holders have almost opposite behaviors, which means that the short- and medium-term holders 
		transfer to each other frequently. Short-term holders tend to be more active when the price decreases. When the price increases, 
		short-term holders tend to keep bitcoins and are less active. In contrast, the activity of medium-term holders is approximately synchronized with the ups and downs of bitcoin price. 
		
		We have then investigated the distribution of book-to-market returns (unrealized profit and loss), which reflects the investors' behavior with price change
		and we have compared them to realized returns. We find that, prior to the peaks of the three major bubbles, the average unrealized loss is vanishing
		and there are distinct peaks in the average unrealized profit. At times of crashes, the average loss sharply increases. 
		We document significant differences between book-to-market and realized returns.
		In the gain domain, realized returns that are mostly positive turn out to be significantly smaller overall than the Book-to-Market returns.
		We also find that realised returns tend to be concentrated on specific values, 
		which can be interpreted as ``taking profit'' decisions of investors who have bought at a specific time in the past.
		
		We have investigated in detail the dependence of transaction flows on the holding times of bitcoins by investors.
		We find that the distribution of holding times is a power law extending over at least 200 weeks with an exponent 
		approximately equal to $0.87$. The distribution of holding times has thus a heavy-tailed structure since the exponent 
		is less than $1$. Taken at face value, this would imply an extremely wild regime, where the distribution is non-normalizable.
		There are thus significant non-ergodic effects that can be expected from this long-lived memory dependence.
		We also report significant sample-to-sample variations of the distribution of holding times, that can be best 
		characterized as multiscaling, with power-law exponents varying between $0.3$ and $6$ depending on time
		and bitcoin price regimes. 
		
		We have complemented this analysis of the distribution of holding times by the study of transaction flows, i.e., 
		what fraction of bitcoins are traded at a given time and that were born at some specific earlier time.
		We document again a very neat power law dependence of the time-averaged transaction flow fraction
		as a function of age, with an exponent close to $1.5$, a value compatible with priority queuing theory.
		
		We also document strong qualitative and quantitative evidence in many instances for the
		preponderance of the disposition effect in the Bitcoin Blockchain data, namely that
		investors tend to realize profits too early and at small gains while, on the contrary, they realize losses too late and thus incur large losses.
		
		Last, we document the existence of multifractality on the measure defined as the normalized number of bitcoins exchanged at a given time.
		Defining structure moments of this measure at different box counting scales and of different orders $q$ allows us to exhibit 
		the scaling behavior of the structure moments as a function of the time-box sizes, with exponents that have a nonlinear
		dependence as a function of the order $q$ of the moment.
		
		\vskip 0.5cm
		\noindent
		{\bf Declaration of competing interest}:
		The authors declare that they have no known competing financial
		interests or personal relationships that could have appeared to influence the work reported in this paper. 
		
		\vskip 0.5cm
		\noindent
		{\bf Data availability}:
		The datasets used and/or analyzed during this research are available from the public Bitcoin ledger.
		
		\vskip 0.5cm
		\noindent
		{\bf Acknowledgements}: A preliminary study was performed in collaboration with Jan-Christian Gerlach. We thanks Cyrille Grumbach for discussions.
		The work of DS is partially supported by the Feature Innovation Project of Colleges and Universities in Guangdong Province (2020WTSCX082), Shenzhen Science and Technology Innovation Commission Project (grant no. GJHZ20210705141805017 and grant no. K23405006), and the Center for Computational Science and Engineering at Southern University of Science and Technology.
		
		\pagebreak
		\clearpage

		\section{Appendix}
		
		\subsection{Coinbase Transactions}\label{sec:coinbase}
		
		The first transaction of any block is always the so-called coinbase transaction that the miner of a block creates to obtain the block reward $ S(t) $ (see Eq.\eqref{eq:N-t-rec})\footnote{The initial block reward was 50BTC. It is halved every 210000 blocks at the so-called ``halving time''. As the block mining times, the halving time is also uncertain but historically has taken place every four years. As the halving period is much longer than the time between two block creations, it can be predicted much more accurately.}, plus any fees accrued overall transactions in the block, which provide the incentive and remuneration for the mining job. The coinbase is commonly pictured as the ``Bitcoin mine''. It is successively depleted over time with the generation of each new block by miners. The initial finite size of the coinbase was 21 Million units of BTC, of which about 19.4 Million have already been depleted in August of 2022. The created coins are then distributed by the miners through transactions. Thus, any current BTC must have been released through a coinbase transaction. The sum of all currently circulating BTC is the total number of coins emitted by the coinbase up to now, as is formulated by the conservation equations in Section \ref{sec:model}.

		\subsection{Dataset and Preprocessing}
		
		\begin{figure}[ht]
			\centering
			\includegraphics[width=0.8\linewidth]{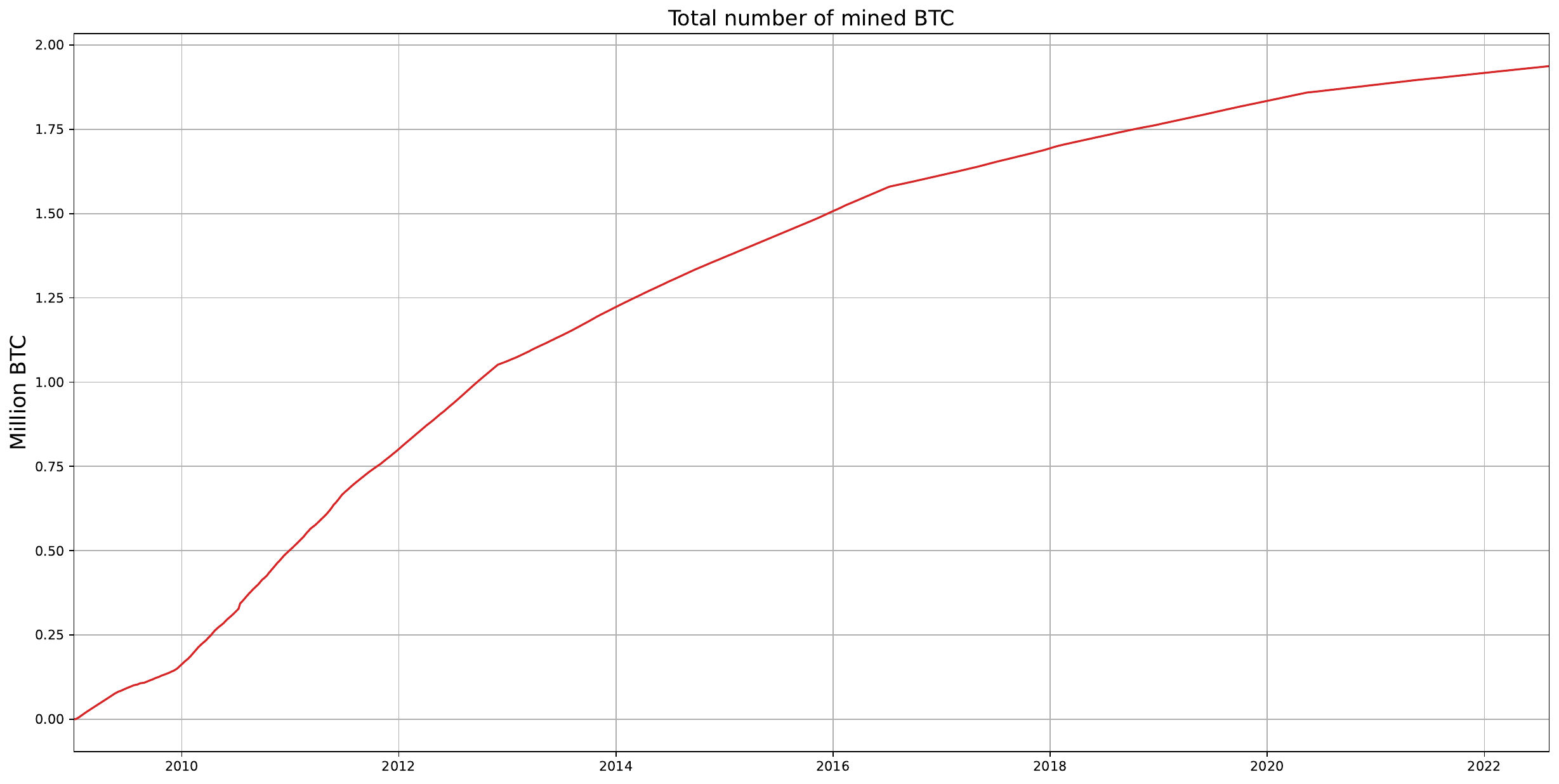}
			\caption{\label{f:number-mined-btc} The total number of BTC in existence (mined BTC) over time. This plot was constructed as a ``sanity check'' of the transaction processing algorithm. Summing up the entire age distribution $n_\tau(t)$ over $\tau\leq t$ yields the total number of existing coins $N(t)$ (red curve). This data accurately compares to online sources, such as blockchain.info.}
		\end{figure}
		
		\begin{figure}[ht]
			\centering
			\subfloat[Subfigure 1 list of figures text][]{
				\includegraphics[width=0.48\textwidth]{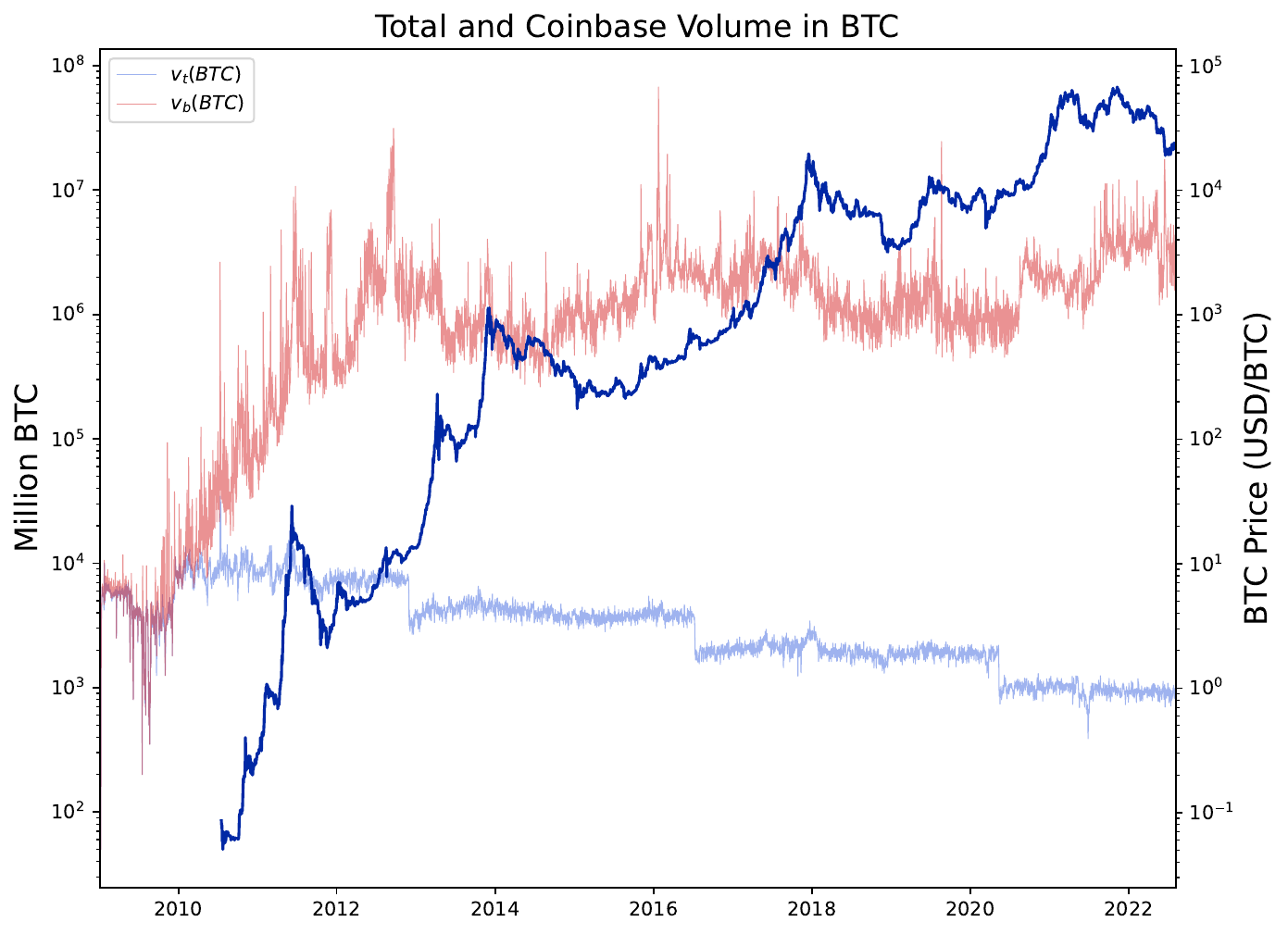}
				\label{fig:subfig1}}
			\subfloat[Subfigure 2 list of figures text][]{
				\includegraphics[width=0.48\textwidth]{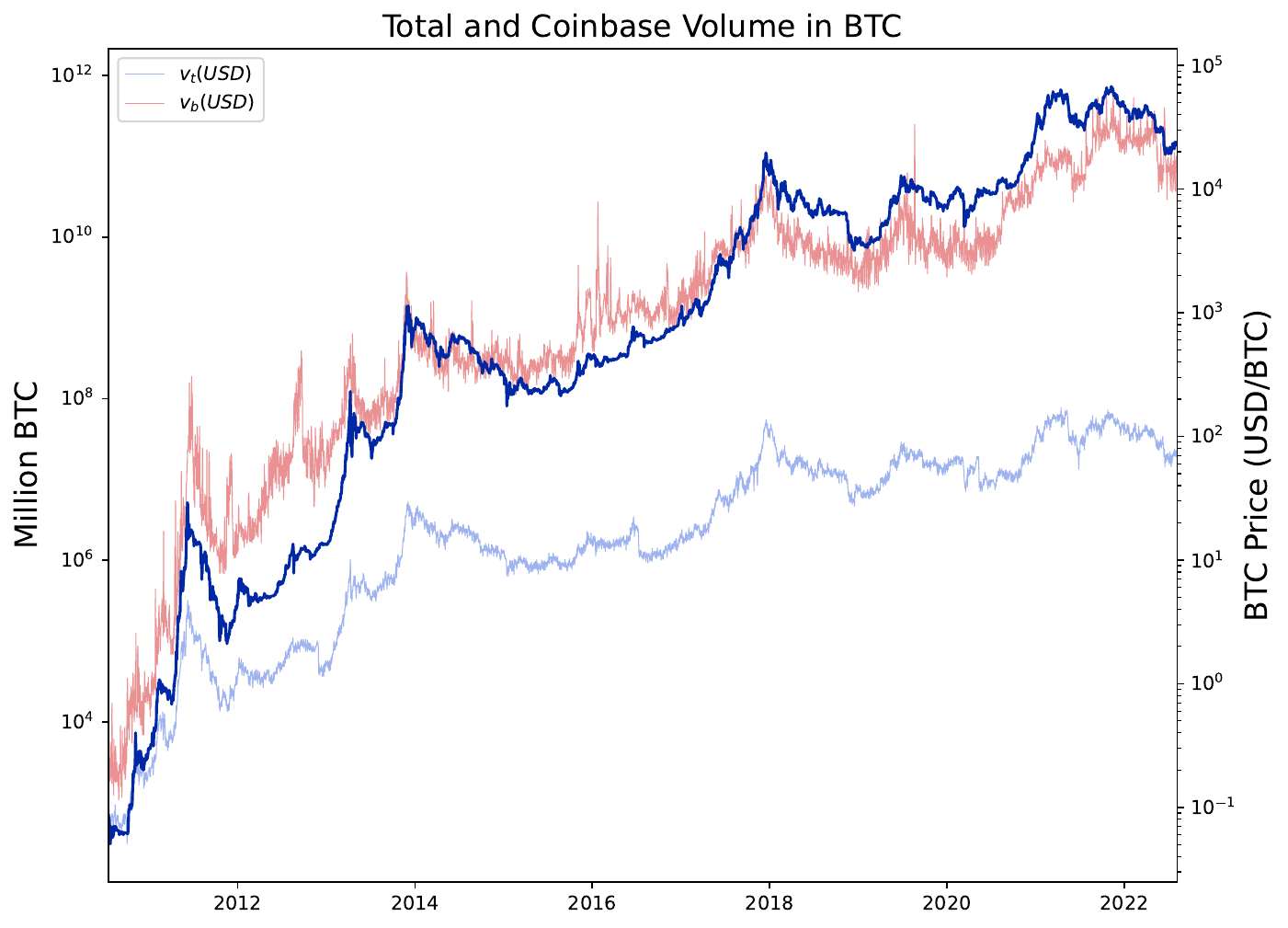}
				\label{fig:subfig2}}
			
			\caption{\label{f:volumes-btc-usd} The total volumes $V(t)$ (red) (see figure \ref{f:btc-age}) and the coinbase volume $v_c(t)$ (light blue) (i.e. the number of newly mined coins) per timestep expressed in BTC (Panel (a)) and US Dollars (Panel (b)). The entire analysis proceeds in timesteps of $\Delta t = 5$ days. In the left panel, the points of "halvings" can be observed throughout Bitcoins history, i.e. dates at which the mining block reward $R(t)$ halves, can be observed. At an average block mining time of $10$ minutes (as defined in the Bitcoin protocol), the theoretically expected number of coins that are emitted per timestep $\Delta t$ is $N_c(t)=144\Delta tR(t)$.}
			
		\end{figure}

		\clearpage
		\bibliographystyle{unsrt}
		\bibliography{reference.bib}
	\end{document}